 \documentclass[rivista]{cimento} 

\usepackage{amsmath}
\usepackage{graphicx}
\usepackage{latexsym}
\usepackage{rotating}
\usepackage{multirow}

\title{Optical atomic clocks}
\author{N.~Poli\from{unifi}\ETC,
C.~W.~Oates\from{nist}, P.~Gill\from{npl} \atque
G.~M.~Tino\from{unifi}\thanks{corresponding author:
Guglielmo.Tino@fi.infn.it}}

\instlist{\inst{unifi} Dipartimento di Fisica e Astronomia and
LENS, Universit\`a di Firenze\\ INFN Sezione di Firenze, Via
Sansone 1, 50019 Sesto Fiorentino (Firenze), Italy

\inst{nist} National Institute of Standards and Technology, Time
and Frequency Division, MS 847, Boulder, CO 80305, USA

\inst{npl} National Physical Laboratory (NPL), Teddington,
Middlesex TW11 0LW, United Kingdom}

\PACSes{ \PACSit{06.30.Ft}{Time and frequency}
\PACSit{06.20.fb}{Frequency standards} \PACSit{37.10.Jk}{Optical
cooling and trapping of atoms} \PACSit{37.10.Ty} {Ion trapping}}

\begin{document}

\maketitle

\begin{abstract}

In the last ten years extraordinary results in time and frequency
metrology have been demonstrated. Frequency-stabilization
techniques for continuous-wave lasers and femto-second optical
frequency combs have enabled a rapid development of frequency
standards based on optical transitions in ultra-cold neutral atoms
and trapped ions. As a result, today's best performing atomic
clocks tick at an optical rate and allow scientists to perform
high-resolution measurements with a precision approaching a few
parts in $10^{18}$. This paper reviews the history and the state
of the art in optical-clock research and addresses the
implementation of optical clocks in a possible future redefinition
of the SI second as well as in tests of fundamental physics.
\end{abstract}

\section{Introduction}\label{intro}

The optical atomic clock, that is, a laser whose frequency is
stabilized relative to that of an optical atomic transition,
represents a revolutionary step forward in the evolution of atomic
frequency and time standards
\cite{Hall2006,Hansch2006,Wineland2013Nobel}. For more than 60
years atomic frequency standards have played a critical role in
basic science, precision measurements, and technical applications.
During this period, the increasing need for more precise timing
and synchronization, for a range of applications including
navigation systems \cite{Dow2009}, telecommunications, VLBI
telescopes \cite{Normile2011}, and tests of fundamental physics
\cite{Rosenband2008} has demanded oscillators with higher
frequencies and higher performance. Because of the experimental
difficulties in counting high optical frequencies (several
hundreds of terahertz in the infrared/visible domains), atomic and
molecular optical standards have long been used mainly as length
standards, but only rarely as frequency references
\cite{Hollberg2005}. Before the advent of optical frequency combs,
the measurement of absolute optical frequencies was, in fact, a
formidable task, requiring resources available only at large-scale
laboratories
\cite{Evenson1972,Schnatz1996,Udem1997,Touahri1997,Bernard1999}.

\begin{figure}[t]\begin{center}
\includegraphics[width=0.8 \textwidth]{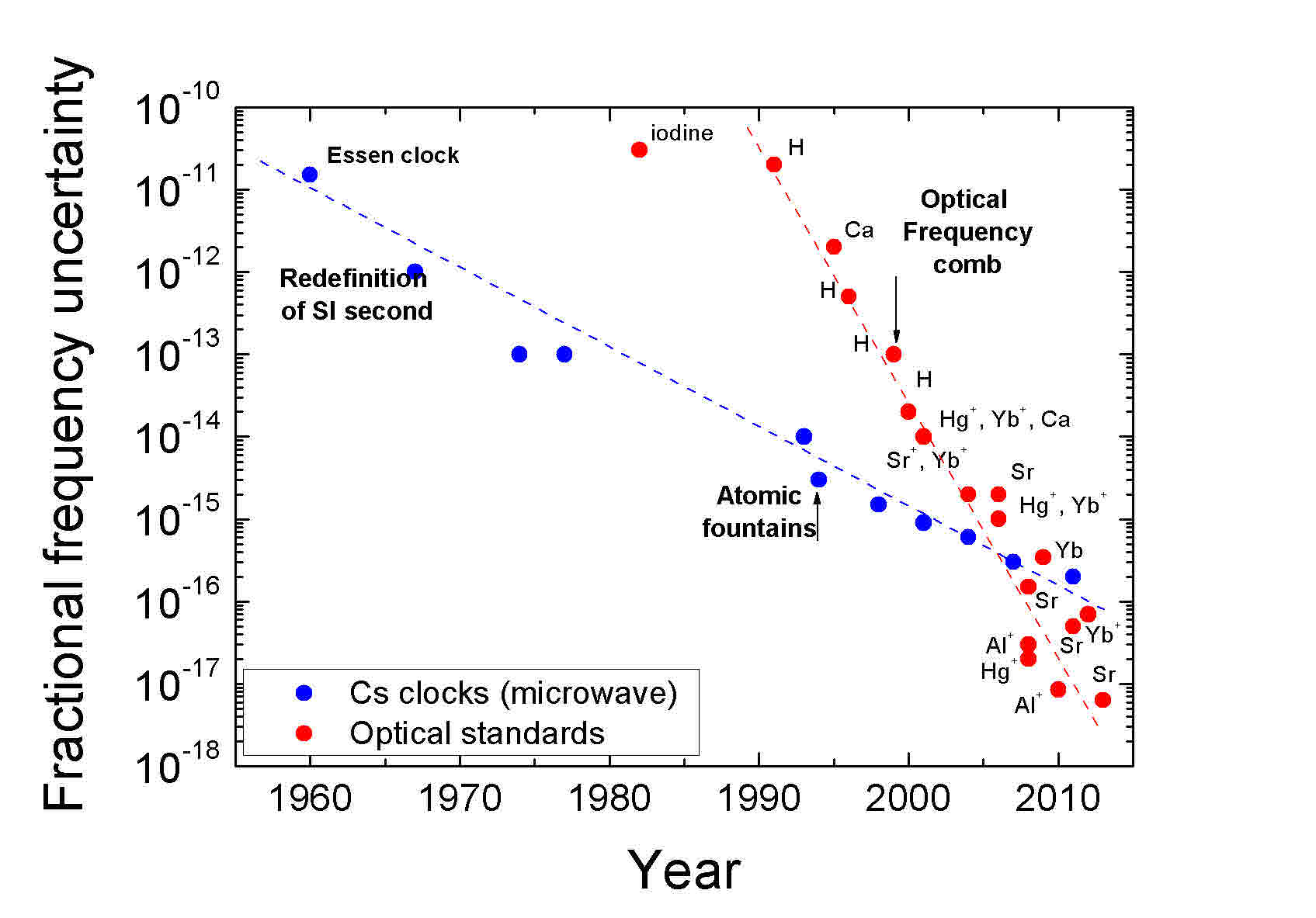}
\caption{Evolution of fractional frequency uncertainties of atomic
frequency standards based on microwave (Cs clocks)\cite{Bize2005}
and optical transitions. A fractional frequency uncertainty in the
$10^{-18}$ region have been reported for two optical clocks
respectively, the $^{27}$Al$^+$ single ion clock at NIST
\cite{Chou2010} and the $^{87}$Sr optical lattice clock at JILA
\cite{Bloom2013}.} \label{fig.accuracyatomicstandards}
\end{center}
\end{figure}

However, recent advances in the field of femto-second laser
technology, with the introduction of optical frequency combs, have
enabled the possibility to cover in a single step the gap between
optical frequencies and countable microwave frequencies
\cite{Udem1997,Udem1999, Udem1999a}.   As a result, the field of
optical frequency metrology has, in some sense, become possible.
Driven also by recent tremendous progress in the fields of atom
manipulation and precise optical frequency control, the field of
optical atomic clocks has, over the past decade, become a hot
research topic.  There are now dozens of labs worldwide working on
various versions of optical atomic systems, based on neutral atoms
or single trapped ions, and optical clocks have already
significantly surpassed their microwave counterparts in terms of
clock performance.  Neutral atom-based and ion-based clocks each
have their own advantages; neutral atom clocks benefit from their
large signal-to-noise ratio due to large atom numbers, while
single-ion clocks ``tick'' in a pristine environment, extremely
well-isolated from external perturbations
\cite{Wineland2013Nobel}.  In fig.
\ref{fig.accuracyatomicstandards} we see that there has been a
tremendous amount of activity and improvement in the field of
optical standards, including both neutral atom and ion clock
systems.  The uncertainty in the ticking frequency of these
standards (red circles) has been reduced by three orders of
magnitude in a little more than a decade of research to the point
where the best sources have absolute uncertainties below one part
in $10^{17}$, more than an order of magnitude below that of the
best Cs microwave clocks (blue dots) that presently define the SI
second.  Additionally, there is spin-off research underway
(especially that based on fs-laser combs), which takes advantage
of much of the technology that was first developed with optical
atomic clocks in mind.  We emphasize, however, that optical clocks
are still primarily research projects, less mature than Rb and Cs
microwave standards, which regularly deliver time to the Bureau
International des Poids et Mesures (BIPM).

In this review article, we will describe the science behind the
explosion of activity and progress shown in fig.
\ref{fig.accuracyatomicstandards} for the field of optical atomic
clocks, beginning with the physics that describes the inherent
advantages of working at optical frequencies, followed by a
description of the three key technologies that have revolutionized
this field.  We will then outline the physics that drives the
design and ultimately limits the performance of optical atomic
clocks, following this with a series of concrete examples of some
of the most successful clock systems.  To highlight the
possibilities enabled by this new frequency/timing technology, we
conclude with a description of how these clocks can be connected
to the outside world, the applications these clocks could enable,
and how such clocks may even one day be put in space to take
advantage of a micro-gravity environment \cite{Tino2007}.

\section{The Optical Clock Revolution}

\begin{figure}[t]\begin{center}
\includegraphics[width=\textwidth]{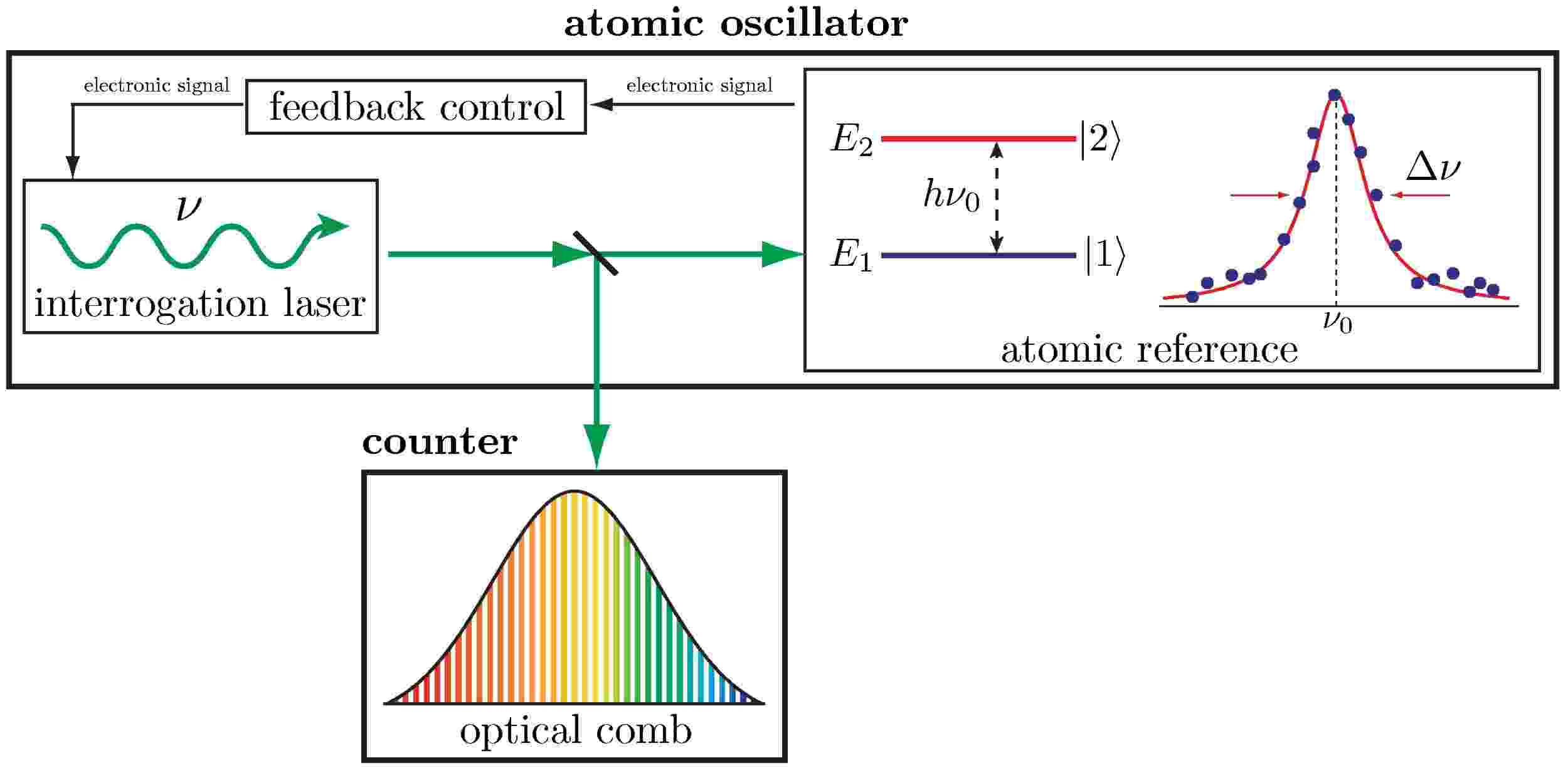}
\caption{Schematic view of an optical atomic clock: the local
oscillator (laser) is resonant with the atomic transition. A
correction signal is derived from atomic spectroscopy that is
fed back to the laser. An optical frequency synthesizer (optical
frequency comb) is used to divide the optical frequency down to
countable microwave or radio frequency signals.
\label{fig.OpticalClockScheme}}
\end{center}
\end{figure}

At the heart of any good clock is a regular oscillatory
phenomenon, whether it be the swinging of a pendulum, the
spring-driven oscillations of a watch, or the voltage-driven
oscillations of a quartz crystal. However, mechanical timepieces
tend to be quite susceptible to environmental effects such as
temperature changes, although ingenious designs have led to some
remarkable devices. Still the best performing clocks are those
that use carefully chosen atomic transitions to steer the
frequency of the oscillator. A typical atomic clock consists of an
oscillator, either microwave or optical (\emph{i.e.}, a laser)
whose frequency is forced to stay fixed on that of an atomic
resonance (see fig. \ref{fig.OpticalClockScheme}). One of the most
important parameters describing such a resonance is the atomic
line quality factor, $Q$, defined as the ratio of the absolute
frequency $\nu_0$ of the resonance to the linewidth of the
resonance itself $\Delta\nu$.

As we shall see, such resonances can have line $Q$'s many orders
of magnitude higher than the best mechanical systems, and they can
be isolated from environmental effects to a much higher degree.
Atomic clocks have the added benefit that atoms are universal, in
the sense that multiple clocks based on the same atomic transition
should have the same oscillation frequency, thereby offering a
degree of reproducibility not possible with mechanical devices.

For these reasons, atomic clocks have ruled the ultra-high
precision timing world for the past 50 years or so.  The famous
9.19 GHz hyperfine transition in Cs has served to define the SI
second since 1967, and as we see in fig.
\ref{fig.accuracyatomicstandards}, the absolute fractional
frequency uncertainty for this transition has been reduced by
about a factor of ten every decade.  We emphasize that this
remarkable rate of progress should not be taken lightly, as it has
been the result of dedicated efforts and several ingenious
advances, particularly in terms of atomic manipulation via laser
cooling techniques.  However, the present day state-of-the-art Cs
fountain \cite{Parker2010}, which uses clouds of laser-cooled
atoms that are tossed vertically through an interaction region, is
nearing its practical limitations  (astoundingly, the best Cs
systems are approaching uncertainties of $10^{-16}$ with a system
whose line $Q$ is only $10^{10}$).  For timing advances to
continue to move forward, however, a new technology is required,
and this is where optical clocks have come to the forefront.

To understand the advantage of going to a higher-frequency clock,
we must first examine more explicitly what it means for a clock to
be good.  There are two principal characteristics that we consider
when evaluating state-of-the-art clocks:  stability and
uncertainty.  Stability is a measure of the precision with which
we can measure a quantity (think of how widely scattered a group
of arrows fired at target might be), and is usually stated as a
function of averaging time since for many noise processes the
precision increases (\emph{i.e.}, the noise is reduced through
averaging) with more measurements.  The stability is usually set
by the combination of the inherent frequency purity of the
physical system (\emph{e.g.}, the line $Q$) and the
signal-to-noise ratio with which we can measure the system.  In
contrast, the (absolute) uncertainty for an atomic clock tells us
how well we understand the physical processes that can shift the
measured frequency from its unperturbed (``bare''), natural atomic
frequency (think of how off-centre our group of arrows might be).
Small absolute uncertainty is clearly an essential part of a good
primary frequency standard and requires extensive evaluation of
all known physical shifts (usually called ``systematic effects'').
We note that in order to be able to compare clocks with different
oscillation frequencies fairly, we usually express the two main
clock parameters, namely,  stability (or its inverse, instability,
as is commonly used in clock comparisons) and the absolute
uncertainty, in fractional units.

With this understanding of the two essential clock properties,  we
can now understand the advantages inherent in moving from the
microwave to the optical domain.  Let us first consider the
formula for clock instability, $\sigma_y$, in the regime where it
is limited by fundamental (as opposed to technical) noise sources,
such as atomic statistics based on the number of atoms
\cite{Itano1993},
 \begin{equation}
\sigma_y (\tau)\approx\frac{\Delta\nu}{\nu_0 \sqrt{N}}\sqrt{\frac{T_c}{\tau}},
\label{qpnstab}
\end{equation}
where $\Delta\nu$ is the spectroscopic linewidth of the clock
system, $N$ is the number of atoms or ions used in a single
measurement, $T_{c}$ is the time required for a single measurement
cycle, and $\tau$ is the averaging period  (see for example,
\cite{Hollberg2005}).  This formula can be understood fairly
intuitively, as $\Delta\nu/\nu_0$ is the inverse of the measured
line $Q$, and the remaining terms combine to yield the
signal-to-noise ratio as a function of averaging period.  As
expected, the instability is reduced for narrower lines
($\Delta\nu$) and shorter measurement cycles, and also benefits
from a higher signal-to-noise ratio ($\sqrt{N}$, known as the
atomic projection noise \cite{Itano1993}).  Moreover, in this
projection-noise regime, which may extend from seconds to hours or
even days and longer, the stability of the clock improves as the
square root of the averaging period, as in the case of Gaussian
noise processes.  However, the term that is most relevant for this
discussion is the appearance of $\nu_0$, the oscillator frequency,
in the denominator, which states that fractionally it is
advantageous to run a clock at higher frequencies. Alternatively,
this can be thought of in terms of line $Q$, for which optical
systems have demonstrated values as high as $10^{15}$, five orders
of magnitude higher than analogous microwave systems, giving
optical clocks a tremendous advantage in terms of potential
short-term and long-term frequency stability. From eq.
\ref{qpnstab}, we see that for a transition linewidth,
$\Delta\nu$, of 1 Hz, measured in an atomic sample of only 10,000
atoms, we can envision fractional instabilities as low as
$10^{-17}$ in 1 second of averaging time, many orders of magnitude
better than in other existing (non-optical) systems.

In terms of absolute frequency uncertainty, the second important
clock characteristic, the advantages of the optical atomic clocks
cannot be summarized in a single formula.  Rather it is necessary
to evaluate the systematic effects for different transitions and
see how they compare with other optical and microwave clocks.
However, the greatly enhanced stability possessed by optical
systems enables the evaluation of the systematic effects much more
rapidly and/or to a much higher precision, thereby enabling
reduced uncertainty budgets.  The most stable optical clock
systems can presently measure shifts at the 1 part in $10^{17}$
level in less than an hour \cite{Hinkley2013}, while the best
microwave systems would take ten days or more, almost
prohibitively long to evaluate clocks systematics at this level.

While the inherent stability advantage of optical clocks was
experimentally demonstrated quite early in the brief history of
optical clock research, the absolute uncertainty took longer to
catch up and surpass that of its microwave counterparts,
principally because of the more painstaking nature of systematic
uncertainty evaluations. However, as we see in fig.
\ref{fig.accuracyatomicstandards}, around the turn of the
millenium, the uncertainty of optical sources rapidly decreased,
culminating in the landmark 2008 demonstration by the ion clock
group at NIST, where they achieved fractional absolute
uncertainties for the Al$^+$ and Hg$^+$ clocks of about one part
in $10^{17}$ \cite{Rosenband2008}.  More recently, a neutral Yb
lattice clock demonstrated a fractional instability (not
uncertainty) of $1.8\times10^{-18}$ for 25,000 s of averaging time
\cite{Hinkley2013}.  As we will see in subsequent sections, many
other optical clock systems are also now rapidly approaching this
level, as this field continues to expand.  So what happened around
the year 2000?  The short answer is that the femtosecond-laser
frequency combs showed up on the scene, thereby providing the
field of optical frequency metrology with an essential and
long-awaited tool.  A fuller answer would recognize that at the
same time optical frequency standards were also beginning to
mature dramatically due to significant advances in laser
stabilization and atom manipulation capabilities.  Thus, these
more precise optical standards were well primed to take full
advantage of the possibilities enabled by the new optical
frequency measuring devices.  Let us now consider in more detail
the three critical experimental capabilities that led to the
revolutionary advances depicted in
fig.~\ref{fig.accuracyatomicstandards}; in this way we will be
able to understand not only how these new timing devices have come
so far, so quickly, but also how much further they might be able
to go in the future.




\section{Key technologies for optical atomic clocks}
\subsection{Optical frequency synthesizers}
For many years the biggest unknown in the field of optical
frequency standards and clocks was how to measure optical
frequencies.  The oscillations of the electric field for visible
frequencies are too fast ($\sim 10^{15}$ per second) to detect
directly, so there was no way to measure the absolute frequency of
the optical clocks (\emph{i.e.}, relative to the SI second defined
by Cs) and no way to compare them to other clocks in the optical
or microwave domains.  There were a few heroic efforts to connect
specific optical frequencies to the microwave domain through an
array of phase-locked oscillators spanning many decades of the
frequency spectrum
\cite{Evenson1972,Schnatz1996,Udem1997,Touahri1997,Bernard1999}.
What was needed instead was a more general and flexible solution
to this problem.  Nonetheless, many groups proceeded with their
development of stable, accurate, optical sources, with the hope
that one day this problem would be solved.  Indeed in 1999,
efforts in Th. H\"{a}nsch's group and in J. Hall's group basically
solved this problem by developing methods to stabilize mode-locked
femtosecond-lasers to such a high degree that they could play the role of a
frequency divider with such versatility that one could use it to
connect optical frequencies both to other parts of the optical
spectrum and to the microwave
domain\cite{Udem1999,Holzwarth2000,Jones2000}.  For their efforts
Th. H\"{a}nsch and J. Hall shared the Nobel Prize in 2005, a
tribute to the significance of this technique
\cite{Hansch2006,Hall2006}.

\begin{figure}[t]\begin{center}
\includegraphics[width=0.7 \textwidth]{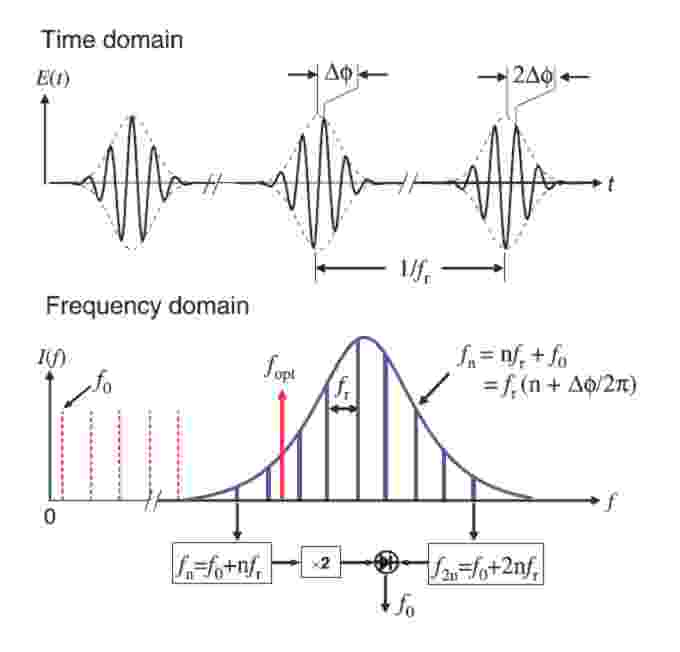}
\caption{Representation of the output of a mode-locked femtosecond
laser in time and frequency-domain. The frequency of emission of
the pulses is $f_{r}$ and due to dispersion in the laser cavity,
the carrier advances with respect to the envelope by $\Delta\phi$
from one pulse to the next. In the frequency domain, the result of
this phase slip is an offset common to all modes of $f_{o} = f_{r}
\Delta\phi/(2\pi)$. $f-2f$ technique for comb self-referencing is
also shown. \cite{Hollberg2005}.} \label{fig.comb}
\end{center}
\end{figure}

To see how a pulsed laser could perform the remarkable feat of
linking the optical and microwave domains, let us first observe
that a pulsed laser indeed contains frequencies in both domains
(see fig. \ref{fig.comb}). The light is visible, but the
repetition rate of the pulses occurs typically in the 100 MHz to 1
GHz regime.  Thus, the trick here is to find some way to
phase-lock the microwave repetition rate to the optical
frequencies contained in the output spectrum. In fact, the method
for doing this becomes clear when we consider the frequency
spectrum for a femtosecond-laser.  The Fourier transform for a
regular sequence of pulses in time is simply a series (``comb'')
of discrete visible frequencies (called ``teeth'') spaced by the
repetition rate.  The range of frequencies is determined by the
output spectrum of the laser, which is in turn determined by
essential laser parameters such as the gain curve for the laser
medium, the laser cavity mirrors, the laser pump source and
intensity, etc. We can describe the resulting values, $f_n$, in
the frequency ``comb'' with just two laser-based parameters:  $f_n
= f_{o} + n \times f_{r}$, where $n$ is an integer (or mode
number), $f_{o}$ is the effective offset of the comb from zero
frequency (physically, it results from the pulse-to-pulse phase
shift of the light pulse relative to its pulse envelope), and
$f_{r}$ is the laser repetition rate.

Thus, simply stabilizing $f_{o}$ and  $f_{r}$ then stabilizes the
frequencies of all the visible modes as well as the microwave
repetition rate.  In order to stabilize these two parameters it is
necessary to measure these quantities with sufficient precision
and use feedback loops to control specific laser characteristics
\cite{Jones2000}.  For $f_{o}$, it is customary to generate the
feedback error signal by comparing one part of the comb to
another.  This is most easily understood and readily accomplished
by achieving a comb spectrum that spans a full octave.  In this
case, we can use nonlinear optics techniques (\emph{e.g.}, a
frequency-doubling crystal) to double the low-frequency end ($n
\sim n_0$)  of the comb and then beat it against the high
frequency ($n \sim 2n_0$).  The resulting beatnote will occur at
the frequency difference, $f_{d}$, and is simply:
 \begin{equation}
f_{d} = 2 (f_{o} + n_0 \times f_{r}) -  (f_{o} + 2n_0 \times
f_{r}) = f_{o}. \label{offsetfreq}
\end{equation}
The challenge, of course, is to generate a full octave (or a
significant fraction thereof) directly out of the laser.  Thus
lasers with very short (femtosecond) pulses are desirable due to their
broad output spectra, but still it is often necessary to complete
the octave by passing the femtosecond-laser output through a highly
nonlinear microstructure fibre \cite{Ranka2000}, which can
significantly expand the laser spectrum.  An $f_{o}$ error signal
can then be generated by comparing the beatnote frequency to that
of a stable microwave frequency (note that since the noise from
this reference oscillator maps directly into the optical signal,
it is fractionally reduced by 4 to 5 orders of magnitude, thereby reducing the requirements for the reference oscillator).  The
frequency of $f_{o}$ can be then controlled in different ways such
as through the pump laser power.  For $f_{r}$, it is
straightforward to measure the repetition rate directly with a
photodetector and then generate an error signal by comparing it
with a stable microwave frequency.  We can use the resulting
signal to stabilize the repetition rate by feeding back to a
parameter that controls the cavity length (\emph{e.g.}, the
position of a cavity mirror). However, in this case, the noise
from the microwave reference is multiplied when we look at signals
in the optical domain.  Instead, the usual configuration for
optical clocks is to generate the $f_{r}$ error signal by
superimposing light from the frequency comb with light from a
stable optical clock frequency on a photodetector.  If $f_{o}$ is
stabilized as described above, then forcing the comb repetition
rate to keep the comb-clock beatnote fixed to a microwave
reference frequency will fix $f_{r}$ and the entire comb.  In this way the stability of the optical
source can be transferred to every comb tooth and the comb's
repetition rate with a fidelity that has been confirmed to one
part in $10^{19}$ \cite{Ma2004}.


Thus the formidable problem of measuring absolute optical
frequencies and comparing optical and microwave frequencies is
elegantly solved by the use of a single, stabilized, pulsed laser.
Moreover, optical sources with very different frequencies can be
compared directly by locking the comb to one source and measuring
the beatnote between the other source and the comb tooth closest
to it in frequency (see for example, \cite{Diddams2001}). Soon
after the first demonstrations of stabilized frequency combs
around 2000, these lasers became the standard tools in frequency
standards labs for providing the clockwork.  Not surprisingly, it
turns out that a stabilized frequency comb also has very regular
pulses in the time domain. As a result, there has been an
explosion of new applications for frequency combs that have been
developed in the past decade for use in both the time and
frequency domains including the calibration of astronomical
spectographs for exo-planet searches, the generation of the
world's lowest-noise microwaves, time-resolved femtosecond and
attosecond spectroscopy, and precision mid-infrared spectroscopy
(see \cite{Diddams2010} and references therein).  In a similar
way, there has been tremendous progress in the development of new
femtosecond-laser frequency comb sources beyond the initial
systems, which were Ti:Sapphire based, including compact
fibre-based systems, and now even whispering-gallery mode-based
``micro-combs'' (for a review of this emerging field, see
\cite{Kippenberg2011}).



\subsection{Laser cooling and trapping of neutral atoms and single ions}
\label{sec.laser cooling}

As we will see in more detail in sect.~\ref{systematics}, the
thermal motion of the atoms and ions used in optical atomic clocks
presents tremendous challenges for clock scientists, because the
effects are intrinsically enormous for the scale on which they
operate. Room-temperature atoms can have a velocity distribution
that yields Doppler shifts of 1 GHz or more - twelve orders of
magnitude larger than the millihertz uncertainties towards which
optical clock scientists aspire.  As a result, the reduction and
control of these effects have had a huge impact on the design of
optical atomic clock systems.  While various ``sub-Doppler''
techniques have been developed to resolve atomic resonances much
narrower than those of the Doppler distributions, these methods
were still plagued with shifts at the kilohertz level or more
\cite{Kersten1999}; additionally, the second-order Doppler effect
(associated with relativistic ``time dilation'' effects)
approaches 1 kHz at room temperature, and so exquisite knowledge
of the atomic velocity distribution is required to characterize
this effect at the sub-hertz level \cite{Shirley1997}. Instead,
through the past three decades or so, scientists have developed an
array of tools using lasers and electric and/or magnetic fields to
cool and trap atoms to very tight volumes with temperatures as low
as $1$ $\mu$K and even below (see \cite{Chu1998Nobel,
Phillips1998Nobel, Cohen-Tannoudji1998Nobel} and references
therein). These techniques have led to amazing advances in atomic
physics, including the generation of Bose-Einstein condensates
\cite{Anderson1995} and ultracold Fermi gases \cite{DeMarco1999},
the trapping/cooling of molecules \cite{Krems2009, Hummon2013},
and a host of fundamental physics experiments using clouds of
atoms as quantum sensors.

\begin{figure}[t]
\begin{center}
\includegraphics[width=9.5 cm]{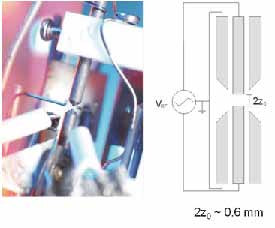}
\caption{A typical single ion end-cap trap employed in high
resolution clock spectroscopy (courtesy of National Physical
Laboratory - NPL).\label{fig.SingleIonEndcapTrap}}
\end{center}
\end{figure}

Many of these techniques were first demonstrated with trapped
ions, due to the long interaction periods and the exquisite
control that the ion systems afford \cite{Wineland1984}.  Due to
their non-zero electric charge, ions can be trapped with
oscillating electric fields (it turns out that static
configurations cannot work for charge-free regions).  There are
several basic designs for ion traps, all of which use some
combination of static and radio-frequency fields. Once confined in
the trap, an ion can be cooled, that is, have its velocity reduced
by illuminating it with laser light tuned just below (red) of a
strong resonance.  In this way, the ion preferentially absorbs
momentum from the laser whose direction opposes its motion, and
its velocity is reduced to the point where it fluctuates around
zero velocity (see \cite{Wineland1987a} for an early review of
laser cooling with trapped ions).  Trapped ions are now routinely
cooled to temperatures of a few millikelvin, putting the ion
predominantly in its motional ground state (see fig.
\ref{fig.SingleIonEndcapTrap}). As a result the Doppler effects
are not only reduced, but in fact the first-order Doppler effect
is completely absent in spectroscopic signals. For this ``tight
confinement'' regime (the so-called ``Lamb-Dicke'' regime
\cite{Dicke1953}), in which the extent of the atomic motion along
the probing direction is much less than the wavelength of the
probe light, residual motional effects are transferred to
spectroscopic sidebands, thereby leaving the carrier virtually
unperturbed \cite{Bergquist1987}. For temperatures at the
millikelvin level, the residual time dilation effect is
fractionally about $10^{-17}$, manageable for even the most
accurate systems.  By about the year 2000, laser cooling
techniques were fairly well developed for all existing optical
clock systems based on trapped ions. More details on techniques
employed in single ion trapping and laser cooling are described in
sect.~\ref{sec.ions}.

\begin{figure}[t]\begin{center}
\includegraphics[width=0.4 \textwidth]{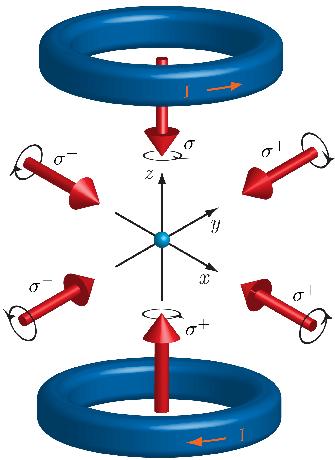}
\caption{Sketch of the Magneto-Optical-Trap (MOT) setup. Three
pairs of retro-reflected laser beams cross each others in the
center of the trap. A pair of anti-Helmholtz coils provide the
necessary quadrupole magnetic field for trapping. The atomic cloud
is collected in the center of the trap (adapted from
\cite{SchioppoPhDThesis}).} \label{fig.mot}
\end{center}
\end{figure}

Because neutral atoms are more difficult to manipulate, it took a
little longer (and more effort) to confine them to the required
degree.  The effects of resonant laser light on atomic motion was
first demonstrated in 1982 \cite{Phillips1982}, but it really
wasn't until the late 1980's that microkelvin temperatures and
magneto-optical traps were becoming routine for alkali atoms
\cite{Raab1987}.  Magneto-optical traps for the species of atoms
used in neutral atom clocks (\emph{i.e.}, two-electron atoms) were
not developed until 1990, due in part to the difficulty of
generating tunable and high-power blue lasers required for cooling
these atoms \cite{Kurosu1990}. Magneto-optical traps (see fig.
\ref{fig.mot}), which confine atoms through the combination of
laser fields and magnetic field gradients, are efficient tools for
collecting large samples of cold neutral atoms from background
vapors or atomic beams. Typical trap geometries contain three
intersecting orthogonal pairs of counter-propagating laser beams
tuned just to the red of a strong cycling transition such as the
$^1S_0\rightarrow^1$$P_1$ transitions in alkaline-earth and
alkaline-earth-like atoms. The region of intersection, the
``trapping region'', provides a three-dimensional optical
molasses, so-called because the trap always supplies a net force
opposing an atom's propagation direction. The addition of a
magnetic field gradient ($\sim$~0.6~T/m for alkaline earth atoms),
supplied usually by a pair of anti-Helmholtz coils, adds a spatial
component to this force, thereby forcing the atoms towards the
centre of the trap. In this way, millions of atoms can be
collected in a fraction of a second, with a resulting atom
temperature of approximately 1 mK. The loading rate can be
increased by slowing a portion of the atomic beam velocity
distribution before the trapping region in order to accommodate
the (usually) fairly low capture velocities of the MOT ($\sim $ 10
m/s).  Such slowing can simply accomplished by sending a laser
beam counter-propagating to the atomic beam. Usually this beam is
red-detuned by several hundred megahertz to compensate the large
Doppler shifts of the oncoming atoms and to reduce asymmetry
distorting the MOT.  More efficient slowing can be accomplished by
adding a magnetic field gradient along the atomic beam, designed
to keep the atoms in resonance with the slowing laser beam
throughout the slowing process \cite{Phillips1985}.

Magneto-optical traps, however, do not present a suitable
environment for precision clock spectroscopy because the trapping
laser beams significantly perturb the clock transition.
Additionally magneto-optical traps do not confine atoms
sufficiently to reach the Lamb-Dicke regime, so residual Doppler
effects still exist, which are at the megahertz level, even for
atoms at millikelvin temperature. The solution was to employ
optical traps \cite{Katori2003} based on red-detuned standing
waves of light (termed ``optical lattices''), which could be
loaded with very cold atoms. The atoms are drawn into the
high-intensity regions of the lattice light and oscillate around
the standing-wave anti-nodes, in a way similar to that of trapped
ions, but with thousands or even millions of atoms confined
simultaneously.  By tuning the lattice light to a carefully
determined wavelength, its effects on well-chosen clock
transitions can be minimized \cite{Takamoto2003}.  In this way
ultra-narrow, minimally perturbed spectra could be generated, and
as we shall see, today such lattice clocks are leading the way for
neutral atom clock research.

\subsection{Frequency stabilization of laser sources}
\label{sec.freqstablaser}
\begin{figure}[t]\begin{center}
\includegraphics[width=0.5\textwidth]{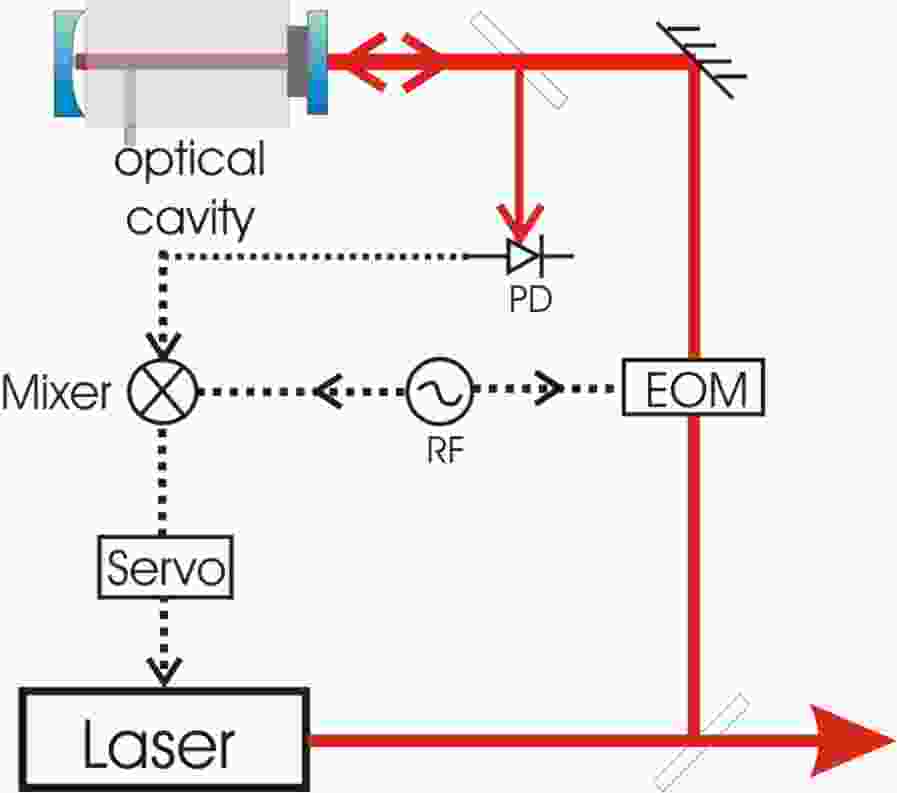}
\caption{Typical scheme used for active laser frequency
stabilization to optical resonators (Pound-Drever-Hall). EOM
electro-optical modulator, PD photodetector, RF radiofrequency
generator. \label{fig.PoudDreverNew}}
\end{center}
\end{figure}

Pushed by the extreme demands in high precision metrology
experiments and optical clocks development, the field of frequency
stabilization of laser sources has been undergoing something of a
renaissance in recent years.  In the early years, efforts in
several labs, including those of J. Hall and Th. H\"ansch, really
tamed the laser through the use of exquisite optical reference
cavities, making hertz-level lasers a reality \cite{Salomon1988}
(see fig. \ref{fig.PoudDreverNew}). An optical cavity consists of
a specialized glass spacer that separates two high-reflectivity
(often as great as 99.999 $\%$) mirrors that provide extremely
sharp resonances that can be used to stabilize the laser on short
to medium time scales (e.g., $<$ 10 s). On longer time scales the
cavities are subject to thermal and mechanical drifts, and the
atomic signals are then needed to provide stable, accurate
references. Early cavity research culminated in the benchmark
result of J. Bergquist and colleagues in 1999, in which they
achieved a laser linewidth of 0.22 Hz by carefully isolating
ingeniously-designed cavities from environmental effects
\cite{Young1999}.  These results stood as the gold standard for
more than a decade, despite the fact that many scientists strove
to push this research further, driven by the needs of the optical
clock community and others.  However in a paradigm-changing paper
in 2004, it was recognized that the specialized cavities that were
used to stabilize the lasers were in reality almost all limited by
fundamental thermal noise processes in the cavity construction
materials themselves \cite{Numata2004}, and that the cavities used
in ref. \cite{Young1999} in fact minimized these effects quite
well. As a result, modest improvements have since been achieved by
choosing proper spacer and mirror material and carefully designing
the cavity shape and spacer supporting points.  However, it seems
the next generation of cavities will need to use novel crystalline
mirror coatings \cite{Cole2013} and perhaps need to reduce the
thermal noise directly by operating cavities at cryogenic
temperatures \cite{Kessler2012b}. Additionally, other approaches
are underway as today's best systems are still chasing the dream
of a millihertz laser linewidth \cite{Salomon1988}.

\begin{figure}[t]\begin{center}
\includegraphics[width=11 cm]{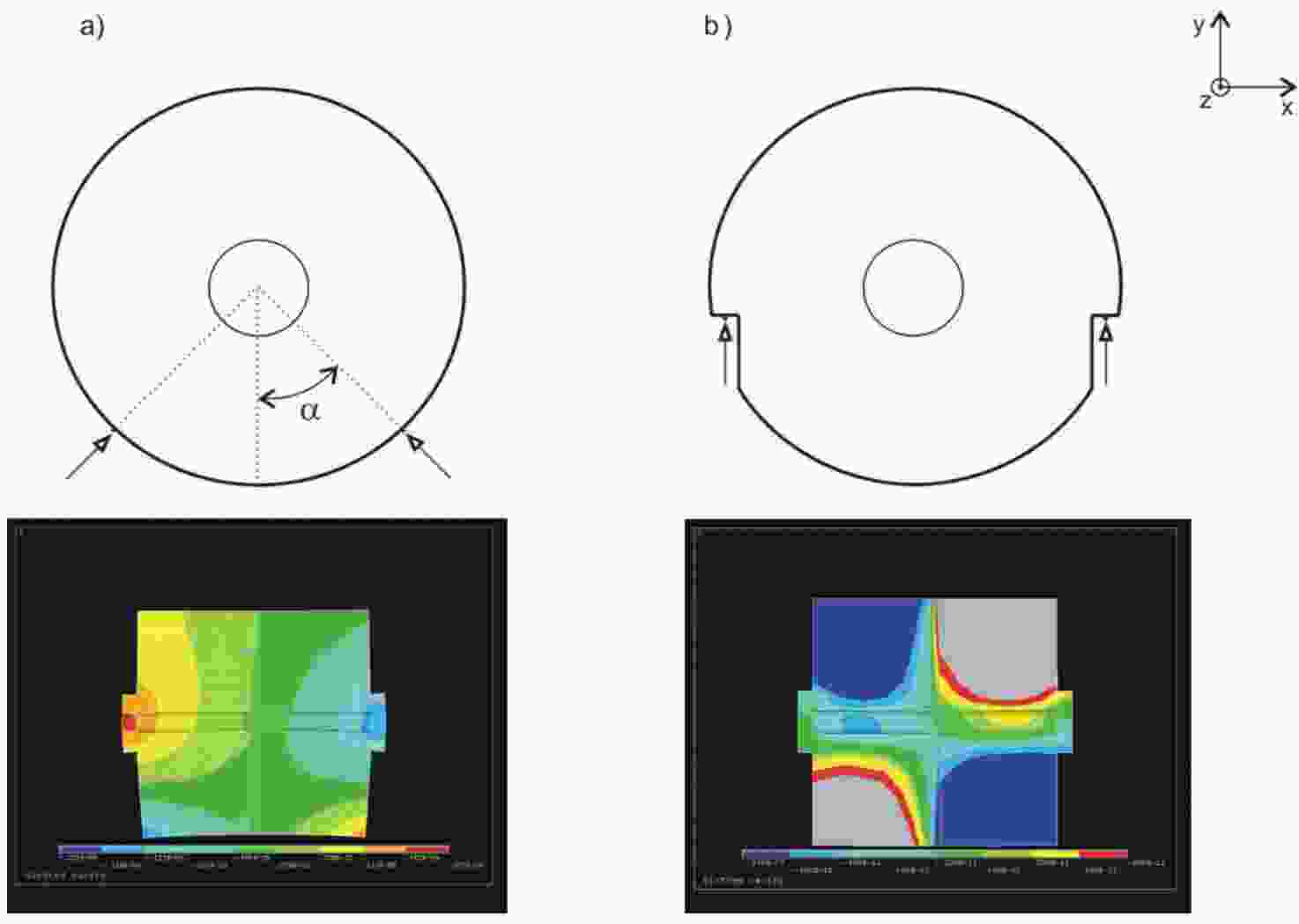}
\caption{Example of FEM static analysis respectively on standard
cavity mount in (a) and ``slotted'' horizontal cavity in (b)  (see
text for further details). The result of the FEM simulation shows
the deformation (in false colours) of the cavity along the optical
axis of the cavity due to a constant acceleration along the
vertical axis. In case (a) the cavity deformation pushes the two
mirrors far apart introducing also a relative tilt of the mirrors.
In (b) the effect is compensated thanks to the modified spacer
geometry and the optimum choice of supporting points.
\label{fig.simul1}}
\end{center}
\end{figure}

\begin{figure}[t]\begin{center}
\includegraphics[width=9 cm]{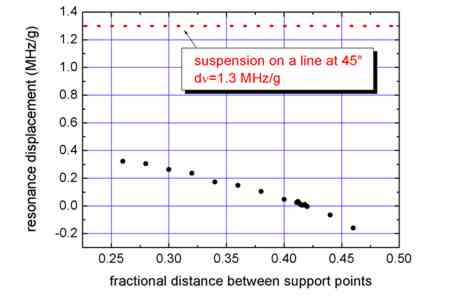}
\caption{Resonance displacement $\delta\nu/\delta a$ for the two
cavity geometries in fig. \ref{fig.simul1} calculated with FEM
static analysis as a function of the supporting points position
along the optical cavity axis. The support points (\emph{i.e.}, 2
mm$\times$ 2 mm areas) for the ``slotted'' cavity are chosen in
order to reduce the sensitivity. With proper chosen support
points, the sensitivity can be reduced by more than a factor of
100. Data from ref. \cite{Poli2007}. \label{fig.simul2}}
\end{center}
\end{figure}

Why do the clocks put such stringent requirements on the laser
stability?  First, in order to perform spectroscopy at the highest
resolutions (\emph{e.g.}, $\Delta \nu < 10$ Hz) needed by many of
today's state-of-the-art clocks, it is necessary to pre-stabilize
the probe laser systems.  Second and equally important is the fact
that in some of today's state-of-the-art optical atomic clocks
(especially those based on large numbers of atoms, for which the
atom projection noise, $1/\sqrt{N}$, is low), the laser frequency
noise actually determines the clock stability.  Think back to Eq.
\ref{qpnstab}, where we saw that optical clocks could enable
stabilities at levels approaching 1 part in $10^{17}$ in a one
second averaging period.  But there we made the assumption that
the clock was limited by fundamental noise sources, and that
technical noise sources (including laser frequency noise) were
reduced below this level.  In fact, this turns out to be a fairly
ambitious assumption.  How the frequency noise limits the atomic
clock stability is a subtle issue, because in principle we can
correct laser fluctuations with the atom signal-based servo loop.
The problem is that the atom signal provides incomplete knowledge
of the laser fluctuations because in virtually all clock systems,
we can probe the atoms for only a fraction (typically $20$ to $50$
$\%$) of the measurement cycle (see the descriptions of the
various clocks, especially Sr, for more details).  The rest of the
cycle is spent preparing the atoms and detecting the excitation
induced by the probe light.  As a result, the clocks suffer from a
sort of stroboscopic effect that aliases higher-frequency noise
into low-frequency noise on the clock signal.  The feedback system
cannot null out the noise, effectively setting a limit for clock
performance.  This stroboscopic effect, known more formally as the
Dick effect \cite{{Dick1987,Santarelli1998,Quessada2003}},
contributes a noise level that is roughly the square root of the
duty cycle times the frequency noise (for white frequency noise).
To support a clock that has a $25$ $ \%$ duty cycle with a
fractional stability that averages down as $10^{-15}\tau^{-1/2}$
requires a pre-stabilized laser with fractional frequency noise
below $10^{-15}$ at 1 second. Thus for realizing better clocks,
one important request is to minimize dead time in the measurement
cycle and to reduce the laser frequency noise as much as possible.

While cycle-time optimization is quite system-dependent, laser
stabilization is less so, thus it is possible to consider quite
generally the challenge of stabilizing lasers at the sub-hertz level.
Although most lasers have narrow fundamental (\emph{i.e.},
Schawlow-Townes limited) linewidths, in reality, their outputs are
quite broad (kHz-MHz) as a result of technical noise sources, such
as vibrations, thermal effects, or noise in the laser pump
sources. To reach the hertz level and beyond, it is therefore
necessary to measure laser frequency fluctuations and correct them
with a feedback loop connected to frequency-changing laser
parameters (such as laser mirror position or the current in a
diode laser). Measuring such fluctuations on a fast time scale
with a high signal-to-noise ratio is a challenging task.  While
several techniques have been developed to meet this challenge
(\emph{e.g.}, \cite{Kefelian2009}), by far the most commonly used
technique is locking the laser frequency to a cavity resonance.
These cavities are typically made of ULE (ultra-low expansion)
glass to minimize thermal drifts, and then have high-finesse
(reflectivity) mirrors attached to either end, usually via optical
contacting, again to minimize drifts.  For typical spacer lengths
of $L$ = 5 to 30 cm, we then find a series of identical, sharp
resonances, spaced by $c/2L\sim$ 1 GHz, with linewidths as narrow
as a few kilohertz.  The laser is usually locked to the fringe nearest
in frequency to the clock resonance and then the remaining gap is
bridged with frequency-shifting devices such as acousto-optic or
electro-optic modulators.

In order to achieve the highest performance levels, it is
necessary to use sensitive techniques for generating the error
signal that contains the laser frequency fluctuations relative to
the cavity resonance.  The preferred technique is to send a small
amount of light ($<$ 100 $\mu$W) to the cavity with modulation
sidebands at $\sim$20 MHz.  We then detect the signal reflected
from the input mirror of the cavity.  In this way, we find that
the error signal contains information on the laser frequency
fluctuations for Fourier frequencies below the linewidth of the
cavity resonance and laser phase fluctuations for Fourier
frequencies higher than this cavity linewidth \cite{Drever1983}.
Moreover, the high modulation frequency encodes this information
at a detection frequency well above that of most technical noise
sources.  When appropriately demodulated, the detected signal
yields a high-bandwidth, high signal-to-noise error signal, which
is then filtered and fed to the frequency correction devices. This
well-known experimental design, the Pound-Drever-Hall technique,
is used in almost all precision laser stabilization systems, and
when used carefully (\emph{e.g.}, minimizing effects such as
residual amplitude modulation), yields laser stabilization results
that are limited only by the reference cavity itself
\cite{Drever1983,Fox2003,Chen2012}. For this reason, current
state-of-the-art research in laser stabilization is really about
improving the reference cavities themselves.  The goals here are
quite extreme when we consider them;  hertz-level laser
performance requires that the cavity length (which maps
proportionally into the laser frequency) be stable to less than 1
fm (about the radius of an atomic nucleus). Nonetheless, this
level of performance is possible, but requires that extreme care
be taken with both the isolation of the cavity from the
surrounding environment and with the material used to construct
the cavity itself.

Adequate isolation usually begins with putting the cavity inside
an evacuated metal can (often with thermal shields) to minimize
sensitivity to refractive index fluctuations in the path between
the mirrors. The vacuum apparatus is then usually surrounded with
a box that reduces coupling of acoustic vibrations to the system.
Although optical cavities utilize a rigid spacer to hold the
mirror spacing constant, no material is infinitely rigid, and as
such is susceptible to elastic deformation. For the case of
optical cavities, acceleration-induced deformation can result in
cavity length changes that can seriously degrade both the short-
and long-time scale frequency noise of the spectrum of the laser
locked to the cavity.  Thus, it is necessary to minimize external
vibrations that can shake the cavity, thereby changing its length
(principally around frequencies ranging from 1 to 100 Hz),
Complicated methods for such vibration isolation have been used
historically, including putting cavities in specialized rooms or
supporting their platforms with long pendula \cite{Young1999}.
However, today one can purchase more compact, active (or passive)
vibration-isolation platforms that can support a cavity and its
associated hardware.  For best performance the cavity geometry and
mounting systems are designed (see figs.~\ref{fig.simul1} and
\ref{fig.simul2}) so as to minimize coupling of seismic vibrations
to the cavity length
\cite{Ludlow2007,Webster2011,Leibrandt2011,Argence2012}.  We note
also an ingenious approach that minimizes the effect of vibrations
on a symmetrically mounted cavity by measuring the vibrations on
the cavity with an accelerometer, then feeding forward frequency
corrections to the light based on the cavity's measured transfer
function \cite{Thorpe2010}.  Additionally, today's cavities are
also designed to minimize a more fundamental limitation: that
imposed by thermal noise due to the Brownian motion of the cavity
itself \cite{Numata2004}, which exists even in a completely stable
thermal environment.  This effect manifests itself in all the
components of the optical cavity:  the spacer, the mirror
substrates, and even the mirror coatings.


\begin{figure}[t]\begin{center}
\includegraphics[width=0.9 \textwidth]{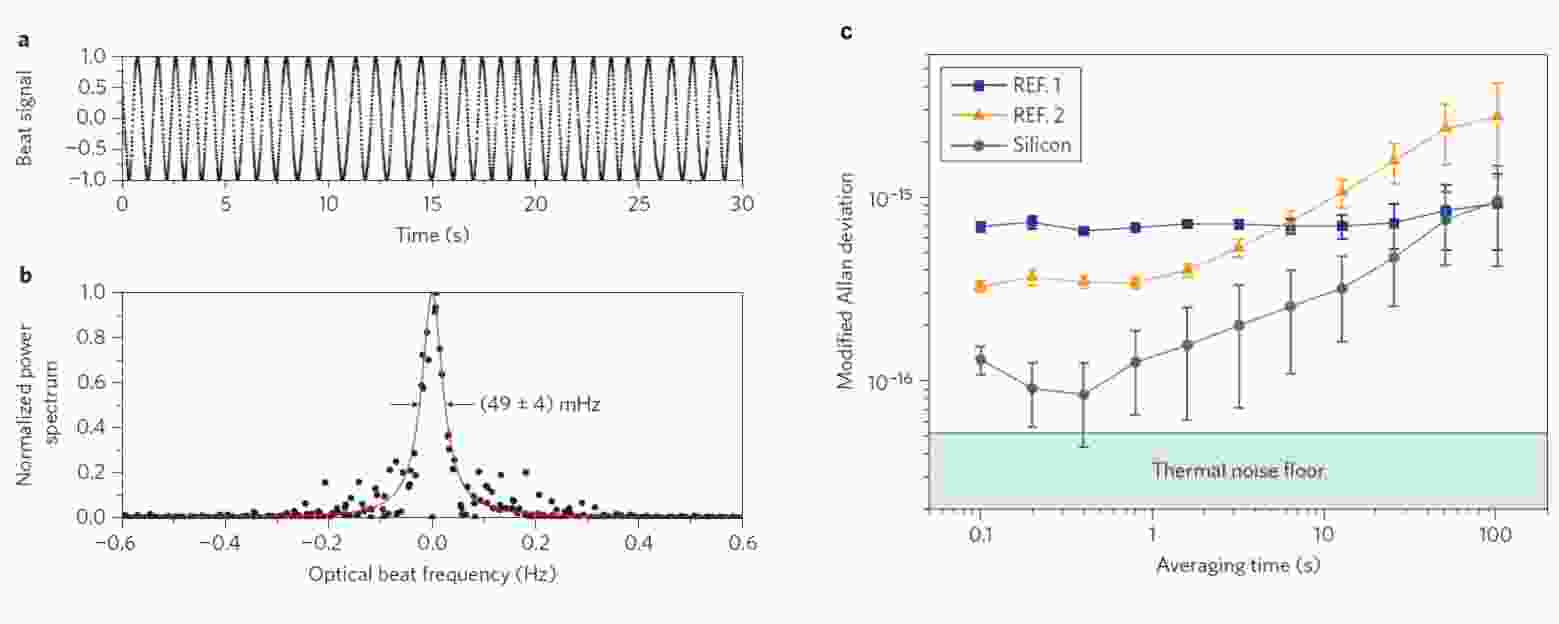}
\caption{Optical heterodyne beat between a 1.5 $\mu$m laser
stabilized to cryogenic silicon cavity and second laser system
stabilized to a 25-cm long ULE-fused silica cavity. a, Beat signal
mixed down close to d.c. and recorded with a digital oscilloscope.
b, Normalized fast Fourier transform of the beat signal. A
Lorentzian fit is indicated by the red line. The combined result
of five consecutive recordings of the beat signal (black dots) is
displayed here, demonstrating the robustness of this
record-setting linewidth. c, Modified Allan deviation of each
cavity-stabilized laser. The predicted thermal noise floor of mod
$\sigma_y\sim 5\times 10^{-17}$ for the silicon cavity system is
indicated by the shaded area. Data from ref.\cite{Kessler2012b}.
\label{fig.KesslerResults}}
\end{center}
\end{figure}

As a result, the materials used in cavity construction are chosen
with thermal noise and thermal-sensitivity properties in mind.
Specifically, this means choosing high mechanical $Q$ materials
(\emph{i.e.}, with low loss angle) with low coefficients of
thermal expansion (CTE). Usually the spacer is made from ULE glass
(maintained near the zero-crossing temperature with residual CTE
$<$ $10^{-9}$/K) because it represents the dominant contributor to
cavity length, while the mirror substrates in modern cavities are
often made of fused silica due to its high $Q$.  Many such
cavities have been constructed over recent years, each with their
own advantages. Overall, for cavities longer than 25 cm, the noise
floor limits the stability of the associated optical atomic clocks
(with typical probe duty cycles) at about $10^{-16}\tau^{-1/2}$, a
level that is already being challenged by the best optical clocks
(see for example, \cite{Jiang2011}). Improving the
cavity-stabilized laser stability significantly beyond this level
is difficult, since at this point the coating noise is the
dominant thermal noise factor. Pushed in part by the
gravitational-wave community, other coating materials are under
test, with some promising crystalline coating materials recently
identified and demonstrated \cite{Cole2013}.

An alternative approach to reducing the Dick effect-limited
instability is to increase the measurement duty cycle. In fact
more stable reference cavities help here as well, since they
enable longer probe times, thereby reducing, fractionally, the
period during which the atoms are not being probed.  Additionally
there has been an original approach that uses non-destructive
probing of the excitation, which enables recycling of the
lattice-trapped atoms \cite{Lodewyck2009}.  This reduces the
reloading period, which is usually the dominant contributor to the
time during which the atoms are not being interrogated (often
called the dead time).  Alternatively, we can probe two
independent atomic samples simultaneously \cite{Takamoto2011}, so
as to make noise common-mode between the measurements.  However,
to make a single system that is virtually immune to the Dick
effect we can use two systems in tandem, each one having at least
a 50 $\%$\ duty cycle, and then interleave their measurement
cycles, so that the probe laser is always under evaluation by one
of the atom systems \cite{Dick1987,Biedermann2013}.

Because of the critical importance of laser stabilization to this
field, other approaches to circumvent the thermal noise
limitations of optical cavities are underway.  Spectral
hole-burning in cryogenically cooled crystals is a route that uses
long-lived spectral holes induced by a narrow laser to serve as
stable frequency references to stabilize further that same laser.
Such systems have demonstrated fractional instabilities of
$6\times10^{-16}$ at one second, with a projected fundamental
limit about another order of magnitude lower \cite{Thorpe2011}.
Other groups are pursuing high signal-to-noise signals in narrow,
but not ultra-narrow (nor accurate), atomic transitions
\cite{Fox2012IEEE}. A different approach is to cool the cavity to
reduce the thermal noise contribution. A set of experiments,
initiated in the mid 90's with cryogenically cooled sapphire
cavities \cite{Richard1991}, obtained a laser instability of
$2.3\times 10^{-15}$ at a temperature of 3 K \cite{Steel1997}.
More recently a demonstration of a silicon cavity at 124 K
reported an estimated thermal noise limit of only
$7\times10^{-17}$ \cite{Kessler2012,Kessler2012b,Katori2012},
limited by mirror coating noise. Experimentally, it has been
demonstrated that a 1.5 $\mu$m laser stabilized to this cavity had
a record laser linewidth of $<40$ mHz and stability of
$<1\times10^{-16}$ at short term (see
fig.~\ref{fig.KesslerResults}). It is worth emphasizing that even
though this result was obtained with a laser at a wavelength
unconnected to an atomic clock transition, it is possible to
transfer this same level of stability to lasers at other
wavelengths with the help of an optical frequency comb. For these
reasons, though they are more complicated experimentally, cooled
cavities may play an ever-increasing role as researchers seek ever
more stable clocks.

\section{Systematic effects in optical clocks}
\label{systematics}

The frequency of an atomic transition of an atom or single ion
(quantum absorber) can be perturbed by many effects such as the
presence of undesired electric and magnetic fields, the fact that
the absorber is not perfectly at rest, collisional effects or even
imperfect vacuum conditions. The influence of such systematic
effects has been treated extensively, both experimentally and
theoretically \cite{Madej2001,Madej2004,Berkeland1998}. In this
section the most important systematic effects that limit the
performance of an optical atomic clock, for both the neutral atoms
and ion cases, are discussed in detail, together with ways
researchers have found to overcome them.

\subsection{Doppler effect}

This effect results from the motion of the quantum system
(\emph{e.g.} neutral atoms or single ion) relative to the
observer. From energy and momentum conservation in relativistic
theory one can find that the frequency shift
$\delta\nu=\delta\omega/2\pi$ of a photon emitted (or absorbed) by
a two-level atom (or an ion) with a mass $m$, moving with a
velocity $\mathbf{v}$ with respect to the laboratory rest frame,
can be written as \cite{VanierJ.andAudoin1989, Demtroder2008}
\begin{equation} \delta\omega=
\mathbf{k}\cdot\mathbf{v} -\frac{1}{2}\omega_0\frac{
\mathbf{v}^2}{c^2}\mp\frac{\hbar{\mathbf{k}}^2}{2 m},
\label{doppler}
\end{equation}
where $\mathbf{k}$ is the wave vector of the photon emitted
(absorbed) and $\omega_0=2\pi\nu_0$ with $nu_0$ unperturbed atomic
transition frequency. The three terms in Eq.~\ref{doppler}
represent respectively, the first-order effect, the second-order
Doppler effect, and the recoil effect.

In a sample of atoms in equilibrium at a certain temperature, due
to the thermal statistical distribution of velocities, the
first-order Doppler effect gives rise to an effective broadening
of the optical transition lines. The Doppler profile of optical
transitions is thus nearly Gaussian with a width related to the
temperature of the cloud. At room temperature this effect produces
a linewidth for an optical transition of the order of several
gigahertz, and even at microkelvin temperatures (nearly the
minimum that can be reached with laser cooling techniques), this
effect still gives a broadening of several hundred kilohertz.

The second term in Eq. \ref{doppler} is the second-order Doppler
shift, which depends on the square of the velocity of the quantum
emitter (absorber).  In spectroscopy schemes adopted in atomic
frequency standards, the asymmetry associated with second-order
Doppler effects generally produces an asymmetry of the line
profile and thus a shift of the line center.  The size of this
asymmetry is then typically evaluated starting from the knowledge
of velocity distribution of the atoms for the spectroscopy scheme
used \cite{Shirley1997}.  For high-performance clocks, the
reduction of the temperature of the atoms by means of laser
cooling techniques is usually necessary to reduce this effect to a
negligible level. For atoms or single ions cooled at millikelvin
temperatures, the shift is small ($\approx10^{-17}$), though not
completely negligible, while at microkelvin temperatures the fractional
frequency shift given by the second-order Doppler effect is below
$10^{-20}$.

The last term (recoil shift) in Eq. \ref{doppler} depends on the
frequency of the photon emitted (absorbed). While in the case of
microwave clocks this term plays no significant role at the
current clock accuracy ($\sim2\times10^{-16}$ for Cs standard at
$9$ GHz), for higher-frequency quantum-based standards this effect
gives a more important contribution (about $10^{-11}$ for Sr
optical clock at 429 THz) and cannot be neglected.

To circumvent or at least minimize these limitations, several
sub-Doppler or Doppler-free spectroscopy schemes have been
developed, which have enabled the observation of electronic
transitions with their natural linewidth:  saturation
spectroscopy, two-photon spectroscopy, Ramsey and Ramsey-Bord\'e
\cite{Borde'1984} interrogation schemes in atomic beams. While
theoretically these schemes all appear to remove at least the
first-order shift, in reality, they don't remove it completely,
due to effects such as imperfect laser beam alignment and
wavefront distortions.  Moreover, these techniques don't address
the large second-order Doppler shifts.  The laser cooling
techniques described in sect.~\ref{sec.laser cooling} seem well
suited to address these problems simply by reducing the velocities
of the atoms.  However, even with laser-cooled samples, Doppler
effects still play a significant role in the uncertainties
\cite{Degenhardt2005,Oates2005}, due to asymmetries in the
interrogation sequence (wave-front curvature and asymmetries in
the intensity of the interrogating laser beams, asymmetries given
by the incoherent signal of the background atoms, etc.), which can
induce asymmetries in the center-line profile and consequently a
systematic shift with respect to unperturbed atomic transition.
Thus, for the most accurate systems, it is necessary for the atoms
or ions to be confined to the Lamb-Dicke regime, for which the
first-order Doppler and the photon recoil shifts disappear
\cite{Bergquist1987} (the recoils are absorbed by the trapping
environment).  Moreover, the second-order Doppler shift can be
made negligibly small by sideband-cooling atoms down to the
vibrational ground state of the trapping potential
\cite{Diedrich1989}.  Under these conditions, whether for neutral
atoms or single trapped ions in an rf trap, Doppler effects can be
reduced to nearly negligible levels, although residual relative
motion between the trapped atoms and probe optics still needs to
be considered and at times compensated \cite{Oskay2006}.

\subsection{Interaction with magnetic fields: Zeeman effect}

Atomic states are coupled to static magnetic fields via their
magnetic moment $\mathbf{\mu}$. The magnetic moment depends on the
atomic state, and the interaction energy $\delta
E=\mathbf{\mu}\cdot \mathbf{B}$ is generally different for the
ground and excited states of the clock transition. The differential
energy shift causes a magnetic field dependency of the clock
transition frequency,
\begin{equation}
\delta\nu (\mathbf{B})=\frac{\delta E_e (\mathbf{B}) -\delta E_g
(\mathbf{B})}{h}.
\end{equation}
The magnetic moment of the atomic state is related to the
corresponding total angular momentum $\mathbf{F}$ (given by the
orbital angular momentum of the electronic cloud $\mathbf{L}$, the
electronic spin $\mathbf{S}$, and the nuclear spin $\mathbf{I}$).
To first approximation, the energy shift is proportional to the
projection of the total angular momentum along the quantization
axis expressed by the quantum number $m_F$. In general, the clock
frequency shift can be written as
\begin{equation}
\delta\nu=(g_e m_{F,e}-g_g m_{F,g})B+\beta B^2=\alpha_{e,g}
B+\beta_{e,g} B^2,
\end{equation}
where $g_e$ ($g_g$) represents the Land\'e factor for the excited
(ground) state. By properly choosing transition states with
$m_{F,g} =m_{F,e}= 0$ it is possible to avoid the first-order
Zeeman effect. However, even for transitions between states with
$m_F=0$ a residual second-order energy shift is always present
(quadratic Zeeman effect) due to interactions with nearby atomic
states with non-zero magnetic moments (see Table
\ref{ZeemanCoefficients}).

In general, fluctuations in the bias magnetic field may limit the
clock stability, and the clock uncertainty can be limited by the
uncertainty in either the Zeeman shift coefficients
($\alpha_{e,g}$ and $\beta_{e,g}$) or in the magnitude of the
magnetic field. The standard solution, as in microwave atomic
clocks, is to choose atomic transitions among states with $m_F=0$,
so that the linear Zeeman effect is absent, or among states with
small $g$-factors, so that the $\alpha$ coefficient is small.
Otherwise, it is common to interrogate sequentially two
transitions with symmetric shift, such as $m_{F,g}=0$, $m_{F,e}
=\pm m$, so that the average frequency is immune to the
first-order Zeeman effect.

For ultimate accuracy, the residual shifts must be properly
accounted for by controlling the magnitude of the magnetic field
and by measuring the Zeeman coefficients. The influence of
magnetic fields on the clock performance is less critical than in
the microwave case, since the absolute Zeeman shift of the atomic
reference is typically of the same order of magnitude, and thus is
proportionally much smaller for the optical case. With proper
magnetic shielding it is possible to reduce the shifts due to
residual magnetic fields and reaching relative accuracies below
the $10^{-18}$ level. Higher-order effects are generally
negligible.

\begin{table}
\caption{Linear ($\alpha$) and quadratic ($\beta$) Zeeman shift
coefficients for some optical clock transitions of neutral
atom/ion.}
    \begin{tabular}{ccccc}
\textbf{Atom/Ion}   &sublevels      &Linear coefficient $\alpha$            & Quadratic coefficient $\beta$  \\
                    &               &(MHz/T)                                & (MHz/T$^2$)                   \\
\hline
$^{24}$Mg           &$m_F=0$        & 0                                     & 164 \\
$^{40}$Ca           &$m_F=0$        & 0                                     & 64\\
$^{87}$Sr           &$m_F=\pm 9/2$  & 1.08                                  & -23.3 \\
$^{88}$Sr           &$m_F=0$        & 0                                     & -23.3  \\
$^{171}$Yb          &$m_F=\pm 1/2$  & 2.1                                   & -7 \\
$^{174}$Yb          &$m_F=0$        & 0                                     & -7 \\
$^{27}$Al$^+$       &$m_F=\pm 5/2$  & -82                                   & -72 \\
$^{88}$Sr$^+$       &$m_J=\pm 1/2$  & $5.6\times 10^{3}$                    & 3.1\\
$^{171}$Yb$^+$quad. &$m_F=0$        & 0                                     & $5.2\times 10^4$ \\
$^{171}$Yb$^+$oct.  &$m_F=0$        & 0                                     & $-1.7 \times 10^3$\\
$^{199}$Hg$^+$      &$m_F=0$        & 0                                     & $-1.9 \times 10^4$ \\

\hline
\end{tabular}
\label{ZeemanCoefficients}
\end{table}

\subsection{Interaction with electric fields}

The influence of electric fields on the clock frequency is,
generally, more important than that of magnetic fields. Electric fields
affect the atomic energy levels through the Stark effect, and a
variety of possible sources of DC or AC electrical fields can
induce such shifts on the clock transition: stray static fields,
motionally-induced electric fields for trapped ions, blackbody
radiation and laser fields. In particular, we need always to
consider the probe light field that excites the clock transition
itself, as its interactions with off-resonant transitions can, at
times, lead to non-negligible frequency shifts, especially for
weak transitions such as those based on octupole or
magnetically-induced moments.  As we shall see in sect.
\ref{sec.opticallatticeclocks}, laser fields play an integral role
in the operation and design of optical lattice clocks, in which
these fields are used to confine the atoms during clock-transition
excitation. Since atoms (or ions) do not possess a permanent
electric dipole in a definite quantum state, the Stark effect is
dominantly a second-order effect. In general, both clock levels
have their energies shifted and the resulting transition frequency
is changed by the differential shift
\begin{equation}
\delta\nu= -\frac{1}{2h}(\Delta\alpha_{DC}
E_0^2+\Delta\alpha_{AC}(\omega) \langle E(\omega)^2 \rangle),
\label{stark}
\end{equation}
where $E_0$ and $E(\omega)$ are the static and dynamic electric
fields applied, respectively,
$\Delta\alpha_{DC}=\alpha_{DC,e}-\alpha_{DC,g}$ is the difference
in static polarizabilities of the excited and ground states and
$\Delta\alpha_{AC}(\omega)$ is the difference in dynamic
polarizabilities for the same states. The general expression for
the atomic polarizability $\alpha_{AC,i}(\omega)$ for a generic
state $i$ can be found from second-order perturbation theory
\begin{equation}\label{eq.stark}
\alpha_{AC,i}(\omega)=\frac{1}{\hbar}\sum_{k \neq i}|\langle
k|\hat{D}|i\rangle|^2\frac{\omega_{ki}}{\omega_{ki}^2-\omega^2},
\end{equation}
in which the sum is performed on all $k$ levels coupled by the
$\hat{D}$ dipole operator to the $i$ level. It is worth noting
that $\lim_{\omega\rightarrow
0}\alpha_{AC,i}(\omega)=\alpha_{DC,i}$.

With the help of these formulas a large number of theoretical
atomic physics studies have been performed on the polarizabilities
of atoms (see Table \ref{StarkCoefficients}). With improvements in
atomic theory these can now be calculated very accurately, with
errors at the few \% level. However, this level of accuracy is not
always sufficient, and recently, for specific clock transitions,
atomic static polarizabilities
\cite{Middelmann2012,Ludlow2012IEEE,Sherman2012} have been
measured directly. Additionally, we re-emphasize that eqs.
\ref{stark} and \ref{eq.stark} pertain to shifts associated with
dipole-allowed transitions.  In fact as we shall see in sect.
\ref{sec.ionclocks}, many single ion optical clock transitions are
based on quadrupole transitions to a D state that contains a
quadrupole moment.  Such moments can interact with residual
quadrupole field gradients of the ion trap in a way analogous to
the DC dipole moment and DC electric fields in Eq. \ref{stark}
(along with a geometric factor based on the angle between the
quadrupole field axis and a bias magnetic field) and lead to
non-negligible shifts that need to be nulled or evaluated, as we
shall see later in the sections on specific ion clock systems in
sect. \ref{sec.ionclocks}.

\begin{table}
\caption{DC Stark coefficients and electric quadrupole
coefficients for some ion clock transitions.}
    \begin{tabular}{cccc}
\textbf{Atom/Ion}  &DC Stark coefficient  & Electric quadrupole shift \\
                   &(mHz/(V/cm)$^2$)       & (mHz/(V/cm$^2$))\\
\hline
$^{27}$Al$^+$      & -0.14                                &~0      \\
$^{115}$In$^+$     & 0.5                                  &~0      \\
$^{88}$Sr$^+$      & -4                                   &14      \\
$^{171}$Yb$^+$quad.& -6                                  &60      \\
$^{171}$Yb$^+$oct. & 1.2                                  &1.5     \\
$^{199}$Hg$^+$     & -1.1                                 &-3.6    \\
\hline
\end{tabular}
\label{StarkCoefficients}
\end{table}


\subsection{Interaction with blackbody radiation (BBR)}

A type of Stark shift that is particularly relevant to
state-of-the-art systems is given by the interaction with
blackbody radiation emitted by the environment in which the the
clock atoms or ion is trapped. The spectral density of energy
emitted by a body maintained at a temperature $T$ per unit volume
has the well known Planck formula,
\begin{equation}
\rho (\nu)=\frac{8\pi\nu^2}{c^3}\frac{h\nu}{e^{h\nu/k_B T}-1}.
\end{equation}
From this it is possible to express the mean square electric field
at temperature $T$ as
\begin{equation}
\langle
E^2\rangle_T=\epsilon_0^{-1}\int_0^\infty\rho(\nu)\mathrm{d}\nu\sim(8.319\times
10^2\, \mathrm{V/m})^2\times(T/300\,\mathrm{ K})^4.
\end{equation}
The distribution of electromagnetic fields can then shift the net
frequency of an optical atomic transition via the quadratic Stark
(and Zeeman effect). The temperature dependence of a given clock
transition depends on the polarizability of its two levels. Since
at room temperature the dominant frequencies of the BBR radiation
are typically much lower than that of the first resonant
transition coupling the excited and ground states, the dominant
contribution comes from the static polarizabilities.  As a result
one can write the net shift, $\Delta\nu_{BBR}$, in terms of a
differential DC Stark term, $\Delta\alpha_{DC}$, with a (much
smaller) dynamic correction, $\eta_T$ \cite{Farley1981},
\begin{equation}
\Delta\nu_{BBR}=-\frac{1}{2h}\Delta\alpha_{DC}\langle
E^2\rangle_T(1+\eta_T) \label{BBRshift}.
\end{equation}
As pointed out in the previous section, it is possible to estimate
both the static polarizabilities and dynamic correction
coefficient $\eta_T$ from ab-initio calculations
\cite{Porsev2006,Safronova2011IEEE}. In table~\ref{blackbody} we
summarize the calculated relative shift at room temperature for
some candidate optical clock transitions.

\begin{table}
\caption{Calculated BBR Shifts at $T = 300$ K for some optical
clock transitions in neutral atoms and ions. In the third and
fourth columns are reported the relative shift and estimated
relative uncertainty with a temperature uncertainty of $\Delta T=1
$ K
\cite{Porsev2006,Safronova2012a,Hachisu2008,Jiang2009,Tamm2009,Hosaka2009}.}
    \begin{tabular}{cccc}
\textbf{Atom/Ion}   & $\Delta\nu_{BBR} \mathrm{[Hz]}$    & $|\Delta\nu_{BBR}/\nu_0|$  & rel. uncertainty\\
\hline
Mg              & -0.258(7)                 & $3.9\times 10^{-16}$    & $1.6\times 10^{-17}$ \\
Ca              & -1.171(17)                & $2.6\times 10^{-15}$    & $7.2\times 10^{-17}$ \\
Sr              & -2.354(32)                & $5.5\times 10^{-15}$    & $1.4\times 10^{-16}$ \\
Yb              & -1.34(13)                 & $2.6\times 10^{-15}$    & $2.9\times 10^{-16}$ \\
Hg              & -0.181                    & $1.6\times 10^{-16}$    & $5\times 10^{-18}$ \\
Al$^+$          & -0.008                    & $8\times 10^{-18}$      & $3\times 10^{-18}$\\
In$^+$          & -0.0173(17)               & $1.36\times 10^{-17}$    & $1.6\times 10^{-18}$ \\
Sr$^+$          & 0.249                     & $5.6\times 10^{-16}$     & $2.2\times 10^{-17}$\\
Yb$^+$quad.     & -0.351                    & $5.1\times 10^{-16}$    & $2\times 10^{-17}$ \\
Yb$^+$oct.      & 0.067(32)                 & $1.1\times 10^{-16}$    & $5\times 10^{-17}$ \\
\hline
\end{tabular}
\label{blackbody}
\end{table}

The BBR shift contributes in two ways to the clock uncertainty.
First, because clock transition frequencies are traditionally
defined at 0 K, it is necessary to extrapolate from the
environmental temperature for the atom/ion system to 0 K.  Such
calculations require good knowledge of the coefficients in Eq.
\ref{BBRshift} and yield the values listed in columns two and
three of table \ref{blackbody}. Second, even with perfect
knowledge of the atomic parameters, there is still an uncertainty
in the environmental temperature, which affects the $\langle
E^2\rangle_T$ term above. Experiments currently estimate around
one degree uncertainty for this term, which yields the relative
uncertainty in the fourth column of table \ref{blackbody}.

An examination of the values of the fractional shifts in the third
column reveals a wide variance.  In particular, the value of
Al$^+$ is anomalously small and contributes in part to its being
the most accurate clock to date. In other cases, the BBR
coefficients are large enough to act as one of the dominant
contributors to clock uncertainty.  For these systems, more
precise (and experimentally verified) knowledge of these
coefficients is required.  In particular, we see that the very
promising lattice clock candidates, Sr and Yb, have comparatively
large fractional shifts at room temperature, $-5.5\times 10^{-15}$
and $-2.6\times 10^{-15}$, respectively.  In both cases, direct,
high-precision experimental evaluations of the static and dynamic
coefficients have been performed (which agreed well with theory)
\cite{Sherman2012, Middelmann2012}, leaving only the uncertainty
in the temperature of the environment itself as the significant
contributing systematic.  This uncertainty in turn is being
addressed in various ways, usually involving a specialized,
temperature-controlled interaction chamber, which, if necessary,
can be cryogenically cooled \cite{Katori2012IEEE}.  Operation at
liquid nitrogen temperatures (77 K) substantially reduces the BBR
shift, due to the T$^4$ dependence of the electric field
intensity.
\subsection{Collision \& Pressure shifts}

Frequency shifts that result from interactions between colliding
atoms have long been a serious concern in atomic standards.  We
note that for Cs fountain-based microwave clocks, these effects
can actually limit clock performance \cite{Jefferts2002}, and
indeed, they are important for optical clocks as well.  While the
idealized case of an atom or ion fully isolated from neighboring
atoms can nearly be realized in single, trapped ion systems (only
infrequent collisions with background atoms exist), for neutral
atom optical clocks the situation is quite different. In both
cell-based and trapped-atom-based clocks the atomic densities can
be high enough that atomic interactions can lead to small, but not
always negligible, shifts of the atomic clock frequencies.  In
cell-based systems these pressure shifts result principally from
collisions with like atoms or molecules (typically around 1
MHz/Torr), but contaminants introduced during cell fabrication
have also been seen to lead to frequency shifts between different
cells (see for example, \cite{Fredin-Picard1989}).  For clocks
using freely expanding atoms, either thermal or laser-cooled, the
density-related collision shifts are too small to be measurable at
their typical uncertainty levels, even at densities as high as
$10^{10}$/cm$^3$ \cite{Sterr2004}. As we will see in more detail
in sect. \ref{sec.srlatticeclock} and \ref{sec.yblatticeclock},
collision effects for lattice clocks are more relevant, primarily
due to their higher sensitivity. Moreover, since a large fraction
of the lattice-trapped atoms are confined to identical motional
states, the atom-atom interaction situation is further complicated
by the quantum statistics of the absorbers (see
\cite{Akatsuka2008} for a comparison between bosonic and fermionic
clocks).


\subsection{Gravitational effect}

A well-known consequence of general relativity is that the
frequency of a clock $\nu(x')$ measured by an observer in the
presence of a gravitational potential $U(x)$ will depend on the
gravitational potential difference between the clock location and
the observer location. This has the direct consequence that a
comparison between two identical clocks operating at different
locations $x_1$ and $x_2$ on Earth will show a difference in
frequency proportional to the potential energy difference
$U(x_1)-U(x_2)$. The change in clock frequency is directly related
to the loss of energy of photons ``climbing'' the gravitational
potential from $x_1$ to $x_2$, which corresponds to a change in
altitude, $\Delta$$h$, on Earth by (see, for example,
\cite{Chou2010a} or \cite{Vessot1980} and references therein)
\begin{equation}
\Delta\nu/\nu_0 \sim g\Delta h/c^2 \label{gravitational_redshift}.
\end{equation}
In other terms, we may say that that clocks closer to massive
bodies (\emph{i.e.}, at lower gravitational potential) run slower,
and for high-precision clocks, these differences can be large.  A
height difference of 1~cm yields an observed frequency shift of a
part in $10^{-18}$, so for clocks located at NIST in Boulder, CO,
(altitude $\Delta h\sim$ 1600 m with respect to the Earth geoid),
the gravitational correction is about 2 parts in $10^{13}$. For
clocks to reach uncertainties at the $10^{-18}$ level,
clearly detailed knowledge of the geoid will be required. This
situation is further complicated by the fact that daily local
fluctuations in altitude and hence gravitational field occur at
the $10^{-17}$ level \cite{Kleppner2006}, so one day it may be
necessary to locate our best clocks far from Earth's surface.  Or
conversely, such clocks could actually be used to perform
precision ``relativistic geodesy'' on Earth's surface
\cite{Chou2010a,Schiller2009}.

\subsection{Locking errors due to technical effects}

In addition to the effects already discussed in this section,
which are mostly of a fairly fundamentally physical nature, there
exists always an array of technical effects that can cause the
probe laser to be stabilized just off the line center of the clock
transition.  Some are quite general to most clock systems, while
others are more particular to a given clock apparatus.  In
general, these details are tucked away in articles on specific
clocks and at times not mentioned at all.  Here we give several
examples of locking error that many clock systems need to
consider.

One of the most important issues involves configuring the
electronics to minimize ``servo'' error.  Typically clock
experiments operate with line quality factors of 10$^{14}$ at
best, yet they strive for performance at the part in 10$^{17}$
level or better. This requires finding the center of the
spectroscopic lineshape to a part in 1000 or better.  In the past
this meant controlling electronic offsets at the milllivolt level or
below, but well-designed digital servo systems have made this
problem much more manageable (the Cs fountain microwave clocks now
achieve a line splitting better than 1 part in 10$^5$!).  A second
source of locking error results from unpredictable drifts of the
probe laser reference cavity.  Servo systems have limited
correction capabilities, so a laser that is drifting too fast or
erratically can have offsets relative to the atomic transition to
which it is stabilized.  For this reason (among others) clock
scientists go to considerable trouble to isolate the reference
cavities (to which the probe laser frequencies are ultimately
stabilized) from environmental effects such as temperature and
vibrations.  A third, less common (but particularly insidious)
source of locking error results from uncompensated vibrations of
the probe-laser optics that send the light into the apparatus
containing the atoms or ions.  If such vibrations are synchronized
with the measurement cycle (for example, through the mechanical
impulse caused by the electronic shut-off of magnetic field
gradient coils), they can cause an effective and repeatable
Doppler shift that leads to an offset in the locked laser
frequency from the spectroscopic line center. Such effects can be
detected (or even zeroed) by changing (or choosing) the time
duration between, say, the coil shut-off and the probe time, but
unfortunately such offsets can drift with time and temperature, so
they can require considerable care.  In one case, such
vibration-induced effects drove one group to suspend their atom
apparatus, pendulum-style, above the main optical table
\cite{Wilpers2007}.  In the end, locking/servo errors rarely limit
ultimate clock performance, but often require significant
attention on the part of the researchers.

\section{Neutral atom optical clocks - general description}

Neutral atom-based optical atomic clocks have attracted great
interest for several reasons including comparative ease of
construction and convenient wavelengths, but their principal
advantage can be seen in Eq.~\ref{qpnstab}. Here we see how the
fractional instability depends inversely on the square root of the
number of quantum absorbers in the system.  For the systems we
will describe in this section, $N$ can be anywhere from 1000 to
50000.  As a result, neutral atom systems have the potential to be
10 to 100 times more stable than systems based on a single quantum
absorber. When used in conjunction with ultra narrow transitions
such as those for intercombination lines of two-electron atoms, as
are commonly used in neutral atom clocks, these large $N$ systems
can, in principle, have fractional instabilities well below
$10^{-16}\tau^{-1/2}$ \cite{Hollberg2005a}.  However, as we
discussed previously in sect.~\ref{systematics}, this potential
advantage can be realized only in systems with frequency noise
lower than this atom shot noise limit (given by Eq.
\ref{qpnstab}). For this reason, neutral atom clocks were slow to
capitalize on their large numbers of atoms, but as we shall see,
recent experiments have demonstrated unprecedented levels of
stability. The absolute frequency uncertainty obtainable with
these systems is a more complicated issue.  As we have seen in
sect.~\ref{systematics}, many effects must be considered when
computing the uncertainty budget for a given clock, and the
combined result depends greatly on the particular system. However,
the dominant consideration, which has driven the design, and
really the evolution, of neutral atom-based systems is how best to
handle Doppler effects.

For this reason, neutral atom frequency standards and clocks have
followed a fairly steady progression from thermal atom-based
systems to laser-cooled atomic systems to those based on atoms
trapped in optical lattices.  Recall that sub-Doppler methods can
greatly suppress these shifts for thermal and laser-cooled atoms;
indeed we will describe in this section several clocks based on
sub-Doppler spectroscopic techniques.  However, these techniques
rely on extreme parallelism between the counter-propagating beams
(and almost perfect wavefronts).  Various experiments have shown
that this is hard to do at the microrad level (implying hertz-level
residual shifts) \cite{Degenhardt2005,Wilpers2007,Friebe2011}.
Even the use of an optical cavity for the probe beams does not
solve this problem, due to the mismatch between the beam waist and
atom cloud size (and wavefront curvature away from the beam
waist).  Researchers thus had to conclude from these experiments
that it was unrealistic to expect sub-Doppler methods used in
combination with freely expanding atomic clouds to achieve
absolute uncertainties of much below 1 Hz (or $10^{-15}$
fractionally), even with laser-cooled atoms.

Thus, scientists have ultimately turned to optical lattice-based
clocks to achieve the highest levels of performance.  With clever
choice of atomic transition and lattice wavelength, perturbations
to the clock frequency induced by the lattice beams can be
suppressed to the sub-$10^{-17}$ level \cite{Katori2003}. Despite
the considerable additional experimental complication required,
optical lattice clocks now dominate the field of high-precision
neutral atom clocks, with 15 to 20 existing worldwide and versions
based on three different atomic species. In less than ten years,
these clocks have gone from being a new idea to having achieved
significant performance milestones, with demonstrated inaccuracy
at the sub-$10^{-17}$ level and instability at the $3 \times
10^{-16}\tau^{-1/2}$ level.

Let us now examine in more detail the most influential neutral
atom optical clock systems, with examples of clocks based either
on freely expanding laser clouds (both at room and laser-cooled
temperatures) or on laser-trapped atoms (\emph{i.e.}, lattice
clocks). In this way we will see how the combination of the atomic
system, atom preparation, and excitation scheme determines clock
strengths, limitations, and future prospects in terms of clock
performance and potential applications.

\subsection{Optical Clocks Based on Freely Expanding Neutral Atoms}

Due to their relative experimental simplicity, optical clocks and
frequency references based on freely expanding (\emph{i.e.}
untrapped, in free fall) atoms have been popular systems for study
for more than four decades.  The atom source can be a cell filled
with atom vapour, a thermal atomic beam, a laser slowed/cooled
atomic beam, or even atoms released from a magneto-optical trap.
With vapor cells, one usually employs  a form of saturation
spectroscopy (or optical heterodyne saturation spectroscopy
\cite{Hall1981}, for higher precision), with overlapped
counter-propagating laser beams to excite the atoms.  Due to
pressure-dependent collision effects and cell contaminants, the
reproducibility between vapor cell frequency references is usually
limited at the kilohertz level, making vapor cells less ideal for the
highest precision optical clock research.  Still, they are very
convenient and compact sources, and fractional frequency
instabilities as low as $5\times10^{-14}\tau^{-1/2}$ have been
reported for a molecular iodine optical clock \cite{Ye2001}.

\begin{figure}[t]\begin{center}
\includegraphics[width=9 cm]{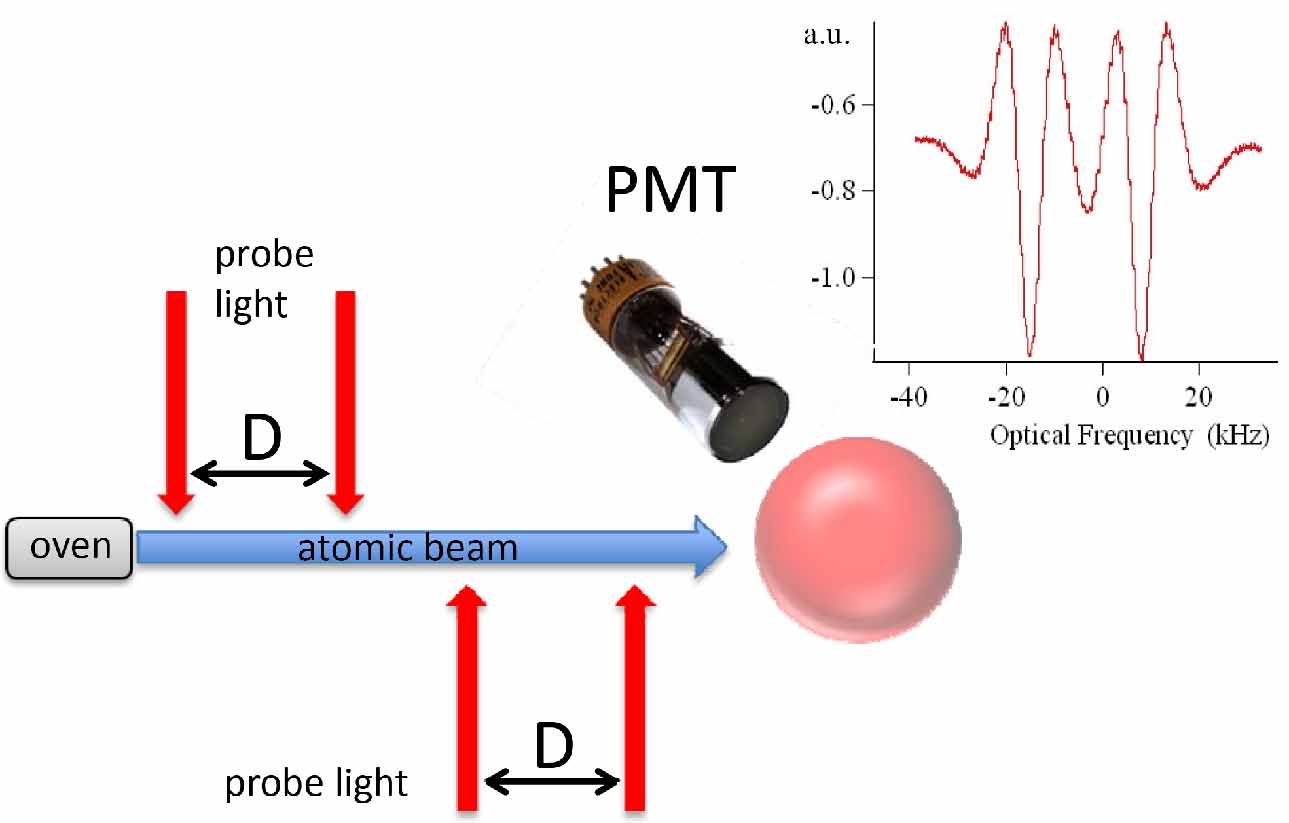}
\caption{Schematic representation of Ramsey-Bord\'e spectroscopy
on atomic beam with spatially separated interaction
regions.\label{fig.Ramsey1}}
\end{center}
\end{figure}

Cell limitations are overcome by using a thermal atomic beam (see
fig. \ref{fig.Ramsey1}), usually fashioned by having an atomic
vapor effuse from a crucible containing a small hole.  In most
cases it is necessary to heat the crucible to a temperature
ranging from 600 to 1000 K, which produces an atomic beam with a
most probable velocity, v$_{prob}$, of $\sim$~500~m/s.  The atomic
beam is typically collimated at the $10$ $\%$  level, so there is
significant Doppler broadening even for probe beams oriented
orthogonally to the atomic beam (the usual geometry). Thus,
sub-Doppler techniques using counter-propagating laser beams are
required to reach sub-MHz spectroscopic resolution. Interestingly,
due to the long lifetime of the excited state ($\sim$ 1 ms) of
these ultra-narrow clock transitions, the counter-propagating
beams can be spatially separated as the $1/e$ distance for decay
from the excited state can be 30 cm or longer. This enables
excitation with travelling waves rather than standing waves (and
its associated spatial intensity dependence), thereby yielding
more uniform excitation of the atoms.  However, a single of pair
of counter-propagating excitation laser beams is not sufficient
for high resolution as the fast-moving atoms pass through a
typically sized laser beam in $\sim 10$ $\mu s$, limiting the
resolution via transit-time broadening (essentially a Fourier
effect that constrains resolved linewidths to be greater than the
inverse of interaction time) to around 100 kHz.

For this reason, high-resolution optical experiments employ instead
a modified version of the Ramsey method of separated excitation
zones \cite{Ramsey1949}, which has been used in microwave clocks
for many decades.  The resolution provided by the two-beam Ramsey
geometry is determined by the separation of the beams, $D$, rather
than the transit time through a single beam.  The first
interaction zone excites the atoms into a superposition of
the excited and ground states.  This superposition evolves in the
dark zone between excitation fields, whose phase is then read out
by the second field.  Mathematically (and intuitively), this
two-zone Ramsey excitation is analogous to the Young's double slit
experiment, in which the excitation zones play the role of the
slits, and the atomic excitation plays the role of the light.
Indeed, the resulting Ramsey excitation is an interference pattern
with a period equal to the inverse of the transit time between
zones.  The optical version uses two pairs of Ramsey beams, with
the second pair counter-propagating to the first, in order to
suppress the first-order Doppler shift
\cite{Letokhov1972,Baklanov1976,Bergquist1977}.  In this case the
period of the Ramsey fringe interference signal is equal to
$v$$_{prob}/2D$, due to the doubled length of the spectrometer.  The
analogy with Young's double slit experiment goes further, as
this optical Ramsey spectrometer is indeed an atomic
interferometer as emphasized first by Ch. Bord\'{e}
\cite{Borde1989}.  For this reason, four-beam optical geometries of
this type are often termed ``Ramsey-Bord\'e spectrometers''
\cite{Borde1984}.  As we will see, such spectrometers can be used
not only to generate high resolution spectra, but can also be used
as atomic interferometric sensors to measure quantities such as
rotation and Stark shifts \cite{Riehle1992}.  This technique
offers the additional benefit of high signal-to-noise ratios as
the large effective Doppler coverage (due to transit broadening)
of the small excitation zones excite a non-trivial fraction of the
atomic distribution.  However, due to the wide range of the
velocities in the atomic beam, the source has a small degree of
coherence, so usually only a few interference fringes are visible,
and these usually come in two sets, one for each of the two
``recoil'' components \cite{Hall1976,Oates2005}.  These recoil
components are a by-product of performing spectroscopy with
counter-propagating beams and yield a pair of resonances
symmetrically located around (and close to) the unperturbed atom
resonance.  In fact only in high resolution work are the two
features usually resolvable, since their separation is usually on
the order of tens of kilohertz.

\begin{figure}[t]\begin{center}
\includegraphics[width=0.9 \textwidth]{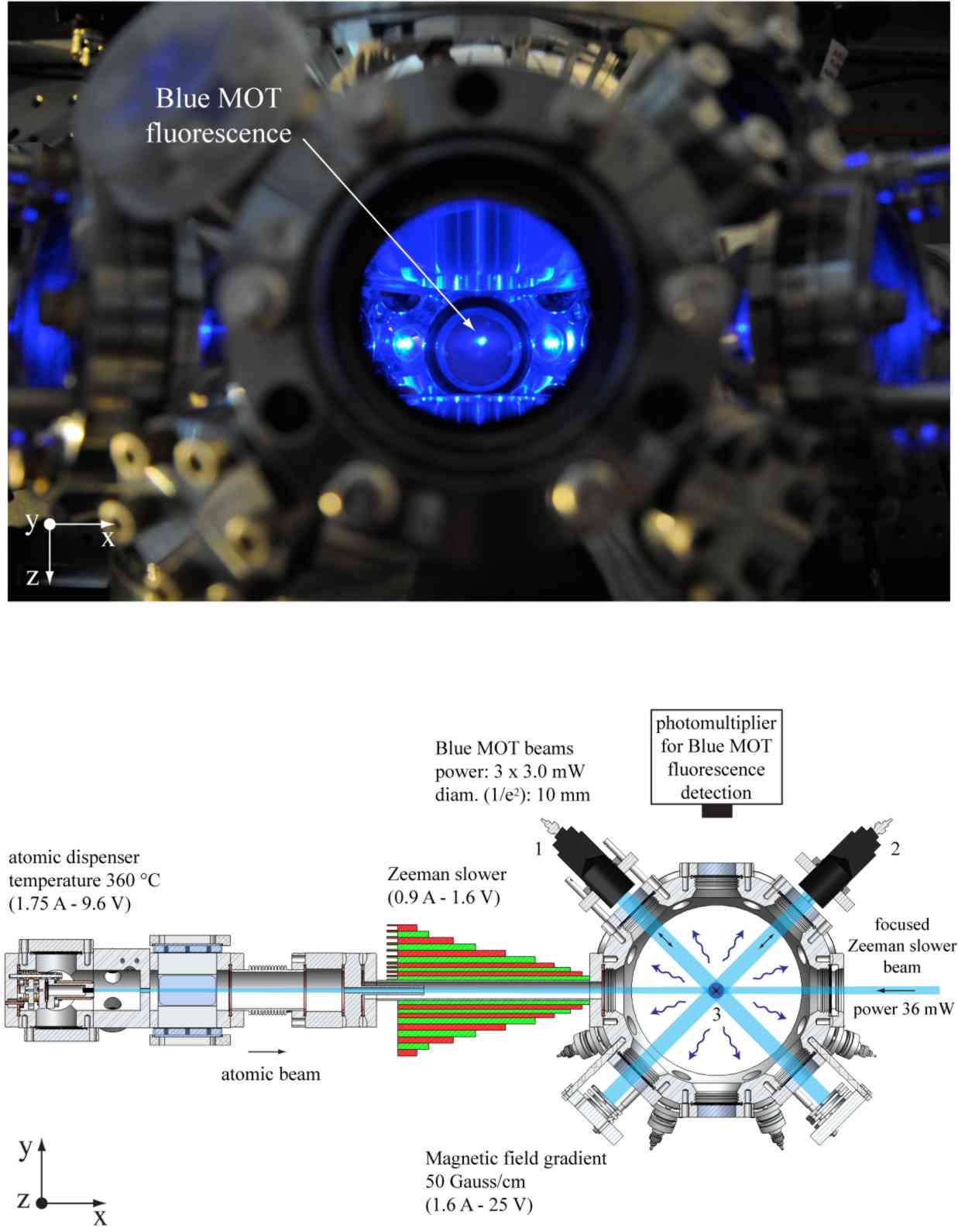}
\caption{Typical alkaline-earth optical clock experimental setup.
Atoms are evaporated in a high temperature oven, successively
slowed longitudinally and finally trapped on the main chamber with
MOT beams [adapted from
\cite{SchioppoPhDThesis}].\label{fig.TypicalSrLens}}
\end{center}
\end{figure}

For accurate work, thermal beam-based Ramsey-Bord\'e
interferometers have several drawbacks, not surprisingly Doppler
related.  First, we recall that the second-order Doppler shift is
on the order of 1 kHz for typical thermal beam velocities, leading
to uncertainties of tens of hertz or more. Moreover, these
spectrometers have a sensitivity to the phase of the laser
excitation laser beams that make them very sensitive to beam
optics, and beam reversal techniques can only partially mitigate
these effects \cite{Morinaga1989}.  For these reasons, the most
accurate experiments employ laser cooling, either simply to slow
the beams, or more commonly to collect the atoms in
magneto-optical traps (MOTs, see fig. \ref{fig.TypicalSrLens})
\cite{Raab1987} before releasing the millikelvin or microkelvin
atoms for spectroscopy (the laser beams in a MOT distort atomic
energy levels via the AC Stark effect too much to permit precision
spectroscopy for the trapped atoms) \cite{Kurosu1990,Kisters1994}.
We note, however, one important exception to this rule: the most
accurate measurement to date of a neutral atom transition with
unconfined atoms was actually performed with thermal atoms and was
made in more of a fundamental physics context than a clock
context.  Due to its calculability, the 1S-2S transition in H is
fundamental in testing quantum electrodynamics and for the
determination of such fundamental quantities as the Rydberg
constant and the proton charge radius. Heroic experimental effects
led by the group of Th. H\"{a}nsch have achieved fractional
uncertainties of $4.2\times10^{-15}$ for this transition through
careful evaluation of the second-order Doppler shift and the use
of two-photon spectroscopy, which can reduce first-order Doppler
effects \cite{Parthey2011, Matveev2013}.

\begin{figure}[t]\begin{center}
\includegraphics[width=9 cm]{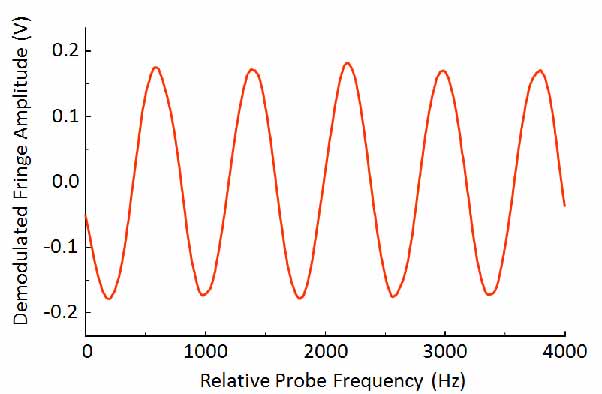}
\caption{Bord\'e-Ramsey fringes with millikelvin Ca atoms.  The
resolution, 400 Hz, is comparable to the natural linewidth for
this transition.\label{fig.RamseyCa2}}
\end{center}
\end{figure}

For MOT-based systems with atomic samples at millikelvin
temperatures, the second-order Doppler shift becomes negligible,
while residual first-order shifts are greatly reduced but still
persist at the MHz level.  Thus, it is still necessary to use
optical Ramsey sequences, this time separated temporally, rather
than spatially, to achieve high resolution signals
\cite{Kisters1994}.  In this case due to the much higher
``coherence'' of the atomic sample, a large number of fringes is
usually visible (with a period of $1/4T$), thereby requiring some
care in the identification of the central fringes (see fig.
\ref{fig.RamseyCa2}). In fact, each recoil component provides its
own set of fringes, so for maximum signal contrast it is necessary
to choose a fringe period that enables the two sets to add
constructively.

Residual first-order Doppler effects (principally those associated
with imperfect wavefronts and imperfect beam alignment) still
limit the absolute uncertainty of systems based on millikelvin
atoms to about $10^{-13}$, so in some cases a second stage of
cooling has been used to reduce the temperature by a factor of
100 to 1000, and the uncertainty in the clock frequency by a factor
of ten or so
\cite{Katori1999,Vogel1999,Binnewies2001,Curtis2003,Wilpers2007}.
But as we will see as with some specific optical atomic
clock systems, it seems prohibitively hard to approach the
accuracy of the best microwave clocks with optical transitions
excited in freely expanding atoms.  Nonetheless, these optical
systems have demonstrated much higher stability than their
microwave counterparts and as a result could find many important
applications due to their relative simplicity in comparison with
lattice-based clocks.

\subsubsection{Ca optical clock}

\begin{figure}[t]\begin{center}
\includegraphics[width=8 cm]{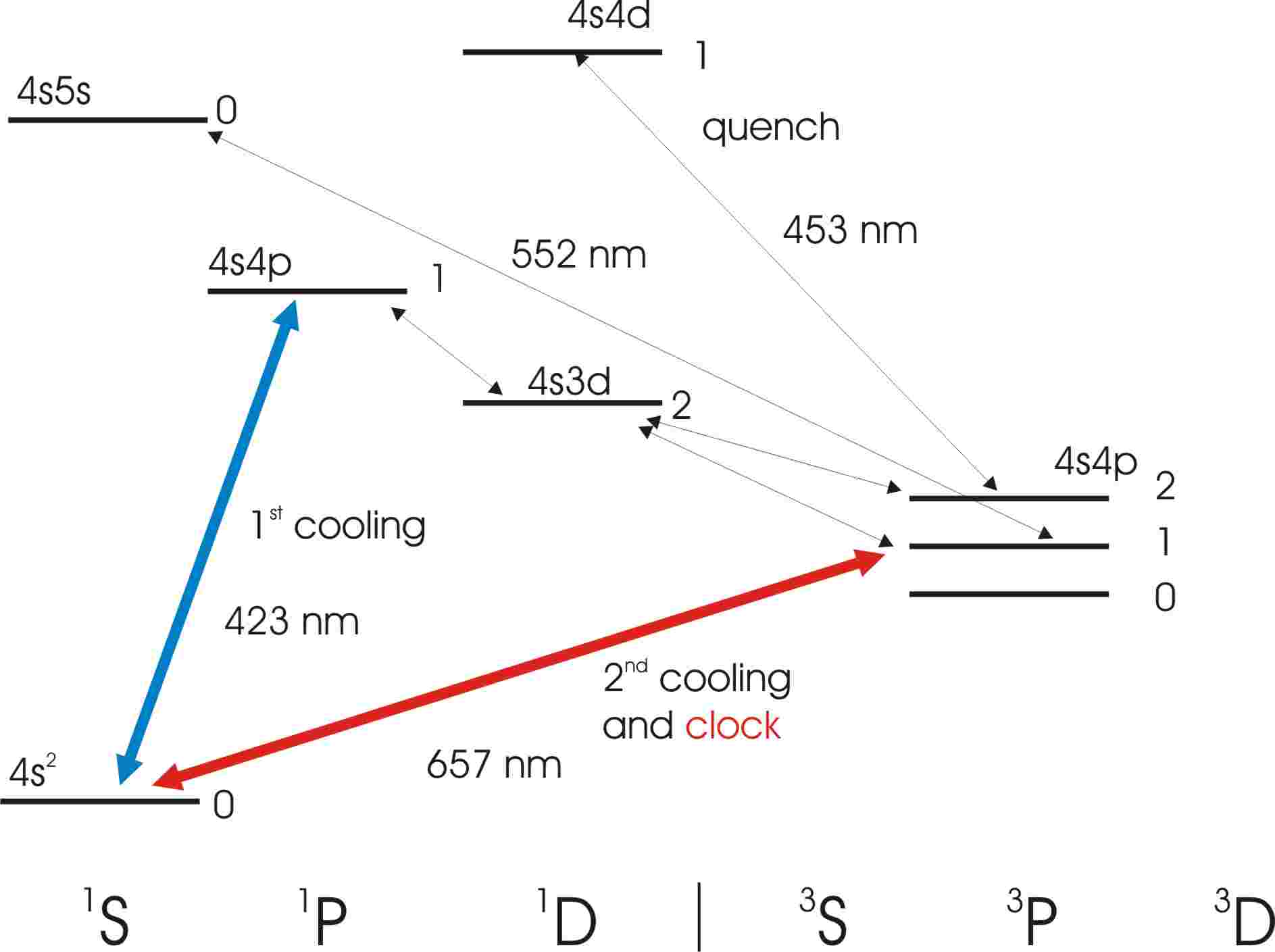}
\caption{Energy levels and relevant optical transition for neutral
Ca. The number near the level is the total angular momentum
$J$.\label{fig.LivelliCa}}
\end{center}
\end{figure}

Calcium was the first element to be probed with kHz-level
resolution in the optical domain.  Calcium's appeal for high
resolution spectroscopy lies in its $^1S_0\rightarrow^3$$P_1$
intercombination line, which has a natural linewidth of only
400~Hz (due to its forbidden nature) at a convenient visible
wavelength of 657 nm (see fig. \ref{fig.LivelliCa}).  Additionally
calcium has a convenient laser cooling transition at 423 nm.
Landmark experiments by Barger and Bergquist used the ``new''
technique of optical Ramsey spectroscopy to resolve spectroscopic
features with linewidths as narrow as 2~kHz (FWHM)
\cite{Barger1979,Barger1980}.

In the following decades, several groups, most notably those at
PTB, Tokyo, and NIST,  performed a large number of experiments
based on this transition in $^{40}$Ca (the most abundant isotope),
some using it as an atom interferometer \cite{Riehle1992}, while
others were focused on its potential as a frequency standard
\cite{Zibrov1994,Morinaga1991,Kisters1994,Wilpers2007}. Early Ca
experiments used thermal atomic beams and specialized vapor cells,
while later experiments employed atoms released from
magneto-optical traps in an effort to reduce Doppler-based
systematic effects.  Such experiments used one and sometimes two
stages of laser cooling, while Ca manipulation techniques
culminated in the observation of a Ca Bose-Einstein condensate in
2009 \cite{Kraft2009}.

Let us first consider atomic beam-based Ca spectrometers.  The
majority of experiments employed an apparatus similar to that used
in the original NIST experiments \cite{Barger1979}.  Figure
\ref{fig.Ramsey1} shows how these experiments were laid out along
the axis of propagation of the atomic beam. The atoms emerge from
an oven held at a temperature of $\sim650$ $^{\circ}$C, which
produces an atomic beam with a most probable velocity of $\sim
640$ m/s. Fluorescence is detected downstream with a
photomultiplier tube. With a magnetic bias field of a few Gauss
generated by a set of Helmholtz coils the first-order
magnetically-insensitive $m_J=0\rightarrow m_J=0$ transition can
be excited.  As discussed before, two recoil components are
visible at resolved linewidths below the 23.1 kHz splitting for
the Ca recoil components at 657 nm.  In fig. \ref{fig.Ramsey1} we
show optical Ramsey fringes taken with a resolution of 5~kHz
(resulting from an excitation beam spacing $2D$ = 9 cm), which
resolves cleanly the two recoil components.  This spectrum is
taken with an extended-cavity diode laser that is pre-stabilized
to a high-finesse optical cavity. The laser has a fractional
instability of $2\times 10^{-15}$ at 1 s, and an effective laser
linewidth of about 1 Hz on a 1 s time scale \cite{Oates2000}.


To lock the laser frequency to the transition, one modulates the
laser at $\sim 300$ Hz with a modulation depth equal to the full-width half-maximum of the spectroscopic feature (here 5 kHz).
Subsequent demodulation yields an odd-symmetry error signal
suitable for locking.  Feedback with a loop bandwidth of $\sim$ 10
Hz can yield a signal that yields a short-term stability of
$5\times 10^{-15}$ at 1 s for a single laser system.  Such systems
could find a variety of precision timing applications, such as
ultra-low noise microwave generation, in cases where stability is
needed for tens or hundreds of seconds.  Other applications arise
from using the system to measure rotations via the Sagnac effect, DC
electric fields via the Stark Effect, and AC Stark shifts via
additional laser beams at different colors strategically located
in the apparatus \cite{Riehle1992}.

As discussed earlier, thermal systems will always be limited in
terms of absolute uncertainty and even long term stability because
of first- and second-order Doppler shifts.  First order shifts
lead to sensitivity to probe laser-beam alignment, while the
second-order effect leads to sensitivity to laser power
fluctuations, oven-temperature fluctuations, and uncertainty in
the velocity distribution.  For these reasons, groups at Tokyo,
PTB, and NIST (among others) used laser-cooling either with atomic
beams or MOTs in an effort to make more accurate standards
\cite{Kurosu1990,Kisters1994,Wilpers2007}.  For Ca, the MOT-based
systems loaded millions of atoms either from Zeeman-slowed beams
or directly from laser-slowed atomic beams.  Initial versions
loaded atoms for a fraction of a second, then probed the atoms for
up to a millisecond with a four-pulse optical Ramsey sequence
\cite{Kisters1994}.  A high signal-to-noise ratio resulted from
normalized shelving detection based on earlier ion detection
techniques (see, for example, reference \cite{Itano1987}), which
collects fluorescence from a strong transition (\emph{e.g.}, the
$^1S_0\rightarrow^1$$P_1$ cooling transition) rather than directly
from the clock transition.    A latter version used a much more
rapid sequence that loaded for 2 ms, probed for 500 $\mu$s, and
then quickly recaptured a majority of the atoms for a subsequent
cycle \cite{Oates1999}.  The advantage of this approach is that
the short cycle can yield high stability ($4\times 10^{-15}$ at 1
s, $4\times 10^{-16}$ at 100 s) \cite{Oates2000}, but the
disadvantage is that the magnetic field gradient needs to be left
on throughout the whole sequence, yielding a large, though fairly
steady systematic shift.  For more accurate performance the longer
cycle was used, for which the magnetic field could be well
controlled and characterized.

Still, the best of millikelvin Ca clocks had absolute fractional
uncertainties of $\sim 10^{-13}$, limited principally by the
first-order Doppler effect, which includes non-parallelism of the
counter-propagating laser beams and movement of the atoms through
non-planar probe laser-beam wavefronts.  Two groups pushed the
performance further by implementing a second stage of laser
cooling based on the $^1S_0\rightarrow^3$$P_1$ clock transition.
However, the cooling force associated with this transition is too
weak for efficient cooling, so these groups accelerated the
cooling cycle with an additional ``quenching laser'', either with
the $^3P_1\rightarrow^1$$D_2$ transition at 453 nm
\cite{Binnewies2001} or with the $^3P_1\rightarrow^1$$S_0$ at 552
nm \cite{Curtis2003} (see fig. \ref{fig.LivelliCa}).  Quenched
cooling for calcium enables reduction of the atom temperature to
10 $\mu$K in three dimensions, and yields a distinctly different
line shape due to the increased coherence of the atomic sample.
With the reduced velocity the clock system attained an absolute
fractional frequency uncertainty of $7.5\times10^{-15}$, still
limited by motion of the atoms in non-planar and non-parallel
wavefronts \cite{Wilpers2007}. Independent absolute frequency
measurements of the Ca optical clock frequency agreed at the 1
part in $10^{14}$ level \cite{Degenhardt2005,Wilpers2007}.  While
lattice clocks have attained a considerably smaller absolute
uncertainty than that of Ca-based clocks, there is still interest
in calcium in the clock world due to its potential as a simple
system that can attain high stability and in even more exotic ``Ca
laser'' experiments \cite{Xie2010}.


\subsubsection{Mg optical clock}

In many ways the development of clock spectroscopy with magnesium
is similar to that of calcium, with several important differences
that result from specific differences in their atomic properties.
Early experiments, pioneered by W. Ertmer's group in Bonn, were
atomic beam-based and used the $^1S_0\rightarrow^3$$P_1$
intercombination line at 457 nm (analogous to the 657 nm
transition in Ca shown in fig. \ref{fig.LivelliCa}).

This transition has a natural linewidth of only 31 Hz, and due in
part to its higher frequency, it has a 5 to 10 times smaller
blackbody shift than its strontium and ytterbium counterparts
\cite{Porsev2006}.  These potential advantages are somewhat offset
by experimental complications associated with more challenging
laser wavelengths and higher atomic velocities due to the
comparatively light mass of $^{24,25}$Mg.  The atomic beam clock
experiments principally used the four-beam optical Bord\'e-Ramsey
geometry with atoms emerging from an effusive oven whose induced
fluorescence was detected downstream.  The atoms fly through the
probe beams with high velocity (v$_{prob} = 900$ m/s), thereby
reducing the interaction periods and spectroscopic resolution.
Nonetheless, spectra have been achieved with Mg thermal
beams with resolutions sufficiently high to reveal clearly the
79.6 kHz splitting between the Mg recoil components
\cite{Sterr1992}.  In 2008 a thermal beam system was used to
perform an absolute frequency measurement of the intercombination
line with a fractional uncertainty of $2.5\times10^{-12}$, limited
predictably by first- and second-order Doppler shifts.

A more precise measurement resulted from using laser-cooled atoms
for atoms released from a MOT based on the strong
$^1S_0\rightarrow^1$$P_1$ cooling transition at 285 nm.  As a
result of a larger Doppler width and a weaker line, only $1$ $\%$ of
the atoms in the cold (3 mK) sample could be excited on resonance,
so a modified detection scheme based on optical pumping the atoms
to the $^3$P$_2$ state from which they could be cycled on the 383
nm transition (with help from repumping from the $^3$P$_0$)
enabled high signal-to-noise spectra to be obtained. In this way
spectral linewidths as narrow as 290 Hz were resolved, and an
absolute frequency measurement with an uncertainty of
$7\times10^{-14}$ was performed \cite{Keupp2005,Friebe2011}.
Second-stage cooling has proved difficult in Mg
\cite{Mehlstaubler2008}, but there remain good prospects for use
of Mg in an optical lattice (the magic wavelength for the
$^1S_0\rightarrow^3$$P_0$ transition is predicted to be 463 nm)
\cite{Ruhmann2011}, where its smaller BBR shift could give it an
advantage over existing lattice-clock systems.

\subsection{Optical Clocks Based on Tightly Confined Neutral Atoms - Optical Lattice Clocks}
\label{sec.opticallatticeclocks} 
\begin{figure}[t]\begin{center}
\includegraphics[width=10 cm]{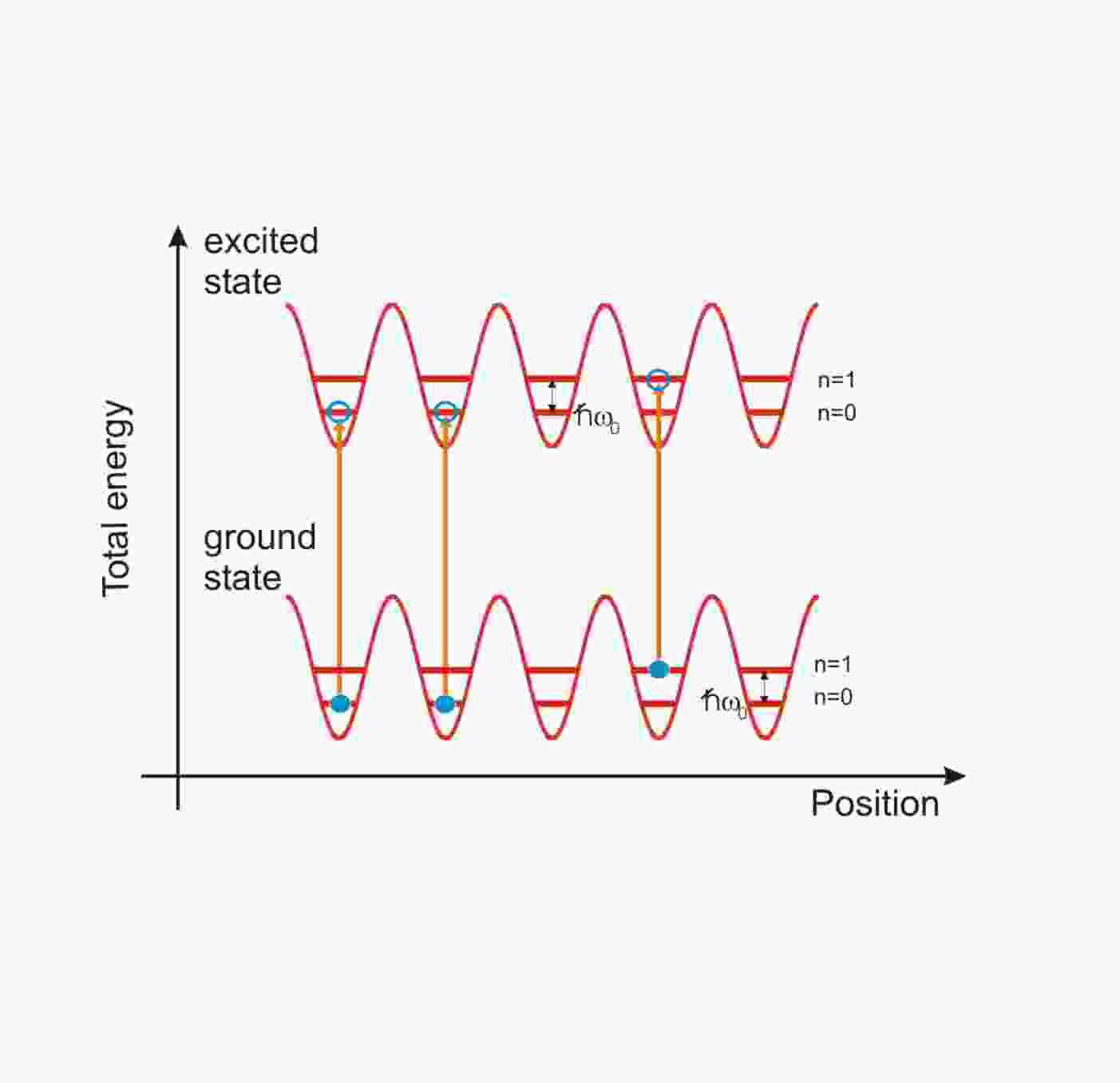}
\caption{Schematic view of the excitation process in an optical
lattice clock. Atoms are trapped in the minima of the periodical
potential created by a standing wave in a region much smaller than
the clock optical wavelength. Atoms are then excited on the narrow
clock transition while the two clock states are equally energy
shifted by the lattice potential. The index $n$ represents the
vibrational states of atoms in the lattice potential. 
\label{fig.LatticeClocks}}
\end{center}
\end{figure}

Despite their increased
complexity, optical lattice clocks have become widespread tools
for new clock research with neutral atoms due to two principal
advantages:  orders-of-magnitude higher resolution and greatly
reduced systematics, especially those related to Doppler effects.
The key aspect of a red-detuned optical lattice clock is the tight
confinement that it provides along the propagation (or
standing-wave) axis as a result of the attraction of the atoms to
high light intensity.  The one-dimensional standing wave (the vast
majority of experiments have used 1-D standing waves, although we
will mention a few that have employed higher dimensionality) forms
a series of potential wells spaced by $\lambda$/2, where $\lambda$
is the lattice wavelength (see fig. \ref{fig.LatticeClocks}).  To
fill the lattice with atoms, it is spatially overlapped with a
magneto-optical trap that typically contains millions of atoms in
roughly a cubic millimeter.  This leads to filling of about 1000
potential wells, with 1 to 20 atoms per well on average.  The
wells contain $\sim$5-10 bound levels, with the $n$ = 0, 1, 2
levels containing the vast majority of the atoms.  Along the axis
of tight confinement, the atoms oscillate at a frequency (50-100
kHz), 10 to 100 times higher than that of the atomic recoil
frequency.  In this so-called Lamb-Dicke regime, broadening and
recoil effects are absent for near-resonant spectroscopy.  Instead
the motional effects are transferred to red- and blue-detuned
sidebands, while the recoil effects are absorbed by the lattice
itself.  Under these conditions, the clock spectra can be as
narrow as the probe laser itself, with several examples of sub-Hz
lattice clock spectra existing in the literature.  Spectra with
widths of a few hertz have been used in conjunction with signal
normalization (that is, normalizing the signal to the number of
atoms in the lattice for the given measurement) to demonstrate
fractional frequency instabilities well below $10^{-15}$ at 1 s
\cite{Jiang2011,Nicholson2012}.

The key challenge in designing a lattice clock is handling the AC
Stark effect associated with the lattice itself.  The lattices
need to have a depth, $U_0/h$, of at least 100~kHz-1~MHz in order
to confine atoms with microkelvin temperatures, the coldest to
which we can straightforwardly cool atoms.  This means the
ground-state energy level in the atom is also shifted by roughly
this amount.  Such a shift seems at first glance unmanageable when
we consider that we are striving to make clocks with millihertz
accuracy.  However, this drawback largely vanishes if we make a
judicious choice of clock transition and tune the lattice
wavelength to the appropriate value.  In 2003, H. Katori and
colleagues made the idea of lattice clocks feasible with their
proposal to use the  $^1S_0\rightarrow^3$$P_0$ transition in Sr
\cite{Katori2003} (soon after, the analogous transition was
proposed for Yb \cite{Porsev2004}).  The spherical symmetry (J=0)
of both clock levels greatly reduces the clock sensitivity to
lattice polarization fluctuations, and tuning the wavelength to
its so-called magic value, at which both clock states experience
identical shifts, suppresses its sensitivity to intensity effects.
As a result the clock transition is left effectively unperturbed,
minimally sensitive to lattice fluctuations.  Perhaps
surprisingly, even the magic wavelength is not that sensitive;
holding the lattice frequency to with a range of 100 kHz is
sufficient to keep fractional clock shifts to the low $10^{-18}$
level.  In practice the magic wavelength requirement leads to
lattice frequencies detuned far to the red, which in turn leads to
shallow lattices even with a watt or more of lattice power focused
to a waist of 30 to 50 $\mu$m . (Alternatively there have been
proposals for blue-detuned, magic wavelength lattices, but these
need to be three-dimensional, since the atoms are repelled by
regions of high intensity \cite{Takamoto2009}).

\begin{figure}[t]\begin{center}
\includegraphics[width=11 cm]{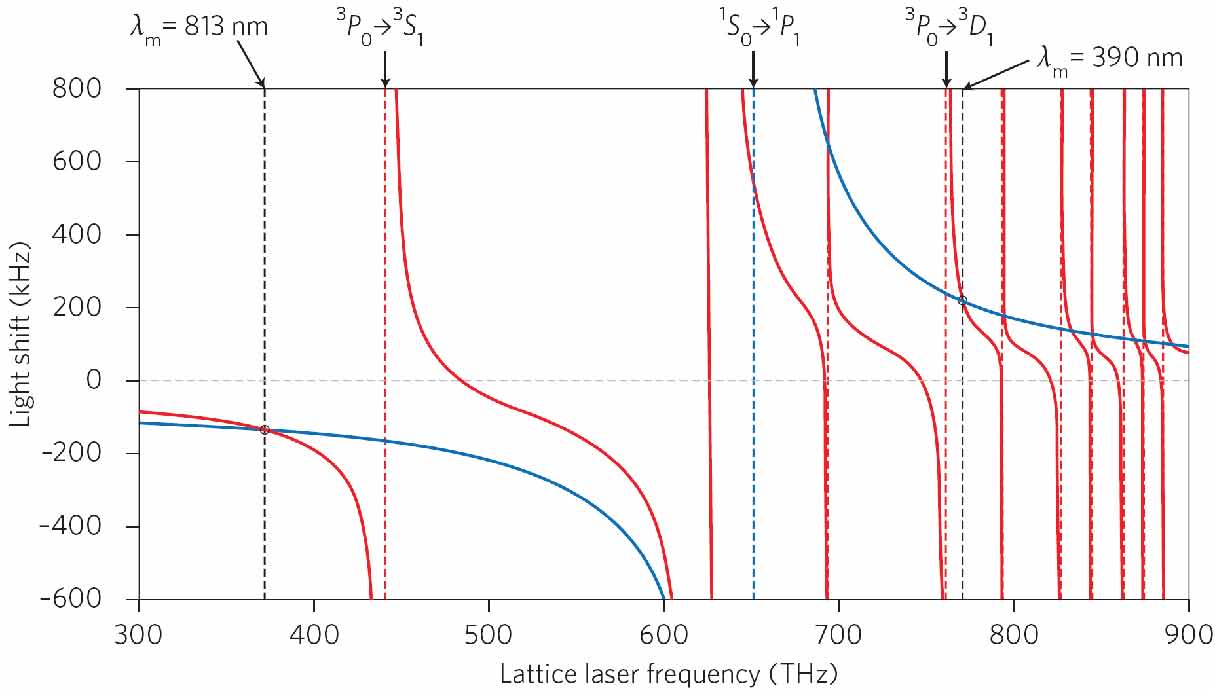}
\caption{Estimated light shifts for the Sr $^1$S$_0$ (blue line)
and $^3$P$_0$ (red line) states are plotted as a function of laser
frequency for a laser intensity of 10 kW/cm$^{2}$. The crossing
points of blue and red lines indicate the magic wavelengths. Data
from \cite{Katori2011}. \label{fig.LightShift}}
\end{center}
\end{figure}

We note that this discussion is correct only to first order in
intensity, for dipole-allowed transitions, which give the dominant
part of the lattice shifts.  We can re-write the lattice-induced
shift in eq. \ref{stark} as:
\begin{equation}
\delta\omega = \alpha_0(\omega)I+\alpha_1(\omega)I^2+ ..... ,
\end{equation}
where $\alpha_0$ is the differential polarizability for the
$^1S_0$ ground and $^3P_0$ excited states (which includes both
static and dynamic term), $\alpha_1$ is the differential
hyper-polarizability for these states, and $\omega$ is the
frequency of the lattice light. By tuning $\omega$ to its magic
value (usually in the near infrared, see fig.
\ref{fig.LightShift}), $\alpha_0$ goes to zero, but the
higher-order terms, while in general very small, still persist
\cite{Brusch2006}.  In particular, the $I^2$ term, or the
hyper-polarizability, which corresponds to two-photon transitions,
can cause non-negligible shifts that need to be evaluated,
especially in the case of Yb. Additionally non E1 transitions,
especially E2 and M1, can potentially contribute, sometimes in
non-intuitive ways.  Yudin et al. showed how in the standing-wave
lattice geometry, such transitions can contribute with a $\sqrt I$
dependence, although possibly at a negligible level
\cite{Taichenachev2008}.


With the lattice-induced shift effectively eliminated via lattice
detuning, let us now consider how to excite what appears to be a
heavily forbidden $^1$S$_0$$\rightarrow$$^3$P$_0$ transition. We
note first that the $^1$S$_0$$\rightarrow$$^3$P$_1$  transitions
are quite strong in heavy alkaline earth atoms due to mixing of
the dipole-allowed $^1$P$_1$ state into the $^3$P$_1$ state (a
result of the breakdown of strict L-S coupling rules).  However,
the $^3$P$_0$ state is still isolated from these allowed
transitions, so further mixing is required.  For even isotopes,
the transition can be excited only with the inclusion of
additional fields, either electric (AC) or magnetic, which mix a
little of the $^3$P$_1$ state into the $^3$P$_0$ state, but also
cause additional (although possibly manageable) systematic shifts
to the clock system \cite{Taichenachev2006}. For the odd isotopes
in Sr, Yb, and Hg, the situation is considerably more favorable.
These isotopes have a nonzero nuclear spin that, through hyperfine
mixing of the $^3$P$_1$ and $^3$P$_0$ levels, yields an almost
ideal natural lifetime (10-100 s) for extreme clock applications.
For such a lifetime, Hz-level spectroscopy is feasible with
sub-microwatt power levels, leading to minimal probe-induced
shifts.  Moreover, the use of systems with non-zero spins yields
multiple clock transitions split very weakly by a linear Zeeman
effect.  The use of transitions based on the two extreme m-values
enables canceling of linear B-field effects (and very weak lattice
polarization effects \cite{Katori2003}) as well as simultaneous
measurement of the B-field to evaluate the quadratic Zeeman shift.
As we will see, while experiments have been performed on both
bosonic (even isotopic) and fermionic (odd isotopic) systems, the
majority of experiments now use fermions since they can be excited
without external fields and have smaller systematic shifts to
date.

The principal remaining systematic effects for lattice clock
systems are (1) collision effects between atoms and (2)
differential blackbody-based light shifts for the clock levels. In
the original lattice clock proposals, collision shifts were
thought to be almost insignificant because one would either use an
under-filled 3-D lattice, which would effectively isolate the
atoms from each another, or use a 1-D lattice with
indistinguishable fermions, which would not collide (at least via
the dominant s-wave mechanism) due to Pauli blocking
\cite{Katori2003}. However, three-dimensional lattices have been
infrequently used due to power limitations and experimental
complexity, and 1-D experiments in both Sr and Yb have in fact
found non-negligible shifts (at the hertz level).  As we shall
see, these effects are now fairly well understood and have been
minimized with experimental design.  The blackbody effects, which
are also at the hertz level, are easier to understand but harder
to control. Control at the 1 K level seems straightforward in
existing apparatus (corresponding to a mid $10^{-17}$
uncertainty), but specialized atom chambers may be required (and
are being designed by several groups) for significant further
reduction.  While numerous atomic systems are under consideration
for lattice clock research, to date three systems, Sr, Yb, and Hg,
have been experimentally realized. In the following sections we
will describe these three systems in more detail.

\subsection{Sr optical lattice clock}
\label{sec.srlatticeclock} 
\begin{figure}[t]\begin{center}
\includegraphics[width=0.7\textwidth]{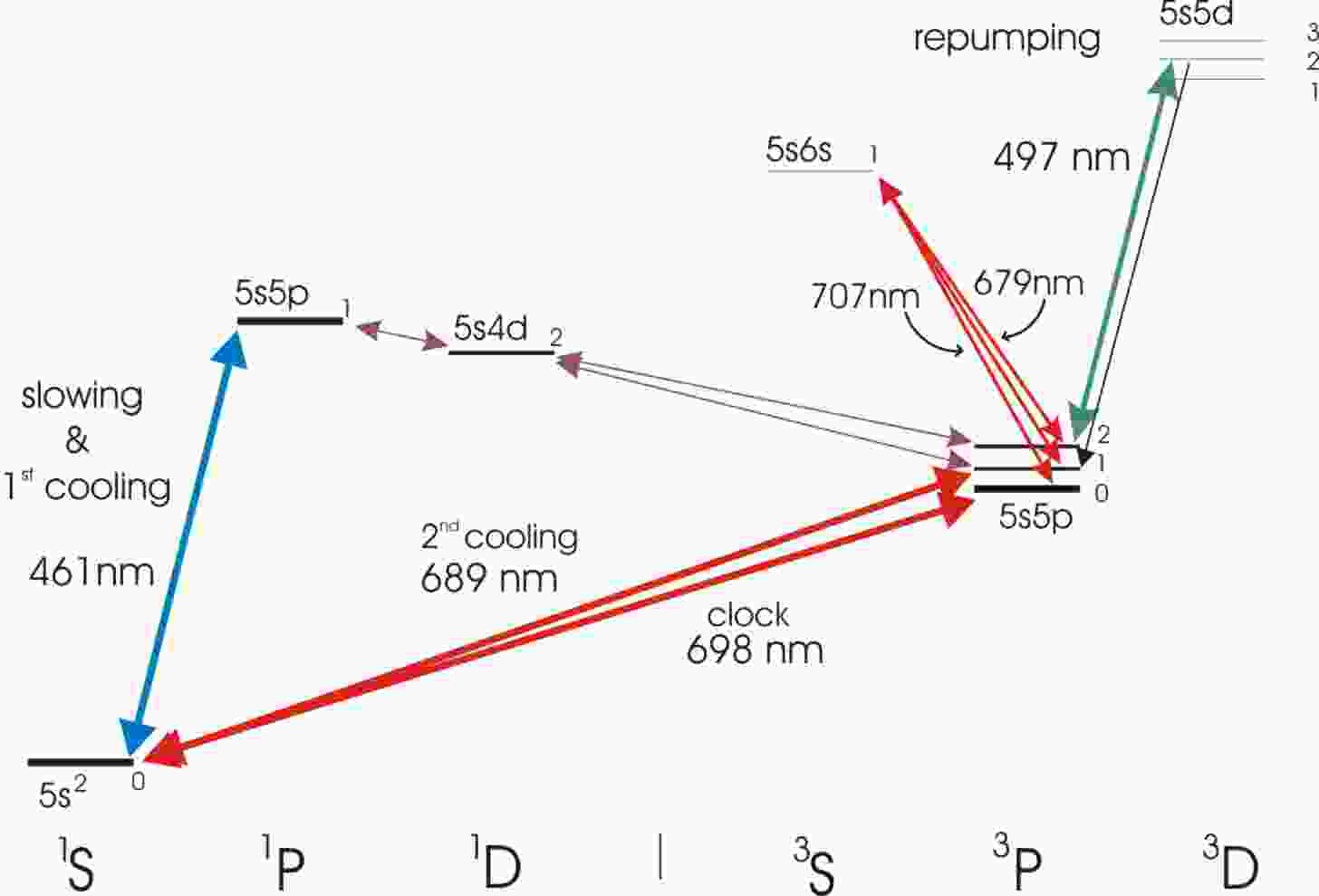}
\caption{Energy levels and relevant optical transition for neutral
Sr.\label{fig.LivelliSr}}
\end{center}
\end{figure}

As with other alkaline earth atoms,
spectroscopy on intercombination lines in Sr began with
atomic-beam and cell-based investigations \cite{Tino1992}. While
there were some precision measurements made on the
$^1S_0\rightarrow^3$$P_1$ transition \cite{Ferrari2003}, these
investigations were somewhat hampered by the shorter lifetime (21
$\mu$s) of the excited state. This lifetime, however, proved to be
ideal for second-stage cooling to temperatures of 1 $\mu$K and
below, which has helped propel Sr to the forefront of optical
lattice clock research. Indeed, Sr was the first atom proposed for
lattice research (by Katori and colleagues \cite{Katori2003}), and
the demonstration of spectroscopy with lattice-confined Sr atoms
soon followed \cite{Takamoto2003}.  As a result, Sr lattice clocks
are widespread throughout the community and have been used to
achieve many of the seminal lattice results.  Sr's suitability for
this type of research comes from its particular electronic level
scheme, presented in fig. \ref{fig.LivelliSr}.

The singlet-triplet structure found in alkaline earth atoms such
as Sr provides strong singlet-singlet transitions for rapid atom
cooling/trapping and weaker singlet-triplet transitions for more
gentle atom manipulation and of course, clock spectroscopy. In Sr
the $^1S_0\rightarrow^3$$P_0$ clock transition occurs at 698 nm,
with a strong cooling transition at 461 nm, and the weak, but
extremely effective, second-stage cooling transition at 689 nm.
There is a convenient magic wavelength for the clock transition at
813 nm, well within reach of Ti:Sapphire and semiconductor laser
systems.  The most abundant isotopes (by far) are $^{88}$Sr (82
$\%$) and $^{86}$Sr (10 $\%$), which have zero nuclear spin, and
$^{87}$Sr (7 $\%$), which has a nuclear spin of 9/2.


Laser-cooled atomic clock systems such as those based on atoms
confined to a lattice or on single-trapped ions must use a
measurement cycle since the atoms/ions need to be loaded and/or
cooled (and perhaps optically pumped) before being probed by the
clock laser.  Moreover, since the detection period following
spectroscopy usually heats or depletes the sample, the full
preparation-excitation-detection cycle must be repeated
continually during clock operation.  Here, to give the reader an
example of how such cycle-based atomic clock systems work, we will
describe the Sr lattice clock system in some detail, in terms of
both its operational measurement cycle and (more briefly) its
typical physical layout.

\begin{figure}[t]\begin{center}
\includegraphics[width=11 cm]{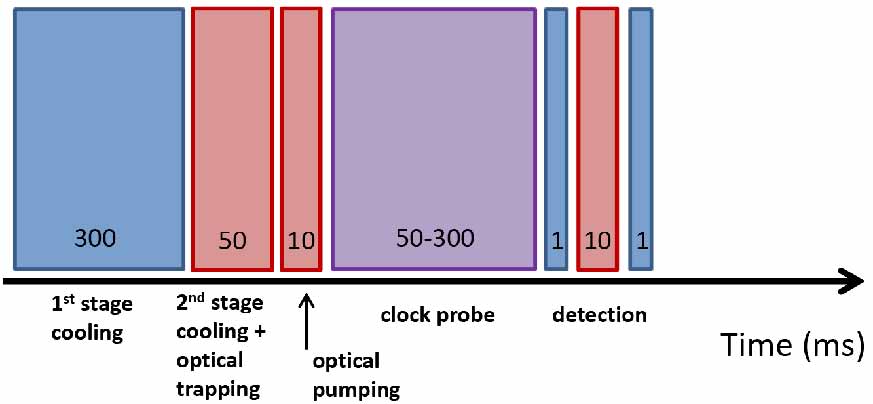}
\caption{Typical optical lattice clock
cycle.\label{fig.clockcycle}}
\end{center}
\end{figure}

The heart of the Sr clock apparatus is the vacuum system that
contains an atomic oven at one end connected by a tube to an
interaction region at the other, which consists of a stainless
steel chamber with multiple laser-grade windows to provide access
for all the laser beams needed to manipulate and probe the Sr
atoms.  The measurement cycle for a typical Sr lattice clock (see
fig.\ref{fig.clockcycle}) commences with $\sim$ 300 ms loading
period, during which a MOT based on the strong 461 nm cooling
transition is loaded from a beam of Sr emerging from a hole in the
Sr oven. The loading rate is usually enhanced with a
counter-propagating slowing beam used in conjunction with a Zeeman
slower.  Transverse heating associated with the Zeeman slower can
be compensated with transverse cooling beams upstream from the MOT
\cite{Loftus2004}. This first-stage MOT can load as many as $10^8$
atoms and cools the atoms to a few millikelvins, a bit above the
Doppler cooling limit for this transition.  A period (typically 50
ms long) of second-stage cooling based on the 689 nm
intercombination $^1S_0\rightarrow^3$$P_1$ line reduces the atom
temperature to near 1 $\mu$K, well below that of typical Sr
lattice depths ($\sim$ 10 $\mu$K).  The low temperature achievable
with a second-stage Sr MOT is principally the result of the narrow
width (7.5 kHz) of this cooling transition, which requires precise
control of the laser detuning, frequency modulation, magnetic
field gradient, and intensity.  Additionally with odd isotopes, a
second frequency source (sometimes called a ``stirring'' laser) is
required to optically pump atoms back into the cooling cycle
\cite{Mukaiyuma2003}. Careful overlapping of the lattice with the
second-stage, red MOT leads to the accumulation of up to $10^5$
atoms in the lattice, predominantly in the $n=0$ and $n=1$
motional states. We note that a different lattice loading
approach, using an atomic ``drain'' has been demonstrated at SYRTE
\cite{Baillard2008}.

\begin{figure}[t]\begin{center}
\includegraphics[width=\textwidth]{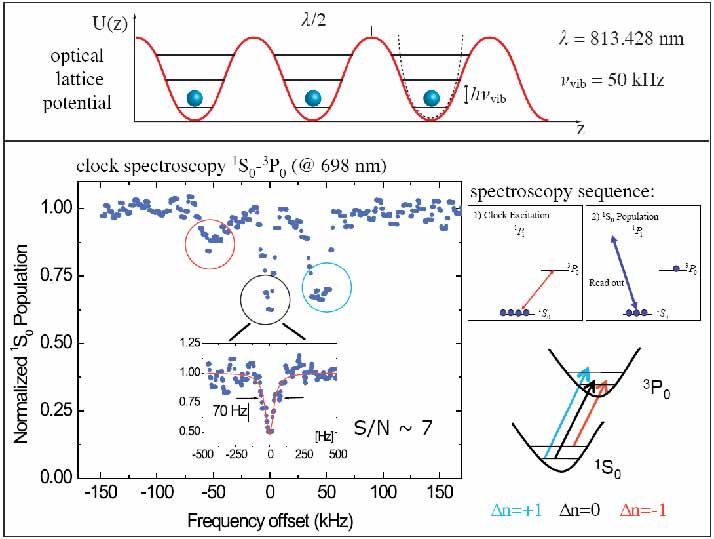}
\caption{Lattice spectroscopy on 698 nm clock $^1$S$_0
\rightarrow^3$P$_0$ transition of neutral Sr. Atoms are trapped in
the minima of the optical lattice potential (upper part) while the
clock transition is probed (lower part). Data from
\cite{Poli2009}. \label{fig.Srbroadscan}}
\end{center}
\end{figure}

Spectroscopy of the clock transition is performed by precisely
overlapping the traveling wave probe beam with the lattice beams
in order to take advantage of the Lamb-Dicke confinement of the
atoms. The frequency of the probe light is pre-stabilized on a
narrow resonance of a high-finesse optical cavity (recall sect.
\ref{sec.freqstablaser}), and an acousto-optic modulator is used
to pulse the light (and stabilize the intensity) for a preset
period of typically 10-200 ms to yield (usually) Fourier-limited
spectra.  The probe light usually excites some fraction of the
atomic sample to the long-lived upper clock state.  Near atom-shot
noise-limited detection of the excitation fraction then results
from the following detection sequence: (1) resonant 461 nm light
(for a few milliseconds) is used to scatter many signal photons
while simultaneously heating the ground state atoms out of the
lattice, (2) lasers that connect the $^3P_0\rightarrow^3$$S_1$ and
$^3P_2\rightarrow^3$$S_1$ transitions (see fig.
\ref{fig.LivelliSr}) are used to pump (in a few ms) the
still-excited atoms back to the ground state, and (3) the
previously excited atom population is read out (for a period of a
few ms) by again cycling on the cooling transition. In this way a
normalized shelving detection signal \cite{Itano1987,Clairon1995}
is generated, which is insensitive to shot-to-shot variations in
the number of atoms in the lattice and can approach atom
shot-noise-limited performance. Alternatively, we can make a
dispersive, non-destructive measurement of the excitation in an
effort to be able to recycle the atoms \cite{Lodewyck2009} more
efficiently.

In any case, the detection period concludes the measurement cycle
and yields a series of integrated PMT pulses, whose effective
areas are read as voltages by a computer and then converted to a
normalized excitation signal.  We then use the computer to step
the optical frequency via a frequency-shifting device such as an
acousto-optic modulator and repeat the measurement cycle.  In this
way, we can generate a spectrum showing the excitation voltage as
a function of laser probe frequency, with the frequency-axis
points carefully referenced to the Fabry-P\'{e}rot cavity
resonance to which the laser is locked. Moreover, since the
computer receives the spectroscopic signals and controls the laser
frequency, it can readily lock the laser frequency (seen by the
atoms) to the clock transition through implementation of a
modulation + digital servo scheme.  Such schemes can be as simple
as modulating back and forth over the half-maximum points of the
spectroscopic signal and feeding back with a digitally integrated
signal, or more sophisticated with a more complicated modulation
protocol involving multiple resonances and/or high-order loop
filters.

As a brief aside, we note that the description of the Sr lattice
clock system up to this point has included a fairly large vacuum
apparatus, at least four laser systems, a computer control system,
a Fabry-P\'{e}rot reference cavity, along with their implied
supporting optics and electronics.  Indeed, state-of-the-art
optical clocks are complex experimental devices, which can easily
fill two 1.5 m $\times$ 2.5 m optical tables, along with several
racks of supporting electronics and often a separate,
vibrationally isolated mini-table for the optical reference
cavity.  Certainly commercial or space-qualified versions will
need to be built in a more compact and robust manner, and groups
are already beginning to address these challenges (see sect.
\ref{sec.practical} for further discussion of this important
issue).

Let us now look at the spectra that result from the Sr clock
apparatus.  A broader scan taken over a range of a few hundred
kilohertz around the clock resonance reveals the key lattice
spectroscopic features (see fig. \ref{fig.Srbroadscan}). On either
side of a narrow carrier, one sees the characteristic motional
sidebands. These sidebands correspond to transitions that change
the motional state (labeled by quantum number $n$) by $\Delta n\pm
1$ motional quantum, and reveal that the motional frequency is
$\sim$ 70 kHz (the splitting between the carrier and sidebands).
From the ratio of the sideband heights we can determine the
effective temperature of the lattice confined atoms, as the red
sideband is reduced in height when more atoms are in the
ground-state motional level ($\Delta n = -1$ transitions are
forbidden from the ground state) \cite{Wineland1987}.  If we focus
on the carrier ($\Delta m = 0$ transitions), we see either a
single feature for $^{88}$Sr or a set of ten features for
$^{87}$Sr split by a weak magnetic field. In typical experiments
with $^{87}$Sr, the atomic sample is pumped to either of the
stretched states ($m=$$\pm$$9/2$) before clock spectroscopy is
performed, and then the drift of the reference cavity is
suppressed by fixing the laser probe frequency to the mid-point
between the two features.  As a result, the first-order magnetic
field sensitivity of the transition and the vector light shift are
suppressed.  Due to the high multiplicity of its ground state, Sr
also has a tensor light shift, which actually manifests itself as
a shift in the magic wavelength that is the same for the two
stretched states (see Reference \cite{Boyd2007} for details).

\begin{figure}[t]\begin{center}
\includegraphics[width=11 cm]{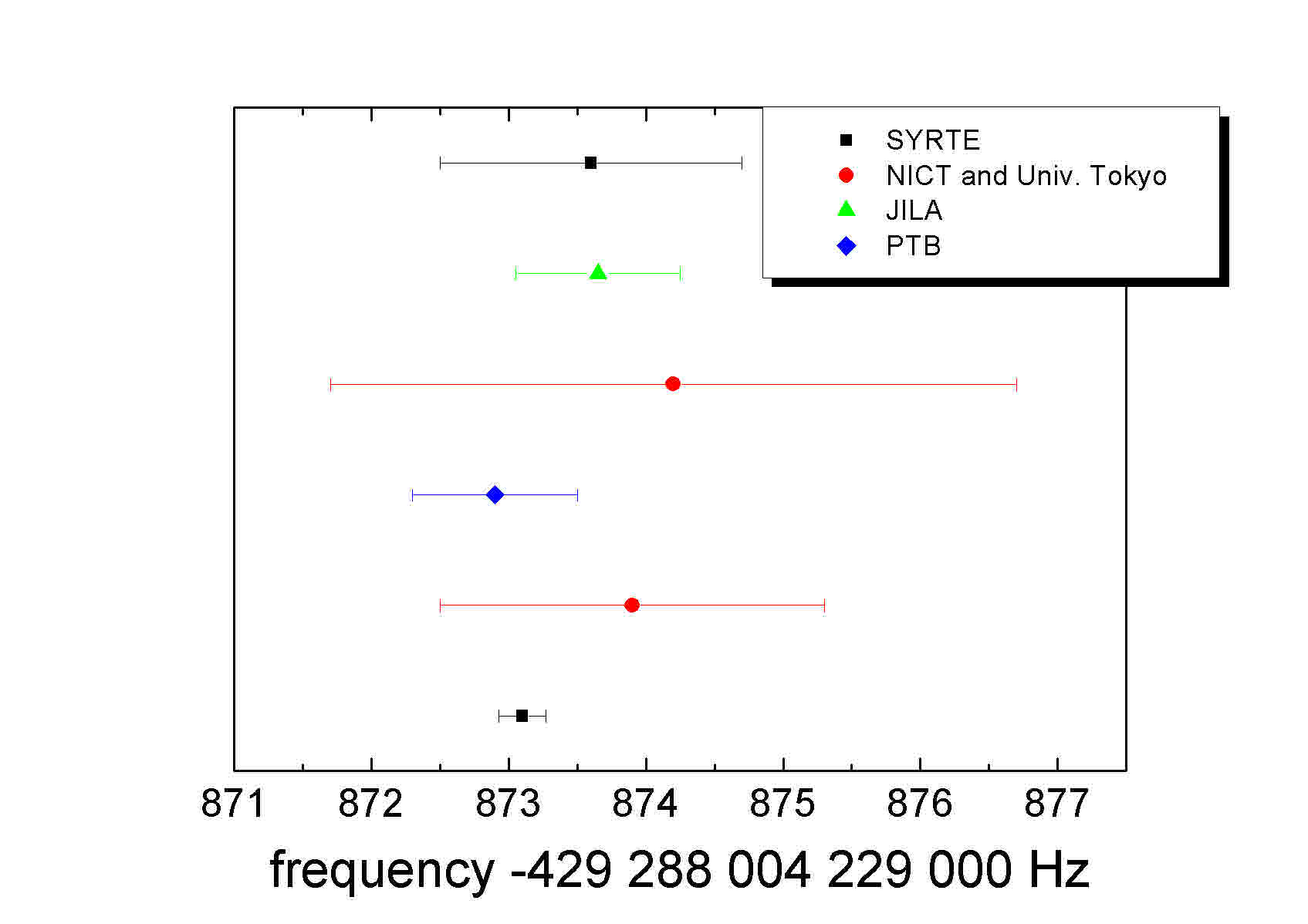}
\caption{Frequencies of the $^{87}$Sr clock transition measured by
different laboratories: SYRTE, Paris (square,
\cite{Baillard2008,LeTargat2013}), JILA, Boulder (triangle,
\cite{Campbell2008}), Tokyo Univ. and NICT (circle,
\cite{Hong2009,Yamaguchi2012}) and PTB, Braunschweig (diamond,
\cite{Falke2011}). \label{fig.SrCs}}
\end{center}
\end{figure}

Evaluations of systems based on both $^{88}$Sr
\cite{Baillard2007,Akatsuka2008} and $^{87}$Sr \cite{Ludlow2008,
Baillard2008,Hong2009,Yamaguchi2011,Falke2011} have been
performed, with $^{87}$Sr looking considerably more promising due
to its smaller quadratic Zeeman and probe light shifts.  As a
consequence the majority of research into systematic effects for
Sr has focused on $^{87}$Sr.  In 2008, a seminal evaluation of
this system was performed at JILA, which found a total fractional
uncertainty of  $1\times10^{-16}$ \cite{Ludlow2008} and was
dominated by lattice light shift, blackbody shifts, and collision
effects.  Since then, the light shifts were investigated in detail
though use of a build-up cavity for the lattice light in order to
enhance (temporarily) the effects for more precise evaluation
\cite{Westergaard2011}.  As a result, the magic wavelength was
determined to higher precision, and the influence of higher-order
effects was shown to be minimal.  In this manner, the uncertainty
for the lattice light shift term is now reduced to below
$2\times10^{-17}$, fractional uncertainty \cite{LeTargat2013}.
Detailed collision studies have been performed on both $^{88}$Sr
\cite{Lisdat2009} and $^{87}$Sr \cite{Campbell2009}, where
systematic shifts as large as 1 Hz have been seen in both systems
as well as distortions of the central feature.  For $^{87}$Sr, in
particular, it was determined that the shifts were a result of
non-uniform excitation of the trapped atoms, due in part to atoms
in different motional states and to divergence in the probe laser.
This non-uniform excitation degrades the indistinguishability of
the atoms and enables s-wave interactions to occur.  Studies with
Sr atoms in two-dimensional lattices have offered further support
for this picture and even showed how at high-enough density,
paradoxically, the atom-atom interaction can be large enough to
suppress the shift altogether \cite{Bishof2011}.  There have been
numerous absolute frequency measurements of the $^{87}$Sr clock
transition performed at the $10^{-15}$ level relative to Cs (see
fig. \ref{fig.SrCs}), and their agreement displays the robustness
of this transition to perturbations.  More recently two $^{87}$Sr
lattice clocks in one lab at SYRTE showed agreement at the
$1\times10^{-16}$ level, and their agreement with three primary Cs
standards was limited at $3\times10^{-16}$ level by the microwave
sources ($\nu_{Sr}$ = 429 228 004 229 873.10 Hz)
\cite{LeTargat2013}.

The most serious systematic remaining for the Sr lattice clock is
the shift due to blackbody radiation.  Sr has the highest
sensitivity to BBR of all of the most accurate optical clock
candidates, with a room-temperature shift of 2 Hz (or
$5\times10^{-15}$, fractionally).  Thus, it is important to know
the coefficient at the 0.1 $\%$ level, and to know and control the
environment at the 0.1 K level for net clock performance in the
mid-$10^{-18}$ at room temperature.  By means of a moving lattice
and a specialized DC Stark chamber, the differential DC Stark
sensitivity was measured with a $28$ ppm uncertainty
\cite{Middelmann2012}, in good agreement with theory
\cite{Safronova2013}.  When combined with knowledge of the dynamic
correction to yield the total BBR sensitivity, the fractional
uncertainty for the BBR shift is now only a result of the
uncertainty in the temperature of the environment, so several
groups are now preparing specialized chambers, which can be
cryo-cooled if necessary.  Due to the $T^4$ dependence of the
shift, operation of the clock at liquid-nitrogen temperature would
reduce the shift to the $10^{-17}$ level with only modest
knowledge of the temperature needed to reduce the uncertainty to
$10^{-18}$ or below.

Recently a clock laser developed for Sr lattice clock spectroscopy
at JILA has demonstrated stability at the $10^{-16}$ level due to
advances in reference-cavity design.  As a result, clean spectra
with sub-Hz linewidths have been resolved, and when locked to the
Sr signal, clock stability of $4\times10^{-16}\tau^{-1/2}$ level
has been observed \cite{Nicholson2012}.  Together with similar
performance demonstrated in Yb \cite{Jiang2011}, these are the
first clear demonstrations of the advantage that a large number of
atoms can give, as previously the frequency noise (enhanced by the
Dick effect) had masked the atom projection-noise limit.  Such
systems are capable of averaging down to less than 10$^{-17}$ in
tens of minutes (see fig. \ref{fig.ClockStability}), making
systematic evaluations at this level feasible, as evidenced by the
record low uncertainty ($6 \times 10^{-18}$) very recently
reported by the JILA group \cite{Bloom2013} (see Table
\ref{tab.Srbudget}). Moreover, still further cavity advances are
on the horizon, and instabilities well below $10^{-16}\tau^{-1/2}$
are anticipated. Sr clock's capabilities and reproducibility have
been noticed outside of optical clock circles, and as a result,
the Sr lattice clock is presently being considered as a prominent
candidate for a possible redefinition of the second
\cite{LeTargat2013} or for use as a remote clock on the
International Space Station
\cite{Poli2009,SchioppoPhDThesis,Schiller2012}.

\begin{table}
  \caption{Uncertainty budget for Sr optical lattice clock at JILA \cite{Bloom2013}.}
    \begin{tabular}{ccc}
\textbf{Sources for shift}          & \textbf{Shift} (10$^{-18}$)   & \textbf{Uncertainty} (10$^{-18}$) \\
\hline
BBR Static                          &-4962.9                        &1.8\\
BBR Dynamic                         &-346                           &3.7\\
Density                             &-4.7                           &0.6\\
Lattice Stark                       &-461.5                         &3.7\\
AC Stark (probe)                    &0.8                            &1.3\\
First-order Zeeman                  &-0.2                           &1.1\\
Second-order Zeeman                 &-144.5                         &1.2\\
Lattice vector shift                &0                              &$<$0.2\\
Line pulling \& Tunneling           &0                              &$<$0.1\\
DC stark                            &-3.5                           &2\\
Background gas collisions           &0.63                           &0.63\\
AOM phase chirp                     &-1                             &1\\
Second-order Doppler                &0                              &$<$0.1\\
Servo error                         &0.4                            &0.6\\
\textbf{Systematic total}           &\textbf{-5922.5}               &\textbf{6.4}\\
\hline
\end{tabular} \label{tab.Srbudget}
\end{table}

\subsection{Yb optical lattice clock}
\label{sec.yblatticeclock}

While Ytterbium has long been an interesting candidate for parity
violation \cite{Tsigutkin2009} and degenerate gas
\cite{Fukuhara2007} research, it became an optical clock candidate
only recently with the advent of optical lattice clocks. This
delay was due to the fact that its $^1S_0\rightarrow^3$$P_1$
intercombination line is much broader (187 kHz) than that of Ca,
Sr, and Mg, making this transition unsuitable for high precision
work, while its $^1S_0\rightarrow^3$$P_0$ transition did not
become practical until the lattice clock proposal of Porsev et al.
\cite{Porsev2004}, soon after the first demonstration of the Sr
lattice clock.  It turns out that for optical lattice clock work,
Yb is an excellent candidate.  While in many ways very similar to
Sr, Yb has some useful characteristics of its own, including an
abundant spin-1/2 isotope and straightforward second-stage
cooling.  Yb is in fact a rare-earth element, but its two-valence
electron structure leads to singlet/triplet manifolds similar to
those seen in the alkaline earth atoms.

\begin{figure}[t]\begin{center}
\includegraphics[width=0.7\textwidth]{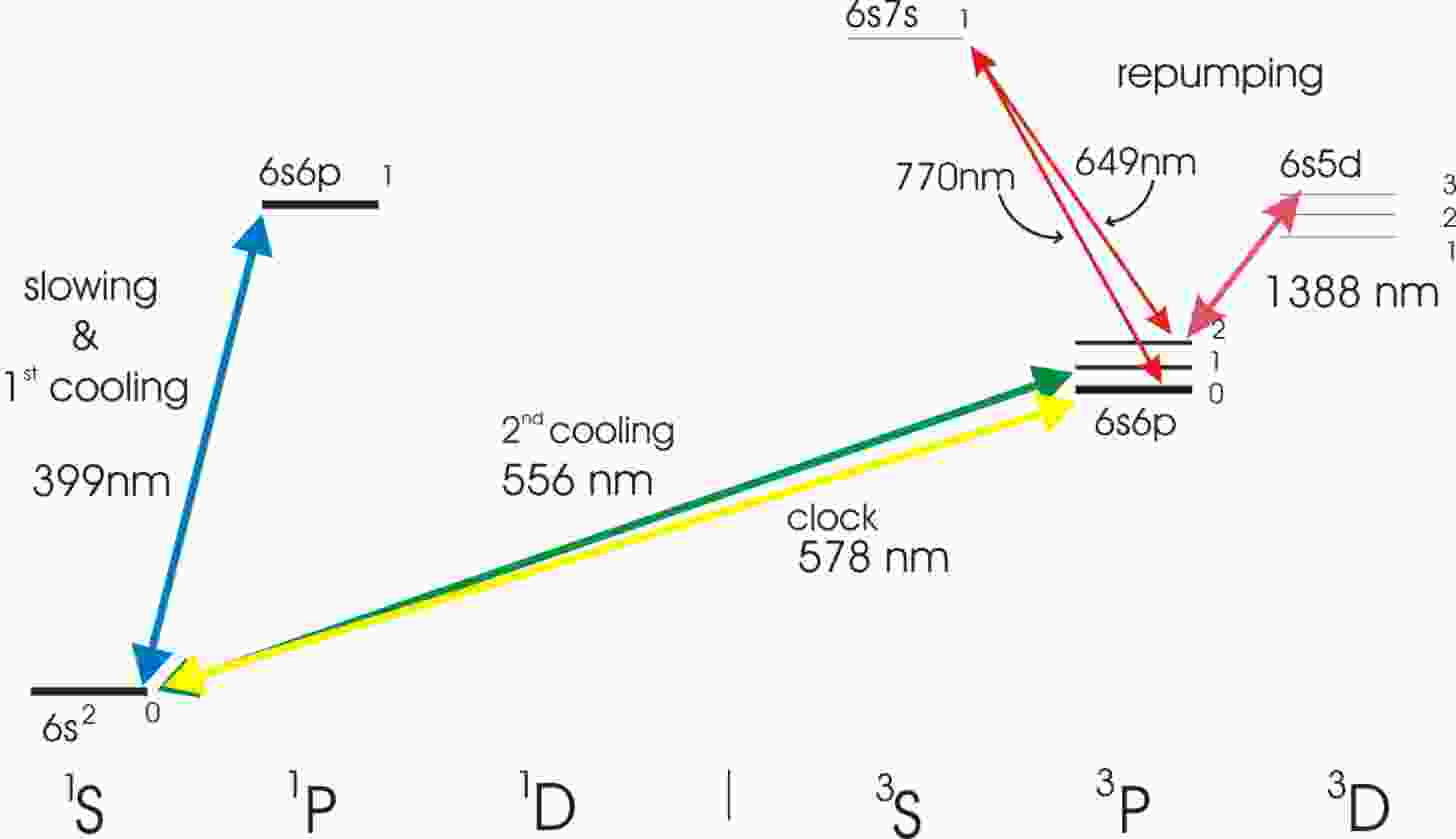}
\caption{Energy levels and relevant optical transition for neutral
Yb.\label{fig.LivelliYb}}
\end{center}
\end{figure}

As shown in Figure ~\ref{fig.LivelliYb}, the Yb
$^1S_0\rightarrow^3$$P_0$ clock transition is at 578 nm, with a
natural linewidth of $\sim$ 10 mHz, practically ideal for a
lattice clock system.  The initial cooling/trapping is performed
on the $^1S_0\rightarrow^1$$P_1$ transition at 399 nm (natural
linewidth $\sim$ 34 MHz), while the previously mentioned
$^1S_0\rightarrow^3$$P_1$ line at 556 nm is used for second-stage
cooling.  This second stage reduces the temperature from a few mK
to 50 $\mu$K, considerably warmer than one obtains with Sr, due to
the 25 times broader linewidth of the second-stage cooling
transition. This has the consequence that the atoms loaded into
the lattice also are warmer (typical temperature $\sim$10 $\mu$K),
thereby putting more atoms into motion levels above the ground
state.  The lattice magic wavelength for the clock transition
occurs at 759.4 nm, and Yb also has repumping transition at 1.388
$\mu$m that is used to pump atoms from the excited state back to
the ground state.

Early research geared toward the development of an Yb lattice
clock took place at KRISS, the University of Washington, and
NIST-Boulder \cite{Park2003,Maruyama2003,Hoyt2005}, while more
recent efforts are underway in Tokyo, D\"usseldorf, Torino,
Shanghai, and other laboratories
\cite{Kohno2009,Nevsky2008,Pizzocaro2012,Jiang2009a}. A typical Yb
lattice clock apparatus looks similar to that of a Sr lattice
clock described above. Atoms are typically loaded from a thermal,
effusive beam (with or without the aid of a Zeeman slower) into a
MOT at 399 nm, with millions of atoms loaded in $\sim$~300~ms.
This can be achieved with only a few milliwatts per MOT beam and
$\sim$10 mW in a counter-propagating slowing beam.  Due to its
broad linewidth, the second-stage cooling transition makes
transferring atoms from a first-stage MOT to a second-stage MOT
very efficient ($>$ 80 $\%$ transfer), and the use of a
``stirring'' laser is not needed for odd isotope cooling.  Some
broadening of the cooling laser spectrum is employed initially in
the transfer; then a single frequency cycle reduces the sample
size to $\sim$ 1 mm diameter and the temperature to $\sim$ 50
$\mu$K \cite{Barber2006}.  This matches the depth of the optical
lattice, which is formed by tightly focusing $\sim$1 W of power at
759 nm into a 30 $\mu$m waist, and then carefully overlapping a
retro-reflected beam.  Up to tens of thousands of Yb atoms with a
residual temperature of 10 $\mu$K can be loaded into about 1000
lattice sites for a 1-D lattice.

The clock transition has been excited in both bosonic ($^{174}$Yb)
and fermionic ($^{171}$Yb) systems for Yb.  While magnetically
induced laser excitation was first demonstrated in $^{174}$Yb
\cite{Barber2006, Poli2008}, the majority of the most recent
research has focused on $^{171}$Yb, due to more straightforward
excitation and reduced systematic effects.  $^{171}$Yb has an
advantage over $^{87}$Sr in that it is a spin-1/2 system, so it
has only two ground-state sublevels, and optical pumping between
these states is extremely efficient.  Moreover, due to its low
multiplicity, it has a zero tensor shift.

\begin{figure}[t]\begin{center}
\includegraphics[width=11 cm]{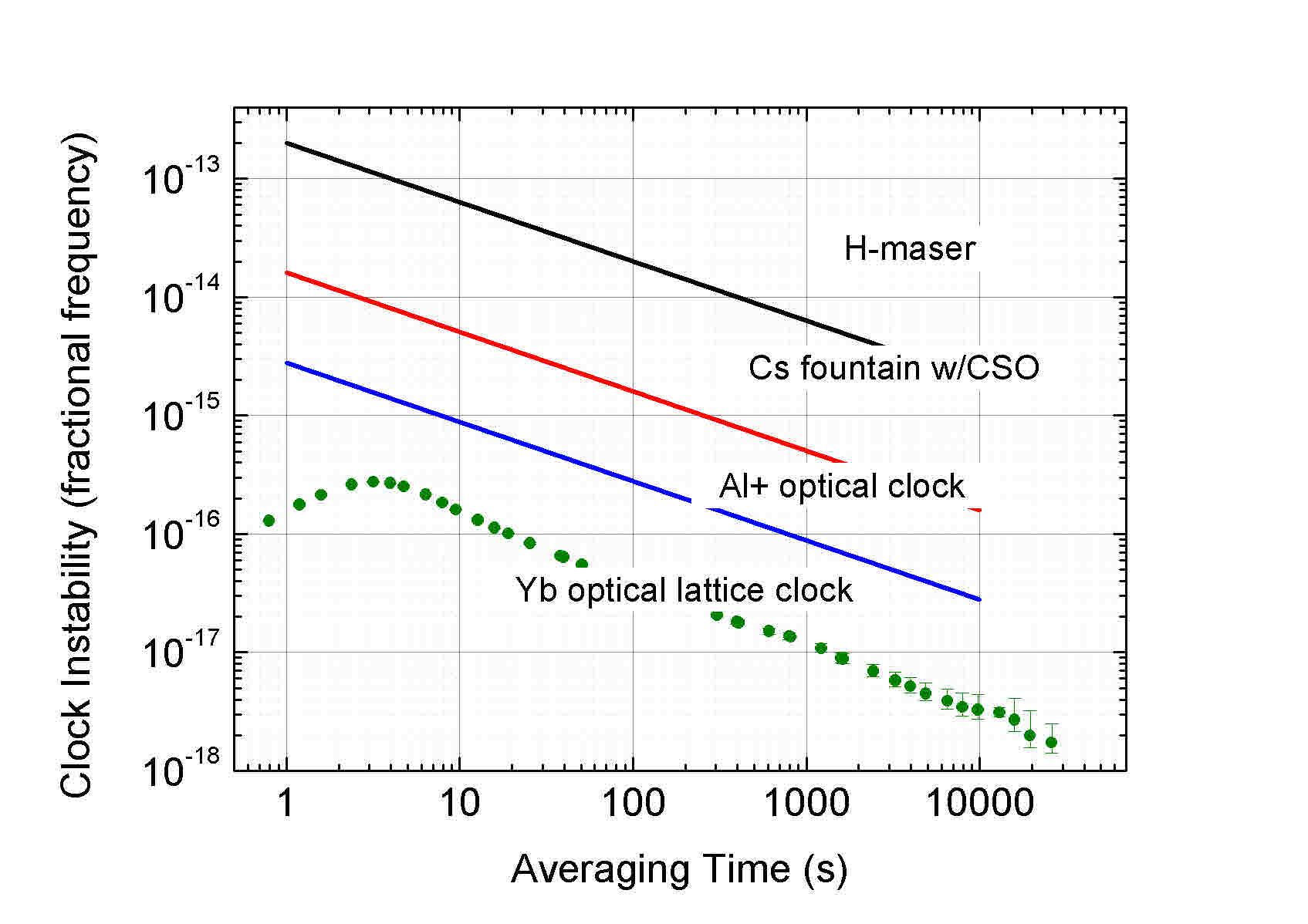}
\caption{Comparison of lowest frequency instabilities achieved
with different clock technologies (red line, high stability Cs
fountain at SYRTE \cite{Santarelli1999}; green line, Al$^+$ ion
clock at NIST \cite{Chou2010}; blue dot, Yb lattice clock at
NIST\cite{Hinkley2013}).\label{fig.ClockStability}}
\end{center}
\end{figure}

As was the case with Sr, it is important to work with an
ultra-pre-stabilized laser to achieve the best results offered by
the Yb system.  Yb clock lasers have been stabilized to cavities
with thermal noise limits as low as $10^{-16}$ and demonstrated
stability at the $2 \times 10^{-16}$ level for up to 5 s.  Such
lasers have been used to perform sub-Hz spectroscopy with the Yb
clock transition, and for the NIST system (which has been at the
forefront of Yb clock research), a $<$ 5 Hz spectroscopic feature
is regularly used for locking the laser in combination with a
normalized shelving detection technique similar to that of Sr
\cite{Jiang2011}. Recent comparisons between two similar Yb
systems at NIST have demonstrated a clock stability of $3.2 \times
10^{-16}\tau^{-1/2}$, which averaged down to $1.6 \times 10^{-18}$
at $25$ $000$ s (see fig. \ref{fig.ClockStability}), the highest
clock precision to date \cite{Hinkley2013}. Precision absolute
frequency measurements of the Yb clock transition relative to Cs
have been performed by three groups with good agreement within the
uncertainties of the clocks \cite{Lemke2009,Kohno2009,Park2012}.
At this point the absolute frequency of the $^{171}$Yb clock
transition, 518 295 836 590 865.2 Hz, is known with an uncertainty
of 0.7 Hz.

The dominant systematic effects for the Yb lattice clock are
basically the same as those for Sr:  lattice light shifts,
collision effects, and blackbody light shifts.  The lattice light
shifts are evaluated by changing the lattice intensity and
wavelength while measuring clock frequency shifts.  The situation
for Yb is complicated by a much larger hyperpolarizability effect
due to two two-photon transitions that are detuned only 0.3-1~nm
from the magic wavelength.  The effect of these transitions has
been evaluated and contributed a fractional uncertainty of $7
\times 10^{-17}$ to the clock uncertainty budget with excellent
prospects for significant further reduction \cite{Barber2008}.
Detailed collision studies have been performed with Yb clocks in
both 1D and 2D lattices \cite{Lemke2011}. These revealed a
significant, even dominant, contribution from p-wave scattering,
which early on was thought to play a small role in lattice clock
atom-atom interactions due to the low atom temperatures.  However
the densities are high enough around the lattice anti-nodes to
enable clock shifts as large as 1 Hz, but it was shown that with a
judicious choice of pulse area Ramsey-style two-pulse excitation
could be used to reduce the clock shift to $<5 \times 10^{-18}$
for 1-D lattice samples of a few thousand atoms \cite{Ludlow2011}.

At this point, the most challenging obstacle to $10^{-18}$
accuracy for Yb optical lattice clocks is the blackbody radiation
shift. While its shift is only half of that of Sr, until recently,
the large uncertainty (10 $\%$) in the knowledge of the blackbody
coefficient led to a $3\times10^{-16}$ contribution to the Yb
uncertainty budget.  A recent high precision measurement (20 ppm)
of the DC differential polarizability for the Yb clock transition
along with a reevaluation of the dynamic correction reduced the
uncertainty in the knowledge of the blackbody coefficient to less
than 0.1 $\%$ \cite{Sherman2012,Beloy2012}.  These results are in
good agreement with recent calculations \cite{Safronova2012}.
Presently, the BBR uncertainty is limited by control and
understanding of thermal environment, conservatively estimated
$\sim$ 1 K, corresponding to $3\times 10^{-17}$ fractional
uncertainty. Progress much beyond this level will most likely
require specialized interaction chambers, and indeed such hardware
is presently under design for Yb clock systems.

\subsection{Hg optical lattice clock}

While the Sr and Yb lattice clocks share many features and
limitations, the third species of lattice clock demonstrated thus
far, one based on the $^1S_0\rightarrow^3$$P_0$ transition at 266
nm in Hg, is quite different in several key ways.  Perhaps its
principal advantage is that, in part as a result of its higher
clock transition frequency (in the UV), it has a BBR shift that is
estimated to be more than an order of magnitude smaller than those
of Yb and Sr \cite{Hachisu2008}.  Additionally, it can be trapped
with a room-temperature source (further reducing BBR effects) and
loaded into a lattice with a single stage of laser
cooling/trapping.  These advantages, however, are offset, at least
in part, by the much-shorter-wavelength radiation required for the
cooling/trapping, lattice, and probe lasers, making them
considerably more difficult to develop, especially at high power.


Hg is a transition metal that has the two-valence-electron
structure common to lattice clock atoms. The strong
$^1S_0\rightarrow^1$$P_1$ transition used in other two-electron
atoms for cooling and trapping is inaccessible for Hg due to its
short wavelength (185 nm), but the $^1S_0\rightarrow^3$$P_1$
intercombination line (at 253.7 nm) that is used for second-stage
cooling of Sr and Yb, is sufficiently strong in Hg (natural
linewidth = 1.3 MHz) to enable trapping, while having a low
Doppler limit to cool the atoms adequately for lattice-loading.
The magic wavelength for the Hg lattice is 362.6 nm \cite
{McFerran2012}.  The radiation for the cooling and clock is
typically generated by quadrupling infrared sources, while the
lattice light requires a build-up cavity to reach the necessary
intensity.  The associated technical challenges have slowed
development of the Hg lattice clock since its proposal in 2008
\cite {Hachisu2008} and, in fact, there are only two groups
aggressively pursuing Hg lattice clocks at this writing \cite
{Hachisu2008,Yi2011}.  Nevertheless, in recent years a successful
demonstration of a Hg lattice clock has been reported
\cite{Yi2011} and impressive results have followed
\cite{McFerran2012}.

In the experimental apparatus of ref. \cite{McFerran2012}, atoms
are loaded from a source held at 233~K (the low temperature is
characteristic of Hg's inherently high vapor pressure) into a 3-D
MOT at 253.7 nm for 1.3 s.  About 2500 atoms are transferred into
a vertical lattice consisting of 3 W of intracavity build up power
(120 $\mu$m waist yields a 9 $\mu$K lattice depth). Clock
spectroscopy at 266 nm reveals a lattice-confined atom temperature
of 4~$\mu$K, and central carrier features as narrow as 11 Hz have
been generated.  The absolute frequency of this transition was
measured to be 1 128 575 290 808 162.0(6.4) Hz, with the
uncertainty dominated by lattice light shifts and limitations in
the precision with which the systematics could be measured
\cite{McFerran2012}.  To date, this system has demonstrated a
fractional frequency instability of
$5.4\times10^{-15}\tau^{-1/2}$, limited by the signal-to-noise
ratio of the system, which has yet to add normalization to the
shelving detection process \cite{McFerran2012a}.  Since these
limitations are basically technical in nature, there are good
prospects for rapid improvement for the system, which, with its
low BBR sensitivity, could lead to Hg being a leading player in
future lattice clock research.

In Table \ref{tab.neutrals} is reported the state-of-the-art of
neutral atom optical-clocks.

\begin{sidewaystable}
  \caption{Neutral atom optical clock species showing their
associated state-of-the-art systematic uncertainties reported in
the literature.}
    \begin{tabular}{ccccccccc}
\textbf{Atom}                   & \textbf{Nucl. Spin}   & \textbf{$\lambda_{cool}$ [nm]}    & \textbf{Clock Trans.}  & \textbf{$\lambda_{clock}$ [nm]}   & \textbf{$\Delta\nu_{nat}$} (\textbf{$\Delta\nu_{obs.}$})   & \textbf{rel. uncertainty} \\
\hline

$^1$H                           &  1/2                  & -                                 & $1S-2S$                              &   243 (2-ph.)                     & 1.3 Hz (2 kHz)         & $4.2 \times 10^{-15}$       \\
$^{24}$Mg                       &   0                   & 285                               & $^1$S$_0$-$^3$P$_1$                      & 457                               & 30 Hz (290 Hz)             & $7 \times 10^{-14}$   \\
$^{40}$Ca                       &   0                   & 432, 657                           & $^1$S$_0$-$^3$P$_1$                      & 657                               & 375 Hz (250 Hz)            & $4.2 \times 10^{-15}$    \\
\multirow{2}[0]{*}{ $^{87}$Sr}  & 9/2                   & 461, 689                           & $^1$S$_0$-$^3$P$_0$                      & 698                               & 1 mHz (0.5 Hz)              &  $6 \times 10^{-18}$   \\
                                &                       &                                   &                                          &                                   & \textit{}                  & \textit{}   \\
\multirow{2}[0]{*}{$^{88}$Sr}   & 0                     & 461, 689                           & $^1$S$_0$-$^3$P$_1$                      & 689                               & 7.6 kHz (14.5 kHz)         & $2.3 \times 10^{-11}$ \\
                                &                       &                                   & $^1$S$_0$-$^3$P$_0$                      & 698                               & $\approx$ 0 Hz  (8 Hz)     & $2.9 \times 10^{-15}$     \\
$^{171}$Yb                      & 1/2                   & 399, 556                           & $^1$S$_0$-$^3$P$_0$                      & 578                               & 10 mHz (1 Hz)              & $3.6 \times 10^{-16}$  \\
$^{173}$Yb                      & 5/2                   & 399, 556                           & $^1$S$_0$-$^3$P$_0$                      & 578                               &  10 mHz (150 kHz)          & $8.5 \times 10^{-11}$ \\
$^{174}$Yb                      & 0                     & 399, 556                           & $^1$S$_0$-$^3$P$_0$                      & 578                               & $\approx$ 0 Hz (4 Hz)     & $1.5 \times 10^{-15}$  \\
$^{199}$Hg                      & 1/2                   & 254                                & $^1$S$_0$-$^3$P$_0$                      & 265.6                             &  100 mHz (11 Hz)           & $5.7 \times 10^{-15}$  \\
\hline

    \end{tabular}%
  \label{tab.neutrals}%
\end{sidewaystable}

\section{Optical clocks based on ions}\label{sec.ions}

A single ion confined in an electromagnetic trap, such as the Paul
trap, has shown to be a powerful way of investigating very weak
(or high-$Q$) clock transitions in the trapped ion, where
environmental perturbations can be effectively controlled. Ions
can be created by electron-beam ionization or photoionization of a
weak atomic flux from an oven within an ultra-high-vacuum ($<$
10$^{-7}$~Pa) chamber whilst the atoms are within the trapping
potential. Trapping of single ions to large clouds of ions
(\emph{e.g.} $N\sim 10^6$ ) can be achieved, dependent on trap
geometries, but in the latter case, space-charge effects create
significant perturbations for the ions. This is not so for the
single trapped ion, where with appropriate parameter control, it
can be isolated as a single particle within a low field
environment, and laser-cooled to $\sim$ 1 mK above absolute zero,
whereby the thermal motion-induced Doppler broadening contribution
to the spectral width of an absorption is minimized. Additionally,
extended confinement times within the trap provide for long
interrogation periods, also allowing environmental fields to be
minimized, nulled and/or controlled. With these arrangements, the
observation of a weak optical (clock) transition with natural
linewidth $\leq$~1~Hz from an ion virtually at rest in
near-unperturbed conditions over extended periods (days) provides
a high accuracy frequency standard.

\subsection{Single-ion clock operation}

\subsubsection{Single ion trap designs}

\begin{figure}[t]
\begin{center}
\includegraphics[width=0.9\textwidth]{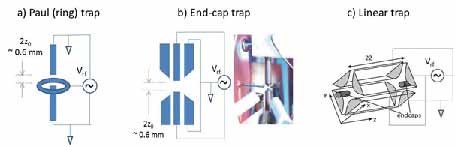}
\caption{Common miniature  rf Paul trap electrode configurations:
a) ring trap, b) end-cap trap, c) linear trap. DC voltages not
shown; typical end-cap electrode separation of 0.6 mm for ring and
end-cap traps; linear trap end-cap separation dependent on 4-blade
electrode length (typically a few mm) .\label{fig.iontrapdesign}}
\end{center}
\end{figure}

It is not possible to create a 3D potential well using just a
static electric potential $\phi$, and different trap designs have
been developed to cope with this (see fig.
\ref{fig.iontrapdesign}). For single ions, these generally
comprise a cylindrically symmetric potential of the form $\phi(r,
z) = A(r^2 - 2z^2)$, where $r$ and $z$ are the radial and axial
coordinates, respectively. The most common of these designs is the
radio frequency (rf) Paul trap, which ideally has electrode
structures with surfaces that match the hyperbolic contours of
constant potential, resulting in a ring of inner radius $r_0$ and
two end-caps separated by $2z_0$. Opposite polarity voltages
applied to the ring and end-caps respectively, give rise to a
saddle potential, which is confining in one direction only.
However, by applying a large oscillating potential $V_0
\cos{\Omega t}$ combined with a small static potential $U_0$
between ring and end-caps, this has the effect of alternating the
confinement direction between the axis and the radial plane,
setting up a time-averaged pseudo-potential well where the
trapping force is directed to trap centre on a timescale that is
longer than the oscillation period, but fast enough to ensure that
the ion(s) remain held within the potential, and strong enough to
ensure that the ion's kinetic energy is unable to surmount the
well. This leads to stable confinement of the ion(s) in the trap
for certain values of applied AC and DC voltages, applied
oscillation frequency (or micromotion frequency) $\Omega/2\pi$,
and ion mass $M$.  The ion motion can be derived as solutions to
the Matthieu equation \cite{Fischer1959}, giving rise to
pseudo-potential oscillation frequencies $\omega_z$, $\omega_r$
(secular motional frequencies) of the ion in the trap, which are
of order of a factor 10 smaller than the drive frequency $\Omega$.

In order to avoid strong perturbations from space charge due to
multiple ions, the ion clock traps are operated with a single
trapped ion. For this, miniature rf traps are used, with ring
diameters of $\sim$ 1 mm, and applied oscillating voltages and
drive frequencies of a few hundred volts at frequencies of 10 - 20
MHz. The ion exhibits 3D simple harmonic motion with secular
frequencies $\omega_r$ and $\omega_z$ in the low MHz range, but
with a high-frequency perturbation (micro-motion) at the drive
frequency $\Omega$. The magnitude of this perturbation depends on
the average distance of the ion from the centre of the potential
well, and small DC voltages are used to tweak the ion's position
close to trap centre to minimise this micro-motion. In fact, these
miniature trap designs have simpler, non-hyperbolic shaped
electrodes. While these shapes are only approximations to the
desired hyperbolic shape, this is not too critical for an ion
located close to the trap centre. In addition to the ring trap,
other common designs used for frequency standards also include the
end-cap trap \cite{Schrama1993}, where ring functionality is
achieved by the use instead of two outer electrodes arranged
co-axially around the end-caps, (see fig.
\ref{fig.SingleIonEndcapTrap}), and the linear ion trap
\cite{Prestage1989} where ions are confined in a linear string
along the z-axis by means of AC voltages applied to two of four
rod electrodes parallel to the axis and a DC voltage applied at
each end. Various linear trap design have been applied to
frequency standards, including the blade trap \cite{Chwalla2009}
and micro-fabricated segmented traps \cite{Chou2010}. More
recently, a micro-fabricated monolithic trap structure built into
a silicon wafer with potential application to future ion clocks
has been demonstrated \cite{Wilpers2012}.

\subsubsection{Single ion generation}

In order to create a single ion within the trapping potential,
electron bombardment of neutral atoms emitted from a simple hot
tubular oven of dimensions $\sim$ 1 mm diameter $\times$ 10 mm
long was originally used to create singly-charged positive ions
for ion trap experiments. This technique, however, was inefficient
as it required both large atom fluxes and electron beam currents, which give rise to excessive coating of trap electrodes and
surrounding materials and charge build up on insulating surfaces.
These, in turn, generate patch potentials on the electrodes and
variation in local surface charges, leading to increased
micro-motion experienced by the ion and anomalous heating rates
experienced by the cold ion. Moreover, the resulting micro-motion
drifts over time, the more so with increased loading frequency,
and requires frequent compensation using DC voltages as described
earlier.

In recent years, photoionization techniques have been developed
for loading single ions in order to mitigate these problems. These
techniques have been shown to dramatically reduce the atom flux
necessary, by a few orders of magnitude, to load a single ion
under near deterministic control. Photoionization loading of ion
traps has been reported for all the common ion species used in ion
clocks. By way of example, we consider the case of
photo-ionisation for strontium. This requires two wavelengths at
461 nm and 405 nm. The 461 nm wavelength raises the atoms from the
ground state to the $^1$P$_1$ resonance level (the same excitation
as used for slowing and trapping Sr neutral atoms), with 405 nm
further exciting the atoms into the continuum. This is simply
achieved in Sr by single pass frequency doubling of DFB laser
diode light at 922 nm in a periodically-poled KNbO$_3$ crystal
(with typical output power of 300 $\mu$W at 461 nm), where
adequate control of a narrow-linewidth laser frequency allows
tuning to the resonance transition frequency. A broad-linewidth
$\sim$ 1 mW 405 nm diode laser is sufficient for ionisation.

A refinement of this technique has demonstrated even more control
\cite{Brownnutt2007}. Here a hot-plate is preloaded with Sr from
the oven and is used as an atom source that can be operated at low
temperatures of $\sim$ 120 $^{\circ}$C. With alternate
photo-ionisation and detection periods of 70 ms and 30 ms
respectively, where the detection is achieved by monitoring the
422 nm Sr$^+$ $^2$S$_{1/2}$ - $^2$P$_{1/2}$ ion fluorescence used
to cool the ion, the hot plate current can be turned off
immediately after the particular detection period that indicates
the presence of the ion fluorescence. This has the effect of
minimising the atom flux used and subsequent generation of patch
potentials, requiring less micro-motion compensation.

\subsection{Laser cooling of single ions}

For simple Doppler cooling, repeated photon scattering of laser
light red-detuned from line-centre of a strong transition of a
single ion confined within the trap, reduces the ion's velocity to
about a metre per second, with corresponding temperature of $\sim$
1 mK, within a fraction of a second after a large number of
scattering events. The rate of scattering is dependent on the
upper level lifetime ($\sim$~10$^{-8}$~s). Ideally, this cooling
would be a closed cycle. In fact, there is generally a branching
decay from the upper level of the cooling transition to a
metastable state, which will terminate the cooling process. In
order to avoid this, a repump laser is used to empty the
metastable state and return the ion to the cooling cycle.  The
Doppler cooling requires only a single cooling beam, since the
driven motion will result in components of the cooling beam acting
in all three orthogonal directions. Both the trap micro-motion and
secular motion frequencies contribute series of sidebands to the
ion's Doppler linewidth.  For a well-cooled ion, the secular
sidebands are reduced to a few weak motional sidebands close to
carrier and separated by the secular frequencies. However, for a
strong cooling transition, the natural linewidth
(\emph{e.g.}$\sim$ 20 MHz) masks these sidebands. For the weak
clock transitions, the sidebands are clearly visible since the
sideband spacing is much larger than the clock transition natural
linewidth. With efficient Doppler cooling, the ion is confined to
a dimension ($<$ 100 nm) much less than the laser wavelength. In
this Lamb-Dicke limit, the spectrum is free from first-order
Doppler broadening, although a small second-order Doppler shift
remains.

In some cases, such as the Al$^+$ quantum logic clock described in
the following, it is necessary to cool well below the Doppler
cooling limit, and close to the ground state of motion in the
trap. In this case, the additional cooling strategies of Raman or
resolved sideband cooling are required.

Confirmation of single ion operation and detection of the state of
the ion is through observation of its cooling wavelength
fluorescence. This is achieved via a high numerical aperture
imaging lens and photomultiplier. With a fluorescence rate of
$\sim$10$^8$ photons s$^{-1}$, solid-angle detection efficiencies
$\sim$ 10$^{-4}$ and photomultiplier efficiencies of $\sim$ 0.1,
single ion fluorescence rates of a few $\times 10^3$ counts
s$^{-1}$ can be achieved for a cold ion at trap centre.

\subsubsection{Quantum jump detection of the clock transition}

High-$Q$ weak clock transitions with $\leq$ 1 Hz natural linewidth
are extremely difficult to detect directly above the background
laser scatter rates. This is surmounted by making use of the
technique of electron shelving or quantum jumps
\cite{Dehmelt1982}. This relies on the widely differing cooling
fluorescence rates observed when the clock probe laser is on or
off resonance with the clock transition, which is connected to the
cooling transition via a common ground state.
 On probing the weak absorption, the ion is driven to
 the metastable upper clock level, remaining ``shelved''
 there for a period corresponding to the long natural lifetime
 of the clock transition, before decaying back to the ground state via
 spontaneous emission. During this period, the ion is not available to
 undergo fast ($\sim$ 10$^{-8}$ s) cooling absorptions, and thus the
 cooling fluorescence level drops to zero until re-established by
 the decay or recovery from the metastable level.
 Weak clock transitions in a single ion can thus be detected with
 virtually 100 \% efficiency. This quantum jump technique is
 key for generating a single trapped ion optical frequency standard.
 The spectral profile of the weak clock transition is built from the statistics
 of the number of quantum jumps as the frequency of the very-high-finesse
 cavity-stabilised clock probe laser (with \emph{e.g.} 1 Hz linewidth) is stepped across it.
 A repeated pulse sequence of cooling, state preparation via optical pumping,
 clock-probing, state detection and clock-state clear-out is undertaken at each
 frequency position, with a single  sequence time of tens to hundreds of milliseconds.
 Ideally, the clock probe pulse should be as long as possible, allowing Hz-wide
 Fourier-limited probe linewidths.  With a knowledge of the experimental
 clock transition profile, the clock laser can be servoed to the profile
 line centre by stepping it back and forth across the profile between
 half-intensity points and generating a frequency discriminant
 by detecting any quantum jump imbalance between these points.


\subsubsection{Ion clock computer control}

\begin{figure}[t]
\begin{center}
\includegraphics[width=0.9\textwidth]{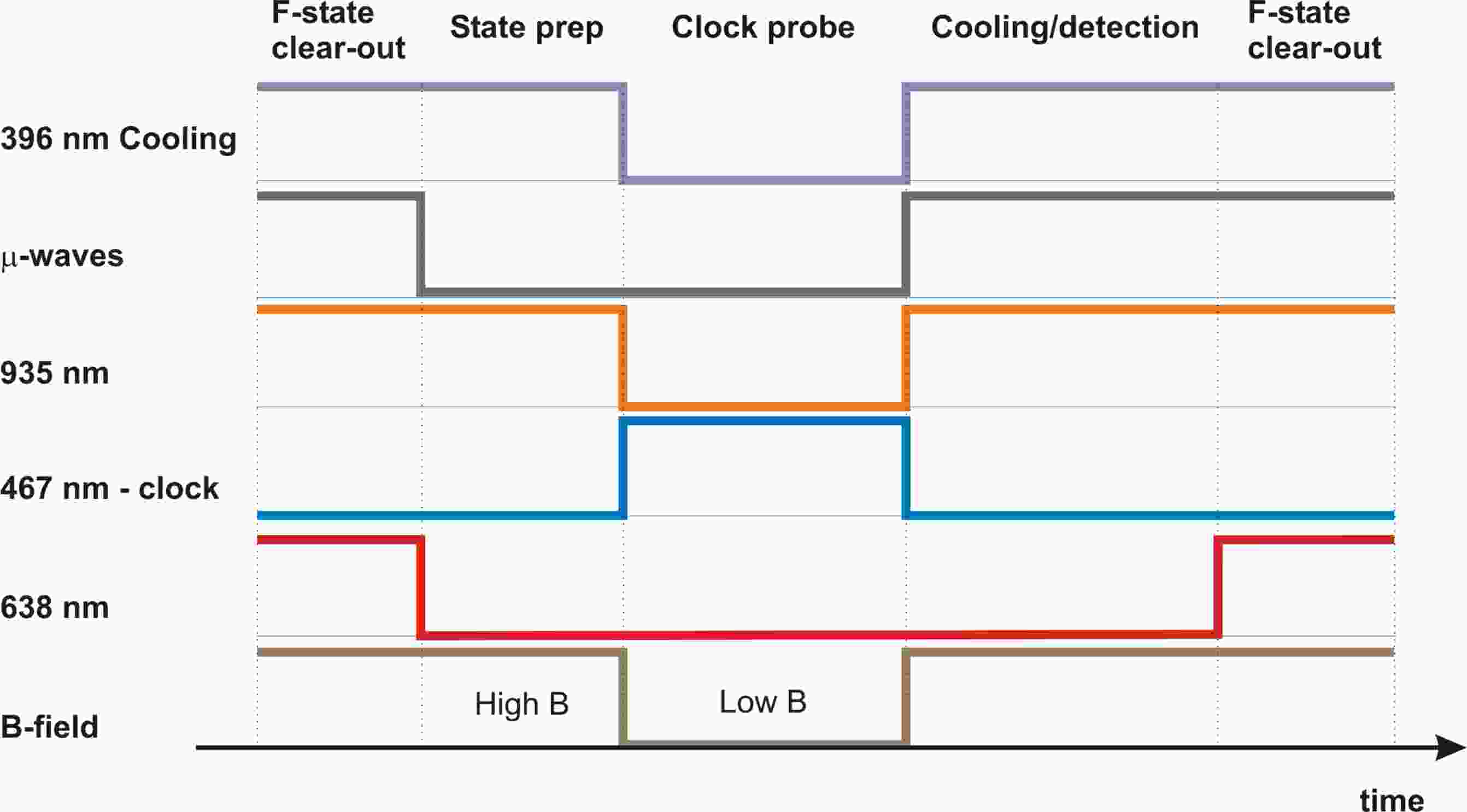}
\caption{Generic pulse sequence showing cooling, magnetic field
state preparation, clock probe and detection and F-state clear-out
pulses for probing the $^{171}$Yb$^+$ 467 nm octupole clock
transition.\label{fig.iontrapsequence}}
\end{center}
\end{figure}

A major requirement of any clock system is to be able to operate
continuously over long periods in order both to achieve clock
stability performance offered by extended averaging periods and to
apply this capability to measurement problems that can benefit
from this increased stability, precision and accuracy. Current
goals for optical clock stability and systematic uncertainty are
of order 10$^{-17}$ to 10$^{-18}$. With typical ion clock
stabilities of a few $\times 10^{-15}  \tau^{-1/2}$, averaging
times of hours to days are needed to reach these values. Thus, it
is paramount that the clock system operates autonomously over such
long measurement campaigns, and auto-control algorithms are being
introduced to provide monitoring and correction of multiple
sub-system processes over a wide range of time scales from the sub
millisecond up to several days. The pulse probe sequence that
forms the basic routine to provide frequency lock requires time
resolution at the 1 ms level (see fig. \ref{fig.iontrapsequence}).
Each probe sequence is repeated $\sim$ 10 times at each side of
the clock transition, resulting in 10 to 20 s to make a correction
to the probe laser frequency. Real-time systematic uncertainty
determination can require B-field or Zeeman-component pair
switching to null the quadrupole (E2) shift, laser intensity
switching for AC Stark shift extrapolation, and 3D micromotion
monitoring and correction on the minute-to-hour periodicity to
allow reaching best performance.

In addition, it is necessary to monitor the various laser
servo-lock signals and automatically re-acquire lock in the event
of loss of lock. It is also necessary to monitor the continued
normal operational characteristics of the ion within the trap (eg
expected cooling rates, fluorescence levels, minimised
micro-motion) and to be able to diagnose the causes of
non-adherence to expected values and initiate the relevant
re-acquisition procedure. Loss of the ion from the trap, for
example, as a result of loss of servolock of the cooling laser due
to a mode hop, and subsequent heating of the ion, will require
reload of a single ion and re-adjustment of compensation voltages.

For auto-control of cavity-stabilised clock and cooling lasers,
monitoring of the cavity fringe, fluorescence signal, quantum jump
rate and wavemeter reading allow fault diagnosis and determination
of the correct re-acquisition procedure for probing and cooling of
the ion. The repumper and clear-out lasers operate at lower
frequency resolution levels, and can be controlled effectively
from the wavemeter readings. Flags are generated between the short
periods of lock loss and re-acquisition to inhibit data-taking
during these periods.  Control algorithms are already in routine
use for comparisons between two ion clock systems over periods of
days \cite{Barwood2012}.

\section{State of the art of ion clocks}
\label{sec.ionclocks}

Here, we examine recent results for a number of single ion clocks.
The most common clock transition across the various ion species is
the quadrupole transition, which has been the subject of research
in $^{199}$Hg$^+$, $^{171}$Yb$^+$, $^{88}$Sr$^+$, $^{40}$Ca$^+$
and $^{135}$Ba$^+$ ions. There has also been a significant amount
of research on the $^{171}$Yb$^+$ highly-forbidden octupole clock
transition, and the $^1$S$_0$ - $^3$P$_0$ transitions in
$^{27}$Al$^+$ and $^{115}$In$^+$.  The $^{27}$Al$^+$ system resulted from the development of the quantum logic clock concept, which has led to leading optical clock performance over recent years.

\subsection{$^{199}$Hg$^+$ quadrupole clock}

\begin{figure}[t]
\begin{center}
\includegraphics[width=0.3\textwidth]{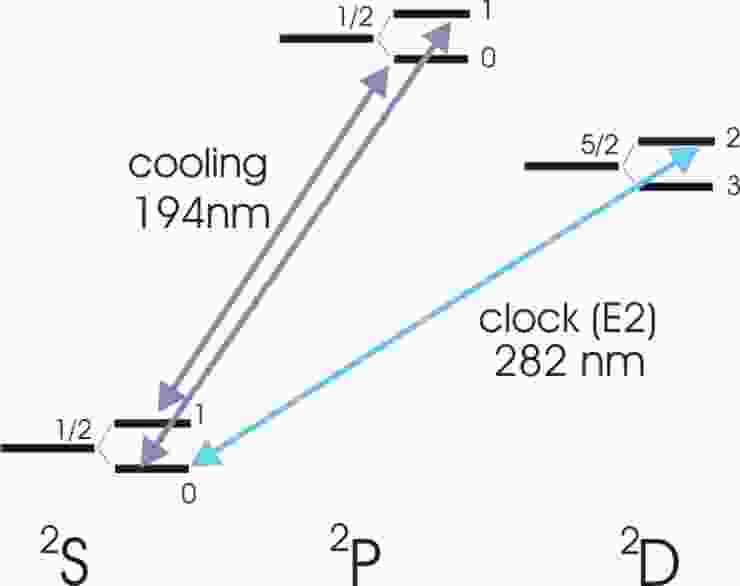}
\caption{Partial energy level scheme for
$^{199}$Hg$^+$.\label{fig.livelliHg+}}
\end{center}
\end{figure}

The $^{199}$Hg$^+$ $^2$S$_{1/2}$ ($F = 0$, $m_F = 0$) -
$^2$D$_{5/2}$ ($F = 2$, $m_F = 0$) electric quadrupole transition
at 282 nm (see fig. \ref{fig.livelliHg+}), with a natural
linewidth of 1.7 Hz, was the first ion clock transition to have a
reported absolute total fractional frequency uncertainty of below
1 part in 10$^{15}$. Laser radiation at 194 nm is used to cool and
prepare the ion, with subsequent probing of the clock transition
with 120 ms pulses at 282 nm. With this arrangement, a
Fourier-transform-limited cold ion linewidth of 6.5 Hz has been
demonstrated, corresponding to a $Q$ of $1.5 \times 10^{14}$ at an
optical frequency of $1.06 \times 10^{15}$~Hz. The absolute total
fractional uncertainty relative to the NIST Cs primary standard
was $7\times 10^{-16}$, of which the Hg$^+$ ion systematic
uncertainty comprised $7.2\times10^{-17}$ \cite{Oskay2006}. This
has since been reduced to $1.9\times10^{-17}$
\cite{Stalnaker2007}, with a reported frequency instability of
$2.8\times10^{-15}\tau^{-1/2}$.

As regards systematic frequency shifts due to external fields, the
$m_F = 0 \rightarrow m_F = 0$ Zeeman component of the
$^{199}$Hg$^+$ quadrupole clock transition is chosen for a
standard, on account of there being no linear Zeeman shift.
However, there are shifts, for example, as a result of the
second-order Zeeman effect and Stark effect. More significant is
the electric quadrupole shift resulting from the interaction of
the D-state atomic quadrupole moment with static electric field
gradients such as those induced by charge build-up on the trap
electrodes. In the recent measurement of the Hg$^+$ to Al$^+$
frequency ratio, the Hg$^+$ quadrupole shift was nulled to the
level of $1\times10^{-17}$ by a series of transition frequency
measurements with the magnetic field oriented in three orthogonal
directions \cite{Rosenband2008}. The Hg$^+$ system is operated
within a cryostat at liquid-helium temperatures in order to
mitigate loss of the ion at room temperature after relatively
short periods (a few minutes) by charge exchange and dimerization.
This has the effect of rendering the blackbody shift negligible,
but at the expense of increased complexity of cryogenic operation.

\subsection{$^{88}$Sr$^+$ quadrupole clock}

\begin{figure}[t]
\begin{center}
\includegraphics[width=0.3\textwidth]{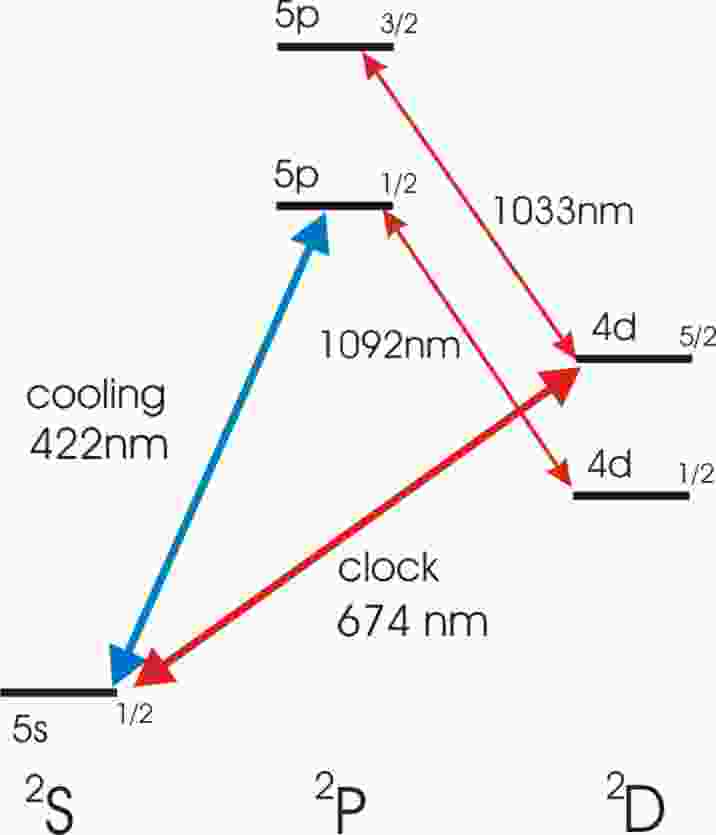}
\caption{Partial energy level scheme for the $^{88}$Sr$^+$
ion.\label{ions1}}
\end{center}
\end{figure}

The $^{88}$Sr$^+$ $^2$S$_{1/2}$  - $^2$D$_{5/2}$ quadrupole clock
transition at 674 nm (fig. \ref{ions1}) has a 0.4 Hz natural
width. The ion is laser-cooled on the strong $^2$S$_{1/2}$ -
$^2$P$_{1/2}$ transition at 422 nm. Two repumpers at 1092 nm and
at 1033 nm are needed repectively to repump the ion back to the
cooling cycle and to recover the ion to the ground state once the
clock transition is driven. Research at NPL \cite{Barwood2007} and
NRC \cite{Dube2013} led to experimental Fourier-limited cold ion
linewidths of 9 Hz (100 ms probe pulse) and 4.4 Hz (200 ms probe
pulse) respectively. One drawback of the $^{88}$Sr$^+$ even
isotope transition is the lack of hyperfine structure and hence a
magnetic-field-insensitive $m_F = 0\rightarrow m_F = 0$
transition. The applied magnetic field of a few $\mu$T splits the
clock transition into five pairs of Zeeman components, with each
pair being distributed symmetrically about the line centre. In
order to take account of this linear Zeeman shift, symmetrical
pairs of Zeeman components are probed, where each component of the
pair experiences an equal and opposite linear shift to the other
component. In this case, the quantum jump imbalance for each
component of the pair is alternately and repeatedly monitored, and
frequency corrections made, after each imbalance determination.
This gives rise to a ``virtual'' magnetic-field-insensitive
transition, and is now used extensively for both ion (\emph{e.g.}
$^{27}$Al$^+$) and atom (\emph{e.g.} $^{87}$Sr) clock transitions
without intrinsic first-order magnetic-field insensitivity.  By
adopting this arrangement, we are then left with a very small
second-order Zeeman shift of order a few mHz/mT$^{2}$.

\begin{figure}[t]\begin{center}
\includegraphics[width=9 cm]{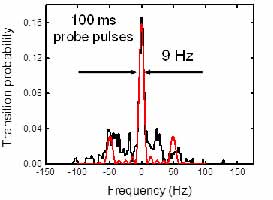}
\caption{$^{88}$Sr$^+$ $^2$S$_{1/2}$ (m=1/2) - $^2$D$_{5/2}$
(m=1/2) 674 nm clock transition, showing a 9-Hz transform-limited
linewidth. Black line, expt data; red line, fit to sinc function
\cite{Barwood2007}.\label{fig.SrIonClockSpectroscopy}}
\end{center}
\end{figure}

Frequency stability measurements have been made between two
independent Sr$^+$ ion clock systems, giving rise to a single trap
stability fitted to $1.6\times10^{-14}\tau^{-1/2}$ (30 s $ < \tau
< $ 5000 s) for interrogation periods of 40 ms and 40 \% peak
excitation probability in the absence of optical pumping
\cite{Barwood2012}. For 90 \% peak excitation and 100 ms probe
periods, this converts to $2.5\times10^{-15}\tau^{-1/2}$ or
$2.5\times10^{-17}$ at $10^4$ s. Cycle periods required to make a
frequency correction are of the order of 20 s. Once the excess
micromotion is fully reduced in all directions, the impact on the
second-order Doppler shift and Stark shift contributions to the
clock frequency uncertainty are in the range $10^{-18}$ -
$2\times10^{-17}$. However, these shifts need to be monitored
during long measurement periods. The major systematic
uncertainties then relate to the electric quadrupole shift and
blackbody radiation Stark shift. The former (together with the
tensor Stark component) can be nulled effectively by making
centre-frequency measurements of three Zeeman component pairs, which
give an average value with the quadrupole shift nulled out
\cite{Dube2005}. With these arrangements, the blackbody shift
becomes the dominant effect. The uncertainty in this shift has
contributions from the effective temperature of the trap
surroundings, the emissivity of these surroundings (particularly
the electrode surfaces in close proximity to the ion) and the
uncertainty in the calculated blackbody shift coefficients
\cite{Mitroy2010}. Temperature rises of a few degrees for the
electrodes are typical for small traps with rf drive voltages in
the low hundreds of volts, but the effect of the electrode
contribution to the blackbody field can be mitigated by use of
highly polished electrode surfaces, where the emissivity is very
low. With current calculated coefficient uncertainties, the
overall blackbody uncertainty is $\sim 2\times10^{-17}$
 \cite{Madej2012}.

The 674 nm $^{88}$Sr$^+$ $^2$S$_{1/2}$  - $^2$D$_{5/2}$ quadrupole
clock transition absolute frequency (see fig.
\ref{fig.SrIonClockSpectroscopy})has previously been measured
using the NPL Cs fountain-referenced femtosecond comb, with a
relative fractional uncertainty of $3.4\times 10^{-15}$
\cite{Margolis2004}. More recently, another absolute measurement
has been made via a comb referenced to the NRC hydrogen maser,
itself calibrated via GPS precise point positioning (PPP) and BIPM
Circular T \cite{Madej2012}. This gave an overall uncertainty of
the clock frequency of $2\times10^{-15}$, with an ion systematic
uncertainty of $2\times10^{-17}$. The offset between the NPL and
NRC values is $\sim 2\times10^{-15}$, well within the combined
uncertainty of the two measurements.

\subsection{$^{40}$Ca$^+$ quadrupole clock}

The $^{40}$Ca$^+$ clock transition is the electric quadrupole
transition 4s $^2$S$^{1/2}$  - 3d $^2$D$_{5/2}$ at 729 nm, which
has a lifetime of 1.17 s and a corresponding natural linewidth of
0.2 Hz. This has been investigated at the University of Innsbruck,
the Wuhan Institute of Physics and Mathematics in China and NICT
in Japan. The Innsbruck trap is a linear rf trap with four blades
separated by 2 mm and two end-cap tips separated by 5 mm. The
Wuhan and NICT traps are miniature ring traps (the Wuhan trap has
a ring radius of 0.8 mm). Ions are produced by photo-ionisation of
neutral calcium via 423 nm + 374 nm, and laser cooled at 399 nm. A
repumper at 866~nm is used to maintain the cooling cycle.

The first measurement of the 729 nm clock frequency was made by
the Innsbruck group. This gave a fractional uncertainty of
$2.4\times10^{-15}$ via a comb measurement referenced to the
LNE-SYRTE transportable fountain clock \cite{Chwalla2009}. A
subsequent measurement at Wuhan with fractional uncertainty of
$3.9 \times 10^{-15}$ agreed with the Innsbruck value at a level
of 5 parts in $10^{16}$. In the latter case, the optical clock
transition was measured by a frequency comb referenced to a
hydrogen maser. The maser was calibrated via UTC (NIM) using GPS
and PPP.

In contrast, the measured Ca$^+$ clock transition at NICT was
observed to be discrepant from the mean of these previous
measurements by 5.3 Hz, where the NICT uncertainty was 1.2 Hz, or
$3\times10^{-15}$\cite{Matsubara2012}. This was derived from UTC
(NICT) and Circular-T via microwave link.  NICT also measured the
frequency ratio between their Ca$^+$ and Sr neutral lattice
standards, which was consistent with the Sr and Ca$^+$ absolute
values. NICT also reported a frequency stability between the Sr
and Ca$^+$ standards of $2.4\times10^{-14}\tau^{-1/2}$ (100 s $<
\tau <$ 8000 s), where the Ca$^+$ frequency is considered to be
the dominant instability.

\subsection{$^{171}$Yb$^+$ quadrupole clock}

\begin{figure}[t]
\begin{center}
\includegraphics[width=8 cm]{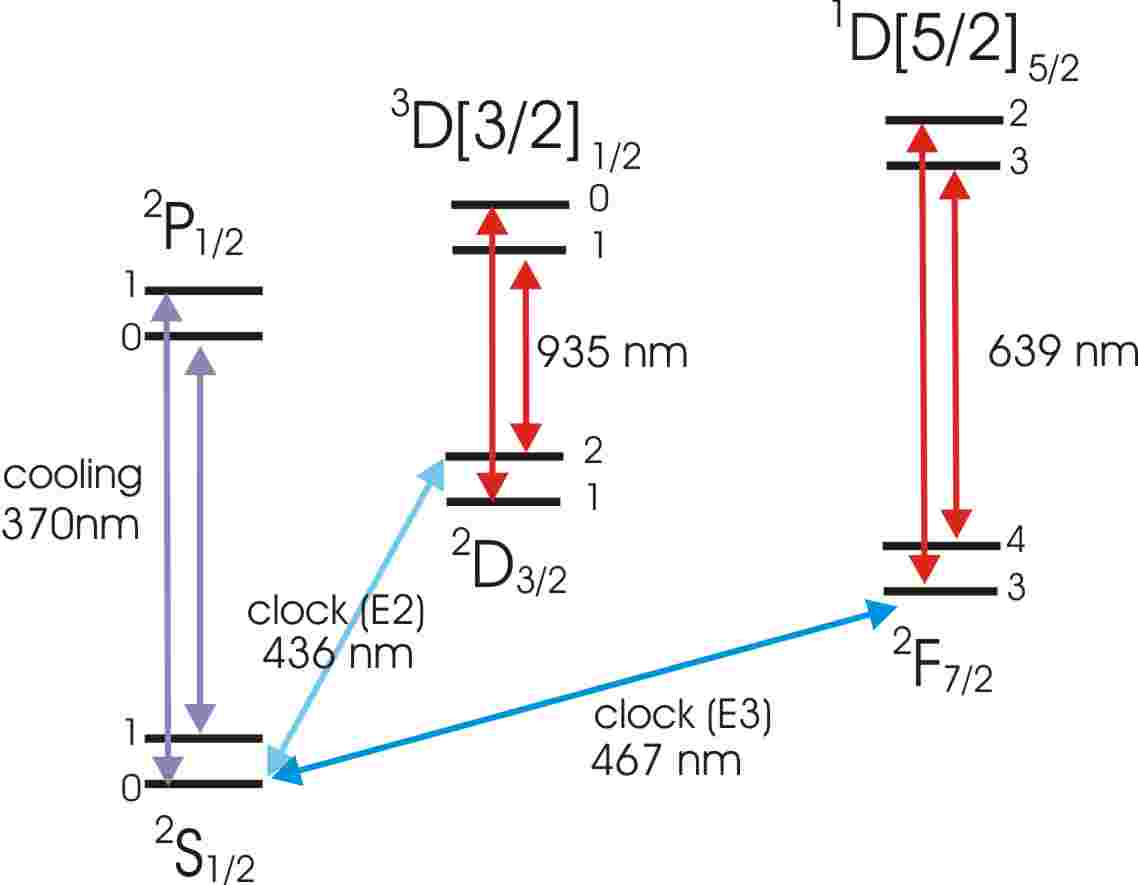}
\caption{Partial term scheme for $^{171}$Yb$^+$.\label{ions2}}
\end{center}
\end{figure}

Similar to the $^{199}$Hg$^+$ case, the odd isotope $^{171}$Yb$^+$
has complicating hyperfine structure, but also has an advantageous
1$^{st}$ order magnetic-field-insensitive Zeeman component $m_F =
0 \rightarrow m_F = 0$ of the quadrupole clock transition. The
$^2$S$_{1/2}$ - $^2$D$_{3/2}$ quadrupole transition at 436 nm,
with a natural linewidth of 3.1 Hz, is chosen as the clock
transition (see fig. \ref{ions2}), rather than the more normal
$^2$D$_{5/2}$ transition. This is on account of existence of the
extremely long-lived  low-lying $^2$F$_{7/2}$ metastable energy
level to which the $^2$D$_{5/2}$ can readily decay.

However, use of the $^2$S$_{1/2}$  - $^2$D$_{3/2}$  quadrupole
transition is also not straightforward in that the $^2$D$_{3/2}$
level is also accessed during the laser cooling cycle. The cooling
is effected by cycling on the $^2$S$_{1/2}$  - $^2$P$_{1/2}$ 370
nm dipole transition via frequency-doubled diode laser at 740 nm
or direct generation of 370 nm with a UV diode laser, with
occasional decays from the $^2$P$_{1/2}$  level into the $F=1$
$^2$D$_{3/2}$ metastable level with a branching ratio of $\sim 0.7
$\%. The ion is recovered to the cooling cycle by driving out of
level this $^2$D$_{3/2}$ level with a 935 nm diode laser. A
frequency-doubled extended-cavity diode laser at 872~nm,
stabilized to a high-finesse reference cavity, is used to probe
the 436 nm $^{171}$Yb$^+$ $^2$S$_{1/2}$ ($F=0$) - $^2$D$_{3/2}$
($F=2$) quadrupole (E2) clock transition, thereby decoupling the
clock transition from the cooling transition to first order.
However, it is also possible for the 935 nm repumper laser to
off-resonantly drive out of the $F=2$ level, so this needs to be
both weak intensity and spectrally narrow to avoid prematurely
emptying the $F=2$ level during the clock-interrogation quantum
jump detection period. Nevertheless, a small background of false
counts is generally encountered. Further, there is a small
probability of the ion accessing the extremely long-lived
$^2$F$_{7/2}$ level by means of collisional interactions, with the
need to clear the ion from this level back to the ion
$^2$S$_{1/2}$ ground state by use of diode laser light at 639 nm
or 760 nm.

The 436 nm $^2$S$_{1/2}$  - $^2$D$_{3/2}$  quadrupole transition
natural linewidth of 3.1 Hz corresponds to an upper state lifetime
of $\sim 52$ ms, which compromises the use of increased
interrogation pulse times to improve the experimental line $Q$ of
the clock transition. Experimental linewidths of 10 Hz have been
demonstrated for the clock transition, with measured frequency
stabilities between two single Yb$^+$ ion systems of $1\times
10^{-14} \tau ^{-1/2}$. The clock transition has been measured to
a fractional uncertainty of $1.1 \times 10^{-15}$ but without
correction for the blackbody shift\cite{Tamm2009}, where the major
contribution to the uncertainty is the systematic uncertainty of
the Cs fountain reference. The contribution to the systematic
uncertainty from the Yb$^+$ ion was 0.31 Hz ($4.5 \times
10^{-16}$).

\subsection{$^{171}$Yb$^+$ octupole clock}

\begin{figure}[t]
\begin{center}
\includegraphics[width=8 cm]{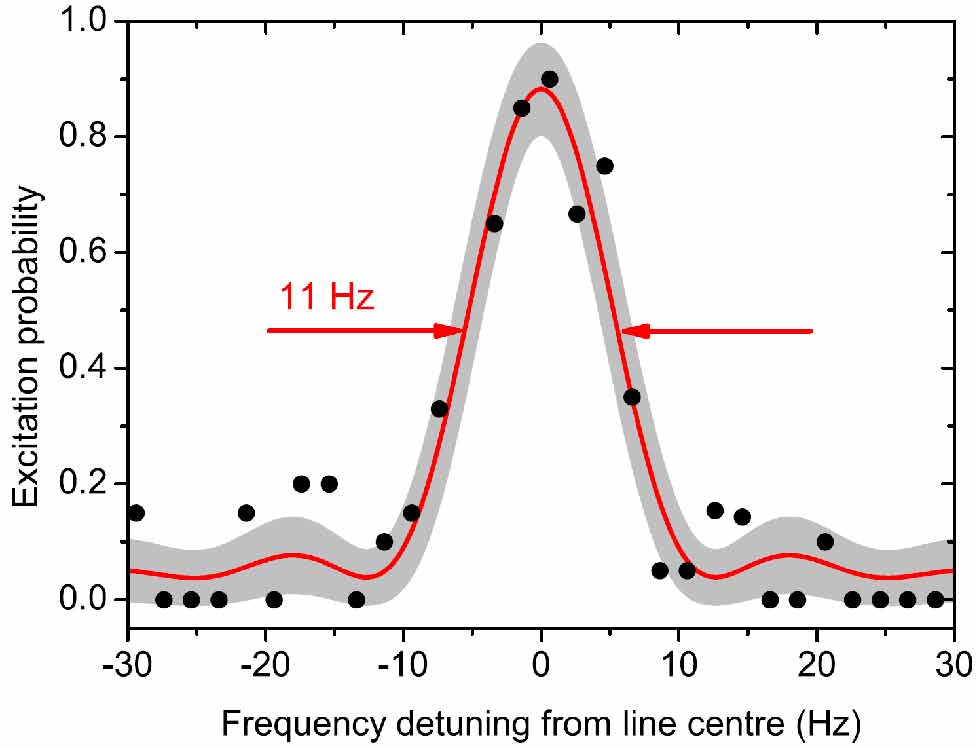}
\caption{Line profile of the $^{171}$Yb$^+$ octupole transition
taken with a 100 ms probe time. The fitted lineshape shows a
linewidth close to the 9-Hz Fourier limit
\cite{King2012}.\label{fig.YbIonOctupoleTransition}}
\end{center}
\end{figure}

The $^{171}$Yb$^+$ single ion is especially interesting in that it
has another optical clock transition, namely the extremely weak
467 nm $^2$S$_{1/2}$  - $^2$F$_{7/2}$ octupole (E3) transition
between the ion ground state and the long-lived $^2$F$_{7/2}$
level with a measured lifetime of $\sim 6$ years
\cite{Roberts2000}, and corresponding natural width in the
nanohertz range and line $Q$ of $10^{23}$. With such a narrow
natural width, the observable clock transition linewidth will be
set by the Fourier-limited linewidth corresponding to the
clock-pulse interrogation period, and this will reflect on the
attainable $Q$ and stability.  This offers an improved quantum
limit compared to those typical for quadrupole transitions where
natural decay can foreshorten interrogation periods, and thereby
lead to reduced stabilities. The advantage of increased
interrogation periods approaching a second is better stability
achievable with shorter averaging periods. This in turn allows
systematic frequency shift investigations to be made at reduced
levels (\emph{e.g.} at the $10^{-17}$ level) at a faster rate.
Cold ion octupole linewidths in the range $\sim 7$ Hz to 11 Hz for
near transform-limited probe pulses have been observed
\cite{Huntemann2012,King2012} (see fig.
\ref{fig.YbIonOctupoleTransition}).

Similar to the $^{171}$Yb$^+$ quadrupole case, laser cooling is
effected by cycling on the $^2$S$_{1/2}$  ($F=1$) - $^2$P$_{1/2}$
($F=0$) 369 nm dipole transition with repumping at 935 nm. It is
also possible to induce spontaneous Raman transitions to
$^2$S$_{1/2}$ ($F=0$) as a result of weakly driving the
$^2$S$_{1/2}$  ($F=0$) - $^2$P$_{1/2}$  ($F=1$) transition in the
wing of the cooling light profile. In order to counter this, a
14.7 GHz sideband is generated by electro-optic modulation to
drive the $^2$S$_{1/2}$ ($F=0$) - $^2$P$_{1/2}$ ($F=1$) transition
to recover the ion from the $^2$S$_{1/2}$ ($F=0$) level to the
cooling cycle. Finally, a further 369 nm sideband at 2.1 GHz is
used to state prepare the ion in $^2$S$_{1/2}$ ($F=0$) in
preparation for driving the $^2$S$_{1/2}$ ($F=0$)  - $^2$F$_{7/2}$
($F=3$) octupole (E3) transition at 467 nm. As with the quadrupole
case, the long $^2$F$_{7/2}$ lifetime is also a problem for the
cooling and probing pulse sequence, in that the metastable level
acts as a sink for the ion. During the stepped frequency
interrogation of the octupole transition, it is necessary to
quickly recover the ion from the $F$-state back to the
$^2$S$_{1/2}$ ground state using 638 nm or 760 nm clear-out
lasers.

The $^{171}$Yb$^+$ odd isotope has a low nuclear spin of $I=1/2$
resulting in a fairly simple hyperfine structure, and the octupole
transition also has an $m_F = 0 \rightarrow m_F = 0$
magnetic-field insensitive transition with no first-order Zeeman
effect. The second-order Zeeman shift has been measured
\cite{Hosaka2005} to be $-1.72(3) \times$ mHz/mT$^{2}$ for the
octupole transition. Operational fields of 20 $\mu$T are typical.
Loss of fluorescence at close-to-zero magnetic fields due to dark
state formation (coherent population trapping) in the ground state
is prevented by fast polarisation switching of the cooling light.
The small magnetic field applied is sufficient to prevent similar
coherent population trapping in the $^2$D$_{3/2}$ level during the
cooling cycle.

Until recently, the dominant shift associated with the
$^{171}$Yb$^+$ octupole transition has been the AC Stark shift
owing to the large 467 nm probe laser intensity needed to drive
the transition, with typical coefficient of \emph{e.g.} 48(10)
$\mu$Hz W$^{-1}$m$^2$ \cite{Blythe2003a}, dependent on the probe
beam wave-vector angle to the quantization axis. Measurement
strategies to deal with this are based on extrapolation of the
probe intensity to zero intensity by accurate measurement of
intensity ratios for two power levels during the probe sequence.
Further improvement could be made by reduction in probe laser
linewidth, giving rise to reduced spectral spread and improved
efficiency of overlap with the octupole transition. In addition, a
technique of hyper-Ramsey excitation has been reported, where the
phase and pulse length of the second Ramsey pulse are varied, with
the effect of suppressing the AC Stark shift by two orders of
magnitude or more \cite{Huntemann2012a}.

The electric quadrupole shift of the octupole transition can be
nulled by measuring the transition frequency in three orthogonal
magnetic fields, whereby the shift averages to zero
\cite{Itano2000}. This nulling is made all the more easier by the
magnitude of the electric quadrupole moment, which has been
measured to be -0.041 $e a_0^2$ \cite{Huntemann2012}. This is
nearly two orders of magnitude smaller than the quadrupole moments
measured for the quadrupole clock transitions. The other major
systematic frequency shift is the blackbody radiation shift.
Calculated blackbody shift coefficients \cite{Mitroy2010} give
rise to uncertainties of a few $\times 10^{-18}$ K$^{-1}$ at room
temperature. Recently, a total systematic uncertainty of
$7\times 10^{-17}$ was reported for the octupole transition
\cite{Huntemann2012}. Independent absolute frequency measurements
at PTB and NPL show excellent agreement within the combined
$1\sigma$ value of the individual uncertainties of
$8\times10^{-16}$ \cite{Huntemann2012} and $1\times10^{-15}$
\cite{King2012} respectively. The agreement is even better when
account is taken of the individual fountain clocks relative to the
International Atomic Time (TAI).

\subsection{$^{27}$Al$^+$ quantum logic clock}

The ion species that has demonstrated the leading optical clock
performance to date is the $^{27}$Al$^+$ ion, which has a
$^1$S$_0$ - $^3$P$_0$ weak clock transition at 267 nm with 8 mHz
natural width corresponding to a 20 s lifetime of the upper clock
state \cite{Chou2010}. It also has extremely low or zero
sensitivities to environmental fields, giving rise to an
advantageous systematic frequency-shift uncertainty budget. For
example, it has no electric quadrupole shift, and its sensitivity
to blackbody radiation is the lowest of all clock species
currently under consideration.  However, the major issue with this
ion is that its cooling wavelength is $\sim$ 169 nm in the vacuum
ultraviolet region of the spectrum, and thus is technically very
difficult to produce and transmit between source and ion trap.
Researchers at NIST have overcome this problem by separating clock
and laser cooling/preparation functionality between two
different-species ions in the same linear rf ion trap. The logic
ion ($^9$Be$^+$ or $^{25}$Mg$^+$ have been used to date) is
directly laser cooled in the conventional way via accessible
cooling radiation (313 nm for $^9$Be$^+$, or 280 nm for
$^{25}$Mg$^+$). The $^{27}$Al$^+$ clock ion is sympathetically
cooled by virtue of the combined motional mode set up via the
coulombic interaction between the two logic and clock ions in the
linear trap. The Al$^+$ ion is then probed on the 267~nm clock
transition. However, there is an additional complexity here in
that there is no Al$^+$ cooling fluorescence available with which
to observe clock transition quantum jumps. This is overcome by
making use of coherent manipulation techniques developed
previously for quantum information research at NIST, whereby the
quantum jump clock transition profile data is mapped back to the
logic ion, again via the collective 2-ion vibration motional
state.

\begin{figure}[t]
\begin{center}
\includegraphics[width=11 cm]{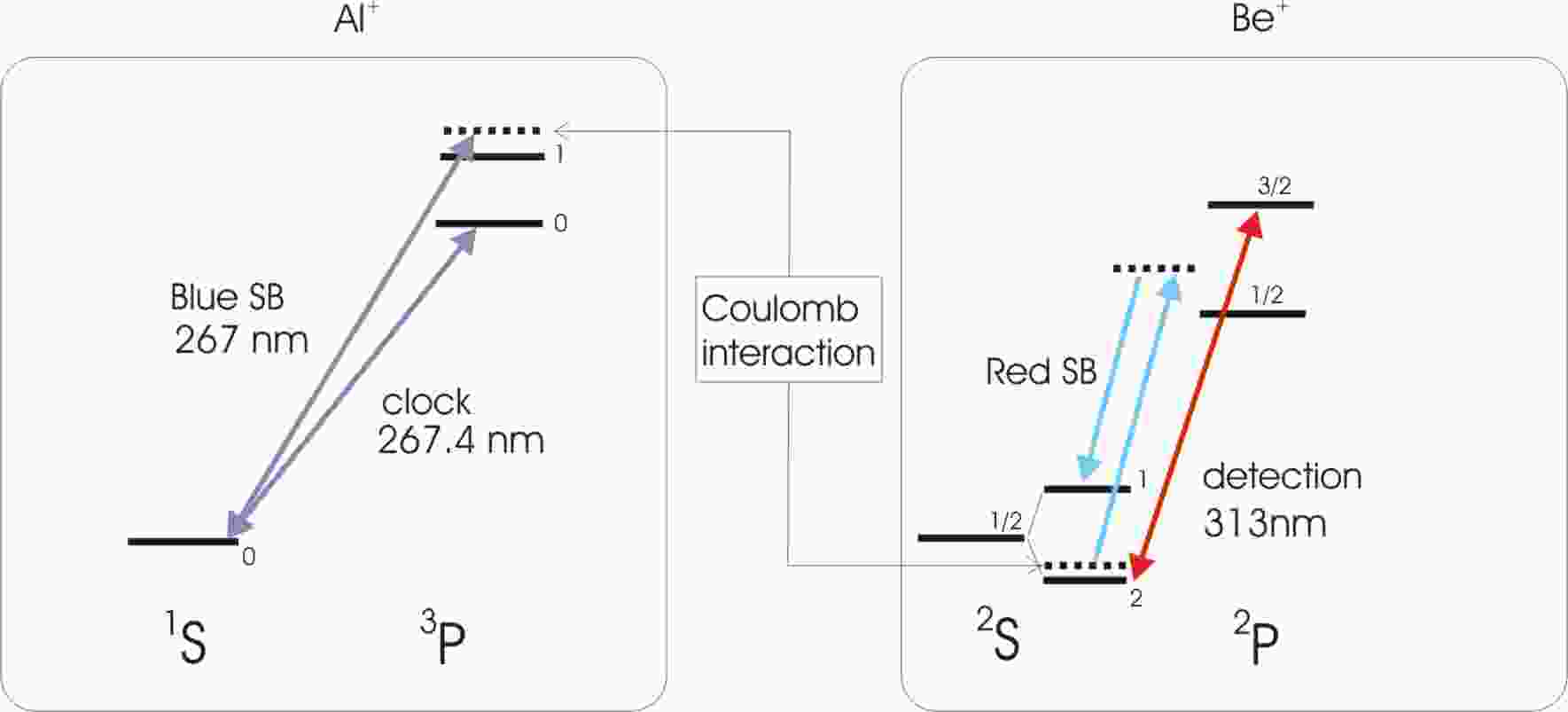}
\caption{Schematic of the quantum logic clock state read-out
algorithm; reprinted from \cite{Rosenband2007}.\label{ions3}}
\end{center}
\end{figure}

In very brief detail, the pulse sequence to achieve this
sympathetic cooling and $^1$S$_0$ - $^3$P$_0$ clock transition
read out is as follows. The combined ion motion is prepared close
to its motional ground state by a combination of Doppler cooling
and sideband cooling via the logic ion cooling wavelengths. The
clock transition is driven (or not), and the result is mapped to
the 267 nm $^{27}$Al$^+$ $^1$S$_0$ - $^3$P$_1$  coupled transition
\cite{Rosenband2007}, as shown in fig. \ref{ions3}.  By means of a
blue sideband $\pi$ pulse on this transition followed by a red
sideband $\pi$ pulse on the logic ion Raman cooling transition,
which are interconnected through their common motional mode, the
clock drive result is then read out on the logic ion fluorescence.
Typical clock probe pulses of 100 to 150 ms have been used, but
the subsequent clock state readout algorithm can be executed
remarkably quickly in $\sim$ 2 ms. Overall duty cycles of 65 \%
have been achieved where the dead-time is primarily due to state
preparation and read-out plus other interleaved parameter
switching in order to minimise micro-motion. Of order ten
interrogation pulse / read-out sequences would be made at each
frequency step across the clock transition profile. This technique
has resulted in an observed 2.7 Hz clock transition FWHM linewidth
\cite{Chou2010}, which corresponds to a $Q$ of $4.2\times10^{14}$,
the highest recorded $Q$ to date.

As mentioned earlier, $^1$S$_0$-$^3$P$_0$ transitions are free
from quadrupole shift, on account of the $J=0$ value for both
lower and upper levels. There is, however, a linear magnetic field
dependence of order several $\times 10$ Hz ($\mu$T$^{-1}$) for the
outermost Zeeman components. This is eliminated in the same way as
with the even isotope quadrupole transitions, by measuring Zeeman
component pair frequencies and taking the average value. The
second-order Zeeman shift is measured to be - 70
Hz/mT$^{2}$\cite{Rosenband2007}.

\begin{figure}[t]
\begin{center}
\includegraphics[width=11 cm]{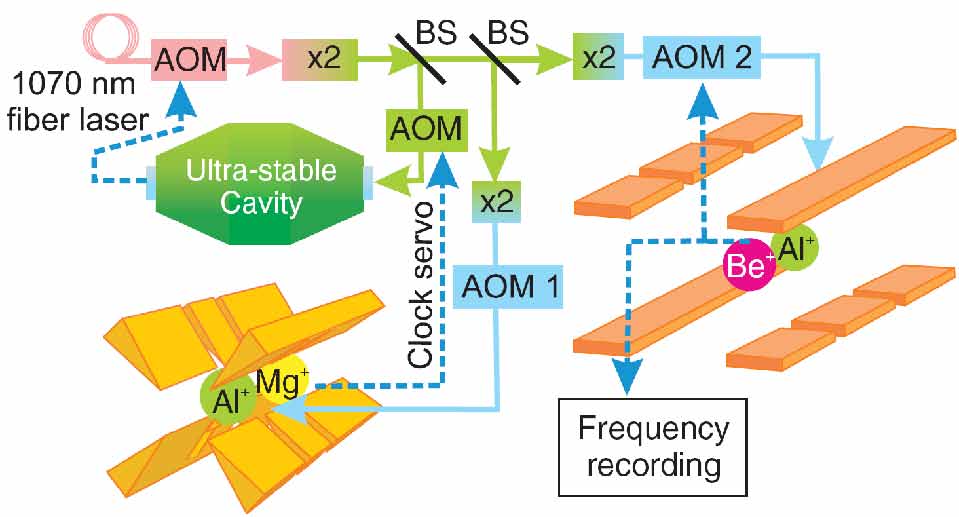}
\caption{Experimental arrangement for comparing two $^{27}$Al$^+$
quantum logic clocks at NIST \cite{Chou2010}.} \label{ions4}
\end{center}
\end{figure}

In a recent comparison between two Al$^+$ clocks (where each clock
made use of a different logic ion species, see fig. \ref{ions4}),
the better of the clock realizations had a total systematic-shift
fractional uncertainty of $8.6\times10^{-18}$ \cite{Chou2010}.
This was dominated by second-order Doppler shifts due to residual
secular motion and excess micromotion, which are likely to be
larger than in the single ion case, partly due to the lighter mass
of the ion (see Table \ref{tab.al+budget}). The relative frequency
stability between the clocks was observed to be $2.8\times
10^{-15}$ $\tau^{-1/2}$ (50 s $< \tau   <$ 2000 s). The relative
frequency difference between the clocks was measured to be
$-1.8\times10^{-17}$.

\begin{table}
  \caption{Uncertainty budget for Al$^+$ clock \cite{Chou2010}.}
    \begin{tabular}{ccc}
\textbf{Effect}   & \textbf{Shift} (10$^{-18}$)   & \textbf{Uncertainty} (10$^{-18}$) \\
\hline
Excess micromotion          &-9                 & 6\\
Secular motion              &-16.3              & 5\\
Blackbody radiation shift   &-9                 & 3\\
Cooling laser Stark shift   &-3.6               & 1.5\\
Quad. Zeeman shift          &-1079.9            & 0.7\\
Linear Doppler shift        &0                  & 0.3\\
Clock laser Stark shift     &0                  & 0.2\\
Background-gas collisions   &0                  & 0.5\\
AOM freq. error             &0                  & 0.2\\
\textbf{Total}              &\textbf{-1117.8}   &\textbf{8.6}\\
\hline
\end{tabular} \label{tab.al+budget}
\end{table}

\subsection{$^{115}$In$^+$ clock}

$^{115}$In$^+$ is another ion species with a weak $^1$S$_0$ -
$^3$P$_0$ clock transition at 236 nm, with natural linewidth of
0.8 Hz \cite{VonZanthier2000}. It also has a relatively small
blackbody coefficient. There is no accessible strong cooling
transition, but direct bi-chromatic cooling has been performed via
the weak $^1$S$_0$ - $^3$P$_0$ intercombination transition at
230~nm \cite{Becker2001}. As a result, this direct cooling is not
greatly efficient, and other possibilities such as sympathetic
cooling eg by Yb$^+$ are under investigation. Previously, the
clock transition has been measured to an absolute uncertainty of
230 Hz ($\sim 2\times10^{-13}$ fractional uncertainty), but this
remains inconsistent with a later measurement
\cite{VonZanthier2000,Liu2007}.

\vspace {0.5cm}

There are other cold ion systems that have been researched as a
possible standard over the past two decades, such as the
$^{135}$Ba$^+$ $^2$S$_{1/2}$  - $^2$D$_{5/2}$  infra-red
transition at 1.762~$\mu$m. In the latter case, no high accuracy
frequency measurements have been made at this time to our
knowledge.

Table~\ref{tab:ions} shows the current trapped ion optical clock
systems where state-of-the-art systematic uncertainties have been
reported. In a number of cases, these uncertainties can be seen to
be lower than the systematic uncertainties achieved for the best
cold caesium fountain primary standards ($\sim 2\times 10^{-16}$).

\begin{sidewaystable}
\caption{Trapped ion optical clock species showing their
associated state-of-the-art systematic uncertainties reported in
the literature. The penultimate column gives the lowest cold ion
clock transition linewidths observed to date; in general, these
observed linewidths are Fourier-limited widths dominated by the
clock laser probe pulse width.}
\begin{tabular}{ccccccccccc}
Ion  & Nuclear  spin & $\lambda_{cool}$ & Clock. trans.             & $\lambda_{clock}$ & Natural linewidth $\Delta\nu$      & obs. linewidth  $\Delta\nu_{obs}$ & rel. uncertainty\\

\textbf{}     & \textbf{}                & (nm)             &                      & (nm)              & (Hz)                               &(Hz)                               &                 \\
\hline
$^{27}$Al$^+$ & 5/2                      & 279.5 ($^{25}$Mg$^+$)     & $^1$S$_0$-$^3$P$_0$           & 267                        & 0.008                                       & 2.7                                        & $8.6\times 10^{-18}$\cite{Chou2010}    \\
$^{40}$Ca$^+$ & 7/2                      & 397                       & $^2$S$_{1/2}$-$^2$D$_{5/2}$   & 729                        & 0.2                                         & 30                                         & $2.4\times 10^{-15}$\cite{Chwalla2009} \\
$^{88}$Sr$^+$ & 0                        & 422                       & $^2$S$_{1/2}$-$^2$D$_{5/2}$   & 674                        & 0.4                                         & 5                                          & $2.1\times 10^{-17}$\cite{Madej2012}   \\
$^{115}$In$^+$& 9/2                      & 231                       & $^1$S$_0$-$^3$P$_0$           & 236                        & 0.8                                         & 43                                         & $2\times 10^{-13}$\cite{VonZanthier2000} \\
$^{171}$Yb$^+$ (oct.)& 1/2               & 370                       & $^2$S$_{1/2}$-$^2$F$_{7/2}$   & 467                        & $10^{-9}$                                   & 7                                          & $7.1\times 10^{-17}$\cite{Hosaka2005}  \\
$^{171}$Yb$^+$ (quad.) & 1/2             & 370                       & $^2$S$_{1/2}$-$^2$D$_{3/2}$   & 436                        & 3.1                                         & 10                                         & $4.5\times 10^{-16}$\cite{Tamm2009} \\
$^{199}$Hg$^+$& 1/2                      & 194                       & $^2$S$_{1/2}$-$^2$D$_{5/2}$   & 282                        & 1.7                                         & 6.7                                        & $1.9\times 10^{-17}$\cite{Rosenband2008}\\
\hline
\end{tabular}%
\label{tab:ions}%
\end{sidewaystable}

\section{Future prospects}
\subsection{High accuracy remote optical clock comparisons}

Throughout this survey we have seen the rapid rate of progress
with optical atomic clocks both in terms of instability and
absolute uncertainty.  The best clocks now achieve an instability
of less than $2 \times 10^{-18}$ in only six hours of averaging
time, while the most accurate clocks have uncertainties of less
than 6 parts in $10^{18}$.  Moreover, we anticipate that these
values will be reduced still considerably more in the near future
as no fundamental limitations loom at this point.  However, there
still exists one potential roadblock that could limit progress on
a global scale, namely, our inability to compare clocks separated
by long distance.  There are now dozens of clocks under
development on five continents, but we are severely limited in our
ability to compare them, which ultimately is a critical step from
the perspective of the development of future standards.  Up to
now, remote comparisons have largely been performed indirectly,
for example, as we saw in the case of the Sr lattice clock,
through comparison to the local Cs microwave standards.  Such
comparisons are usually limited, however, by the microwave
standards, and as the optical standards move well beyond their
microwave counterparts in terms of performance, other techniques
will be required that can be used in both shorter-range and
longer-range comparisons.  The challenge here is to find
dissemination techniques that can transfer the clock signal
without significantly compromising its performance.

This problem has been first solved over short distances (e.g.,
less than tens of km) by the use of phase-noise-compensated
optical fibres \cite{Ma1994}.  This technique is used in almost
all optical labs today to transmit light from one lab to another
(or even from one part of the experiment to another) by means of
optical fibre without adding phase noise. As a result optical and
microwave clocks within one institute or even between institutes
in the same city can be readily compared \cite{Ludlow2008}.  For
longer-distance frequency comparisons, however, today links
mediated by satellites are employed. The exchange of coded
microwave signals for TWSTFT links (Two Way Satellite Time and
Frequency Transfer), or the use of the phase of the received
signals in the GPS carrier phase method, enables frequency
comparisons on the global scale at a fractional uncertainty of
$10^{-15}$ for a measurement period of several days, which is far
noisier than required by present-day (and near-future) atomic
clocks. Even though these techniques may improve to the $10^{-16}$
level in a few years, it is already clear that the achievable
stability of a microwave link will not be competitive with the
current demonstrated stability of optical clocks. One new approach
has been to extend the phase-noise-compensated fibre technique to
longer distances with repeater stations as needed
\cite{Daussy2005}.  This all-optical method has already been
demonstrated on scales up to 920 km (as realized, for example, in
Germany between the PTB and Max Planck Institute laboratories),
and the best results have shown frequency instabilities as low as
$5\times 10^{-15}$ at 1 s (reaching $10^{-18}$ in less than
1000~s), with an estimated accuracy within $4\times10^{-19}$
\cite{Predehl2012}. This method today is the only known way to
make comparisons between optical clocks at high stability, and
there is a growing interest in realizing optical-fibre links
between European metrology institutes to test the link
instabilities over distances of more than 1000 km
\cite{Predehl2012}.  Ultimately this approach could lead to all of
the optical clocks in Europe being directly connected by optical
fibre. Similar efforts are also underway in Japan, where optical
clocks spread throughout multiple institutes have been connected
by optical fibre \cite{Fujieda2011} and in Italy, where a
630~km-long fiber link between INRIM laboratories in Torino and
LENS-UNIFI laboratories in Firenze is currently under test
\cite{Levi2013}.

While this approach will bear considerable fruit since it enables
true remote optical clock comparisons, it is not clear how it
could be extended further to enable intercontinental clock
comparisons.  Such comparisons will be necessary as the field
moves inexorably towards a redefinition of the SI second and will
enable a variety of experiments including tests of fundamental
physics (see the subsequent section).  The most likely solution
will be some type of optical version of the microwave TWSTFT
links, but issues such as atmospheric turbulence and limited
connection time (e.g., due to cloud cover), which are much more
serious for optical signals, will need to be addressed.
Nonetheless, experiments along these lines are already underway
with some encouraging initial results (see \cite{Giorgetta2013}
and references therein).  We can envision that one day such
optical links in combination with a high-performance clock located
in space (e.g., on a dedicated satellite or the International
Space Station) to minimize terrestrial effects could enable a
global high-performance clock environment for realization of the
SI second, relativistic geodesy, and tests of fundamental physics
\cite{Schiller2012,Schiller2009}. Indeed, feasibility projects
such as the European Space Agency's ``Space Optical Clock''
Program (SOC) \cite{Schiller2012} and ``Space-Time Explorer and
Quantum Equivalence Principle Space Test'' (STE-QUEST)
\cite{stequest} or the National Aeronautics and Space Agency's
``Deep Space Optical Clock Mission'' \cite{dsac} are already
underway, although it may be a decade or more before we see
optical clocks in Space.

\subsection{Practical applications and tests of fundamental physics with high accuracy/stability optical clocks}
\label{sec.practical} To this point the bulk of the research in
the field of optical atomic clocks has focused on the development
of optical clock systems and the advancement of their performance
rather than on the construction of field-able systems.  Indeed,
given the relative youth of the field (the first optical clock
demonstrations complete with femtosecond-laser combs occurred
about 12 years ago, at the time of this writing), it is reasonable to anticipate that the most
powerful applications of optical clocks still remain to be
identified and developed, much the way the microwave
atomic-clock-based GPS positioning revolution followed decades
after the demonstrations of the first Cs clocks.  Nonetheless, we
expect that the most powerful applications will lie in fields that
have traditionally exploited precision clocks, such as navigation,
communications, and signal synchronization and timing.  Given
their high performance level, these clocks will probably be used
in more extreme cases such as deep-space navigation and ultra-low
noise secure communications.  In fact, we already have seen
glimpses of such applications on the horizon.  The world's
quietest microwave signals are now generated from down-converted
optical signals \cite{Fortier2011}, and referenced optical
frequency combs are used for precision calibration of astronomical
spectrographs in searches for exo-planets \cite{Ycas2012}.
Moreover, discussions are taking place at national standards
institutions about a possible redefinition of the SI second in
terms of an optical transition (either in an ion- or neutral
atom-based system). Finally, the pre-eminence of the second, in
terms of being (by far) the most precisely measurable physical
quantity, has already enabled significant tests of fundamental
physics, with more possibilities on the horizon.

Historically, physicists have sought the best ways to use time,
perhaps our most sensitive probe of nature, to test some of our
most fundamental theories including Quantum Electrodynamics and
General Relativity.  Given the more recent challenges posed by
dark energy and dark matter, there is renewed urgency to find ways
to exploit the possibilities raised by the advances in clock
performance.  Since these clocks can offer eighteen digits of
performance at energies orders of magnitude lower than those of
particle physics experiments, we see that the clocks offer the
possibility of making measurements that would be complementary to
those of high energy physics.  The first contributions of optical
clock measurements to our understanding of fundamental physics
have come from direct comparisons between optical-clock transition
frequencies over time periods of several years.  Since different
transitions can have different dependencies on fundamental
physical constants such as $\alpha$, the fine-structure constant,
tests for drifts of transition frequencies can test for drifts in
the fundamental constants themselves (note that such tests require
small absolute uncertainties for the clock systems).  The most
sensitive test to date was a comparison between two trapped ion
clock systems, based on Hg$^{+}$ and Al$^{+}$, over the period of
a year, which constrained the present day fractional drift in
$\alpha$ to be $-1.6 \pm 2.3 \times 10^{-17}$/year
\cite{Rosenband2008}. Similarly by tracking such a ratio over
space (e.g., as the Earth orbits the sun), one can put limitations
on spatial dependencies of gravitational coupling constants.  A
comparison between Sr lattice clocks and Cs microwave clocks
spanning a year of Earth orbit was able to test local position
invariance and set the tightest limits on gravitational coupling
constants for the fine-structure constant, the electron-proton
mass ratio, and the light-quark mass \cite{Blatt2008a}.

Another way to perform tests of gravitation would be to make
precision tests of the gravitational redshift via its effect upon
the clock frequency (recall sect. 4.6).  An early rocket-based
experiment demonstrated the power of this technique to track
changes in the acceleration of gravity \cite{Vessot1980} (through
the use of a microwave H maser), and a more recent terrestrial
demonstration by Chou et al. \cite{Chou2010a} used optical clocks
to resolve altitude differences of less than 30 cm.  Eventually we
anticipate clock-based terrestrial altitude sensitivities at the 1
mm level, which would enable the prospect of using a portable
clock in conjunction with a fixed clock (maybe even one in a
microgravity environment) to map out the Earth geoid with high
sensitivity over large areas, thereby taking advantage of both the
high stability and low uncertainty of optical clocks.

Methods developed for optical clocks can be extended also to other
fields. A noticeable case is atom interferometry that has strong
conceptual and technical connections with atomic clocks (for a
review on atom interferometry see \cite{Cronin2009,Tino2013}). For
example, an atom interferometry scheme was recently proposed
\cite{Graham2013} based on the clock transition of alkaline-earth
atoms with possible applications ranging from gravimetry to the
detection of gravitational waves \cite{Tino2011}

\subsection{Clocks of the future}

Here we close this review article with a look toward the future of
atomic clock design.  To be sure, the remarkable progress (and
lack of obvious barriers to further improvements) of optical
clocks has not kept scientists from seeking what might be the next
breakthrough in atomic clock technology.  From the discussions in
sect. 2 and 3, the clear path forward would be to base new clocks
on atomic transitions with higher frequencies and smaller inherent
environmental sensitivities.  Higher frequency electronic
transitions could be accessed in highly charged atoms
\cite{Dzuba2012,Berengut2012}, perhaps excited by higher harmonics
of stabilized femtosecond frequency combs
\cite{Kandula2010,Cingoez2012}.  For reduced sensitivity to the
environment, however, we might be better served to look to nuclear
transitions, for which the inherent shielding of the nucleus could
yield a clock with extremely small uncertainty. While precision
spectroscopy of typical nuclear transitions will have to wait for
stabilized MeV sources, scientists have identified one anomalous
nuclear transition in the VUV part of the spectrum.  This
transition, the isomeric 3/2$^+$ (631) $\rightarrow$ 5/2$^+$(633)
magnetic nuclear dipole transition in $^{229}$Th$^+$ has a
transition energy of 7.8 $\pm$ 0.5 eV, which corresponds to a
wavelength around 160 (10) nm, making it, in principle, accessible
with existing spectroscopic tools (e.g., VUV frequency combs).
This transition has a quality factor, Q $\sim$ 10$^{20}$ and could
support a fractional frequency uncertainty of $\sim$ 10$^{-19}$
\cite{Campbell2012}. Several groups have proposed to excite this
transition in various ways
\cite{Peik2003,Rellegert2010,Kazakov2012}, and very recently one
group claimed the first direct de-excitation of the $^{229}$Th$^+$
transition, thereby confirming the first indirect wavelength
estimate and measuring a transition half-life of $6\pm 1$ h
\cite{Zhao2012}. While this result is still to be confirmed, it
will be interesting to see how it compares with other possible
excitation schemes including observation in a doped solid-state,
high-band-gap crystal or via NEET, that is, nuclear excitation
through an electronic transition \cite{Porsev2010}. Although the
challenges facing the construction of a clock based on this
transition in thorium are formidable, the transition offers a
sensitivity to fundamental constant variation six orders of
magnitude greater than do electronic transitions, and thus is
likely to continue to attract considerable attention as an
extremely inviting target for scientists.

\begin{acknowledgments}
N. P. acknowledges support from Italian Ministry of Research and
Education (under contract PRIN 2009 - prot.2009ZJJBLX). We also
acknowledge financial support from the European Union Seventh
Framework Programme (FP7/2007-2013 grant agreement 263500, project
``Space Optical Clocks''). C. O. acknowledges support from the
Defense Advanced Research Projects Agency Quantum Assisted Sensing
and Readout program, NASA Fundamental Physics, and NIST. We thank
N. Hinkley and N. Phillips for their careful reading of the
manuscript.


\end{acknowledgments}


\bibliography{cimento,Chris_OC_v11,newbib}

\end{document}